\documentclass[seceq]{ptptex}

\usepackage{graphicx}
\usepackage{wrapft}

\def\mbf#1{\boldsymbol{#1}}
\def\eq#1{Eq.\ (\ref{#1})}

\def\bK{{\mbf K}}

\def\be{{\mbf e}}
\def\bk{{\mbf k}}

\def\bp{{\mbf p}}
\def\bq{{\mbf q}}

\def\bt{{\mbf t}}

\def\bftau{{\mbf \tau}}

\def\CC{{\cal C}}

\def\CO{{\cal O}}

\def\CS{{\cal S}}

\def\H{{\scriptstyle \frac{1}{2}}}
\def\3H{{\scriptstyle \frac{3}{2}}}
\def\5H{{\scriptstyle \frac{5}{2}}}
\def\7H{{\scriptstyle \frac{7}{2}}}

\title{
Quark-Model Baryon-Baryon Interaction Applied to the Neutron-Deuteron
Scattering (III)
}

\subtitle{Breakup Differential Cross Sections}

\author{Yoshikazu \textsc{Fujiwara} and Kenji \textsc{Fukukawa}
}

\inst{
Department of Physics, Kyoto University, Kyoto 606-8502, Japan
}

\abst{The low-energy breakup differential cross sections 
of the neutron-deuteron ($nd$) scattering are studied by
employing the energy-independent version of the
quark-model baryon-baryon interaction fss2.
This interaction reproduces almost all the breakup differential cross sections
predicted by the meson-exchange potentials
for the neutron incident energies $E_n \leq 65$ MeV.
The space star anomaly of 13 MeV $nd$ scattering is not improved
even in our model. Some overestimation of the breakup differential 
cross sections at $E_n=22.7$ - 65 MeV implies that 
systematic studies of various breakup configurations
are necessary both experimentally and theoretically.}

\PTPindex{205}  

\begin{document}

\maketitle

\section{Introduction}

The three-nucleon ($3N$) system is a good place to study the
underlying nucleon-nucleon ($NN$) interaction, since many techniques
to solve the system exactly are well developed nowadays.\cite{PREP,bench01}
Ample experimental data are already accumulated especially for the
low-energy neutron-deuteron ($nd$) and proton-deuteron ($pd$) scattering 
and extensive studies to detect the $3N$ force have been carried out 
based on the modern meson-exchange potentials,\cite{Ku02a,Ku02b}
and more recently, on the chiral effective field theory.\cite{Ep01,Ep02} 
Most of the researches to such a direction
are concerned with higher energies than 100 MeV for the nucleon
incident energy $E_n$ in the laboratory (lab) system, 
since the $3N$ force effect is expected to be revealed 
more prominently than in the low energies.\cite{Ku02a}
On the other hand, the discrepancies of various $3N$ observables
between the theory and experiment in the $E_n \leq 65$ MeV region,
are not resolved even by the recent accurate treatment 
of the Coulomb force.\cite{De05a,De05b,Is09}
This is particularly true for the nucleon-induced deuteron breakup
processes. It is therefore worth while reexamining the $NN$ interaction
itself if the present-day realistic force is the most
appropriate one to start with.

In previous papers,\cite{ndscat1,ndscat2} referred to as I and II hereafter,
we have applied the quark-model (QM) baryon-baryon interaction fss2 to the
neutron-deuteron ($nd$) elastic scattering.
This interaction model, fss2, describes available $NN$ data
in a comparable accuracy with the modern meson-exchange 
potentials.\cite{PPNP}
By eliminating the inherent energy dependence of the resonating-group
kernel, fss2 was found to yield a nearly correct triton 
binding energy, the $S$-wave $nd$ scattering length, 
and the low-energy eigenphase shifts
without reinforcing it with the three-body force.\cite{ren,KF10,scl10} 
The predicted elastic differential cross sections
have sufficiently large cross section minima at $E_n=35\,\hbox{-}\,65$ MeV 
and $\theta_{\rm cm}=130^\circ\,\hbox{-}\,135^\circ$, 
in contrast to the predictions by the standard meson-exchange
potentials.\cite{ndscat1} 
The so-called $A_y$ puzzle at low-energies $E_n \leq 25$ MeV 
is largely improved in this model.\cite{ndscat2}
In this paper, we continue these studies by examining the $3N$ breakup
processes with various decaying kinematics for the energy
range $E_n \leq 65$ MeV.
The main motivation is to find if the quite different off-shell 
properties, originating from the strong nonlocality of the QM
baryon-baryon interaction, give some influence to the $3N$ breakup
differential cross sections.
In contrast to the elastic scattering amplitude, the breakup amplitude
covers a wide momentum region of the three-body phase space.
It will be found unfortunately that the fss2 gives predictions similar
to the meson-exchange potentials and does not improve much
the discrepancies between the theory and the experiment.

The organization of this paper is as follows.
In $\S\,2.1$, the formulation of the breakup differential cross
sections is given in terms of the direct breakup amplitude.
Various kinematical configurations for the three-body decay
are introduced in $\S\,2.2$. A minimal description of
the three-nucleon breakup kinematics is given in Appendix A.
The isospin factors for the breakup amplitudes are derived in Appendix B.
The comparison with the experimental data is presented in $\S\,3$
for energies $E_n=8,~10.3,~10.5,~13,~16,~19,~22.7$ and 65 MeV.
The difference from the predictions by meson-exchange potentials
are discussed in detail. We close this paper with a summary
of this series of investigations in $\S\,4$.

\section{Formulation}

\subsection{Breakup differential cross sections}

Following the notation of Refs.\,\citen{PREP}, I and II,
the three-body breakup amplitude is given by
\begin{eqnarray}
U_0|\phi \rangle =(1+P)T|\phi \rangle=(1+P) t \widehat{Q} |\phi \rangle\ .
\label{fm1}
\end{eqnarray}
In order to derive the breakup differential cross sections, we start from
the Fermi's golden rule
\begin{eqnarray}
d\,N=\frac{2\pi}{\hbar} |\langle \bp \bq|U_0|\phi \rangle|^2
\int^\infty_0 p^2\,d\,p~\delta(E-E_{pq})~q^2\,d\,q
~d\,\widehat{\bp}~d\,\widehat{\bq}\ ,
\label{fm2}
\end{eqnarray}
and divide it by the incident flux $j=(3 \hbar q_0/2M)/(2\pi)^3$.
Here, $E_{pq}=(\hbar^2/M)(p^2+3 q^2/4)$, $M$ is the nucleon mass,
and $q_0$ is the incident momentum related to the
energy, $E=(3\hbar^2/4M){q_0}^2+\varepsilon_d$, in the
center-of-mass (cm)~system. We obtain in the cm system
\begin{eqnarray}
\frac{d^5\,\sigma}{d\,\widehat{\bp}~d\,\widehat{\bq}~d\,q}
& = & \frac{1}{j}\frac{d\,N}{d\,\widehat{\bp}~d\,\widehat{\bq}~d\,q}
=(2\pi)^4 \frac{2M}{3\hbar^2}\frac{1}{q_0}
\int^\infty_0 p^2\,d\,p~\delta(E-E_{pq})~q^2
~|\langle \bp \bq|U_0|\phi \rangle|^2\nonumber \\
& = & (2\pi)^4 \left(\frac{2M}{3\hbar^2}\right)^2
\frac{3}{4} \frac{p_0 q^2}{q_0}
\sum_\Gamma |\langle \bp \bq \Gamma|U_0|\phi \rangle_0 |^2\ ,
\label{fm3}
\end{eqnarray}
where Eq.\,I(2.88)\footnote{In the following, we cite equations
of the previous paper I (or II), with adding I (or II) in front
of the equation number.} is used to perform the $p$-integral.
In \eq{fm3}, $\Gamma=\Gamma_\sigma \Gamma_\tau$ is
the spin-isospin quantum numbers in the $LS$-coupling scheme
and the subscript 0 in the matrix element implies the on-shell
condition $|\bp|=p_0=\sqrt{(3/4)({q_M}^2-q^2)}$ with $q_M=\sqrt{{q_0}^2
-{\kappa_d}^2}$. Here, $\kappa_d$ is related to the deuteron binding
energy $|\varepsilon_d|$ through $|\varepsilon_d|=(3\hbar^2/4M){\kappa_d}^2$.
In this paper, we use the notation 
$\Gamma_\sigma=(s \H)SS_z$ and $\Gamma_\tau=(t \H)\H T_z$ to
specify the quantum numbers in the $LS$-coupling scheme: i.e.,
\begin{eqnarray}
& & |\bp, \bq; 1 2 3 \rangle
=\sum_\gamma |p, q, \gamma \rangle
~\langle \gamma|\widehat{\bp}, \widehat{\bq}; 1 2 3 \rangle\ ,\nonumber \\
& & \langle \widehat{\bp}, \widehat{\bq}; 1 2 3|\gamma \rangle
=\left[ Y_{(\lambda \ell)L}(\widehat{\bp}, \widehat{\bq})
~\xi_{\Gamma_\sigma}(12,3)\right]_{J J_z}
~\eta_{\Gamma_\tau}(12,3)\ ,
\label{fm4}
\end{eqnarray}
with $\gamma=[(\lambda \ell)L \Gamma_\sigma]J J_z; \Gamma_\tau$,
and $\xi_{\Gamma_\sigma}$ and $\eta_{\Gamma_\tau}$ being
the three-particle spin and isospin wave functions, respectively.
In \eq{fm4}, $Y_{(\lambda \ell)LM}(\widehat{\bp}, \widehat{\bq})
=[Y_\lambda(\widehat{\bp}) Y_\ell(\widehat{\bq})]_{LM}$ are
the angular functions.
For the initial state, we use channel-spin representation 
as for the elastic scattering. We take the sum of \eq{fm3}
over all the spin and isospin quantum numbers and divide by the initial
spin multiplicity 6. The selection of the detected particles in the
final state is controlled by the isospin projection operator $\CO_\tau$,
the explicit form of which will be specified later.
The breakup differential cross sections of the $nd$ scattering are therefore
calculated from
\begin{eqnarray}
\frac{d^5\,\sigma}{d\,\widehat{\bp}~d\,\widehat{\bq}~d\,q}
=(2\pi)^4 \left(\frac{2M}{3\hbar^2}\right)^2
\frac{3}{4} \frac{p_0 q^2}{q_0}\frac{1}{6} \sum_\Gamma \sum_{S_c S_{cz}}
|\langle \bp \bq \Gamma|\CO_\tau (1+P) T|\phi_{\bq_0}; S_c S_{cz}
\rangle_0 |^2\ .\nonumber \\
\label{fm5}
\end{eqnarray}

Let us first consider the spin-isospin sum
$I=\sum_\Gamma |\langle \bp \bq \Gamma|(1+P) f \rangle |^2$
by neglecting the initial spin quantum numbers for the time being.
The effect of the permutation ${P_{(123)}}^\alpha$ in
$(1+P)=\sum^3_{\alpha=1}{P_{(123)}}^\alpha$ is
defined by 
\begin{eqnarray}
\langle \bp \bq|{P_{(123)}}^\alpha f\rangle
\equiv {P_{(123)}}^\alpha f(\bp, \bq)=f(\bp_\alpha, \bq_\alpha)\ ,
\label{fm6}
\end{eqnarray}
if the function $f(\bp, \bq)$ does not contain the spin-isospin
degree of freedom. In fact, we should use
\begin{eqnarray}
\langle \bp \bq \Gamma|{P_{(123)}}^\alpha f\rangle
=\langle \Gamma |{P^{(\sigma \tau)}_{(123)}}^\alpha 
f(\bp_\alpha, \bq_\alpha)\rangle\ ,
\label{fm7}
\end{eqnarray}
where $P^{(\sigma \tau)}_{(123)}$ is the permutation operator
in the spin-isospin space and the bra-ket notation
is used for the spin-isospin degree of freedom.
Using these notations and the completeness relationship
in the spin-isospin space, $\sum_\Gamma |\Gamma\rangle \langle \Gamma|=1$,
we find
\begin{eqnarray}
I=\sum^{3}_{\alpha, \beta=1}
\langle f(\bp_\alpha, \bq_\alpha)|
{P^{(\sigma \tau)}_{(123)}}^{3-\alpha}
{P^{(\sigma \tau)}_{(123)}}^\beta |f(\bp_\beta, \bq_\beta)\rangle\ .
\label{fm8}
\end{eqnarray}
Here we separate the $\alpha$, $\beta$ sum into the diagonal
part ($\alpha=\beta$) and the off-diagonal part ($\alpha \neq \beta$).
In the off-diagonal part, we specify $\alpha$ and $\beta$ by
the cyclic permutations of (123) 
($(\alpha \beta \gamma)$=(123)-cyclic).
For these terms, the $\alpha$-$\beta$ term and the $\beta$-$\alpha$ term
are complex conjugate to each other. Thus we obtain
\begin{eqnarray}
I=\sum^{3}_{\alpha=1}
\langle f(\bp_\alpha, \bq_\alpha)|f(\bp_\alpha, \bq_\alpha)\rangle
+2 \sum^\prime_{(\alpha \beta \gamma)}
{\rm Re}\,\langle f(\bp_\alpha, \bq_\alpha)|P^{(\sigma \tau)}_{(123)}
|f(\bp_\beta, \bq_\beta)\rangle\ ,
\label{fm9}
\end{eqnarray}
where $\sum^\prime$ implies the sum over the three cyclic permutations
of $(\alpha \beta \gamma)=(123)$.

The extension to $I=\sum_\Gamma |\langle \bp \bq \Gamma|\CO_\tau
(1+P) f \rangle |^2$, incorporating the isospin projection operator
$\CO_\tau$, is rather easy. Here, $\CO_\tau$ is specified as
\begin{eqnarray}
\CO^{pp} & = & \frac{1+\tau_z(1)}{2} \frac{1+\tau_z(2)}{2}\ \ ,\qquad
\CO^{nn} = \frac{1-\tau_z(1)}{2} \frac{1-\tau_z(2)}{2}\ ,\nonumber \\
\CO^{pn} & = & \frac{1+\tau_z(1)}{2} \frac{1-\tau_z(2)}{2}\ \ ,\qquad
\CO^{np} = \frac{1-\tau_z(1)}{2} \frac{1+\tau_z(2)}{2}\ ,\
\label{fm10}
\end{eqnarray}
depending on the species of particles 1 and 2 detected.
We use ${\CO_\tau}^2=\CO_\tau$ and the notation
$\langle \bp \bq \Gamma|f \rangle=f_\Gamma(\bp, \bq)$. Then, by
defining
\begin{eqnarray}
\CO^{\alpha \beta}_\tau=\left(P^{\tau}_{(123)}\right)^{3-\alpha} \CO_\tau
\left(P^{\tau}_{(123)}\right)^\beta
\qquad (={\CO^{\beta \alpha}_\tau}^\dagger)\ ,
\label{fm11}
\end{eqnarray}
we obtain
\begin{eqnarray}
I & = & \sum^3_{\alpha=1} \sum_{\widetilde{\Gamma}, \Gamma}
f^*_{\widetilde{\Gamma}}(\bp_\alpha, \bq_\alpha)
\,\langle \widetilde{\Gamma}|\CO^{\alpha \alpha}_\tau |\Gamma \rangle
\,f_\Gamma(\bp_\alpha, \bq_\alpha)\nonumber \\
& & + 2 \sum^\prime_{(\alpha \beta \gamma)} \sum_{\widetilde{\Gamma}, \Gamma}
{\rm Re}\left\{
f^*_{\widetilde{\Gamma}}(\bp_\alpha, \bq_\alpha)
\,\langle \widetilde{\Gamma}|P^\sigma_{(123)} 
\CO^{\alpha \beta}_\tau |\Gamma \rangle
\,f_\Gamma(\bp_\beta, \bq_\beta)\right\}\ .
\label{fm12}
\end{eqnarray}
The spin-isospin factors in \eq{fm12} are calculated by separating
the spin-isospin state $|\Gamma \rangle$ to the spin and isospin parts,
$|\Gamma \rangle=|\Gamma_\sigma \rangle |\Gamma_\tau \rangle$.
We find
\begin{eqnarray}
& & \langle \widetilde{\Gamma}|\CO^{\alpha \alpha}_\tau|\Gamma \rangle
=\delta_{\widetilde{S}, S}\,\delta_{\widetilde{s}, s}
~\langle \widetilde{\Gamma_\tau}|\CO^{\alpha \alpha}_\tau|\Gamma_\tau \rangle
\ ,\nonumber \\
& & \langle \widetilde{\Gamma}|P^\sigma_{(123)} 
\CO^{\alpha \beta}_\tau |\Gamma \rangle
=\delta_{\widetilde{S}, S}\,(-1)^{1+s} X^S_{\widetilde{s},s}
~\langle \widetilde{\Gamma_\tau}|\CO^{\alpha \beta}_\tau|\Gamma_\tau \rangle
\ ,
\label{fm13}
\end{eqnarray}
where $\widetilde{\Gamma}_\sigma=(\widetilde{s} \H)\widetilde{S}
\widetilde{S}_z$ and $\widetilde{\Gamma}_\tau=(\widetilde{t} \H)\H T_z$,
and \eq{b1} is used for the spin part. We also extend the definition in
\eq{b1} for the spin part to the isospin part as in \eq{b2}.
Using the definition of $X^{\tau (\alpha \beta)}_{\widetilde{t},t}$
in \eq{b2}, we can write the matrix elements in \eq{fm13} as
\begin{eqnarray}
& & \langle \widetilde{\Gamma}|\CO^{\alpha \alpha}_\tau|\Gamma \rangle
=\delta_{\widetilde{S}, S}\,\delta_{\widetilde{s}, s}
~X^{\tau (\alpha \alpha)}_{\widetilde{t},t}\ ,\nonumber \\
& & \langle \widetilde{\Gamma}|P^\sigma_{(123)} 
\CO^{\alpha \beta}_\tau |\Gamma \rangle
=\delta_{\widetilde{S}, S}\,(-1)^{1+s} X^S_{\widetilde{s},s}
~(-1)^{1+t} X^{\tau (\alpha \beta)}_{\widetilde{t},t}\ ,
\label{fm14}
\end{eqnarray}
for $(\alpha \beta \gamma)=$ a cyclic permutation of (123).
Thus we find
\begin{eqnarray}
I & = & \sum^3_{\alpha=1} \sum_{\widetilde{\Gamma}, \Gamma}
\delta_{\widetilde{S}, S}\,\delta_{\widetilde{s},s}
\,X^{\tau (\alpha \alpha)}_{\widetilde{t},t}
~f^*_{\widetilde{\Gamma}}(\bp_\alpha, \bq_\alpha)
\,f_\Gamma(\bp_\alpha, \bq_\alpha)\nonumber \\
& & + \sum^\prime_{(\alpha \beta \gamma)} \sum_{\widetilde{\Gamma}, \Gamma}
\delta_{\widetilde{S}, S}\,(-2) X^S_{\widetilde{s},s}
\,X^{\tau (\alpha \beta)}_{\widetilde{t},t}
~{\rm Re}\left\{
f^*_{\widetilde{\Gamma}}(\bp_\alpha, \bq_\alpha)
\,f_\Gamma(-\bp_\beta, \bq_\beta)\right\}\ .
\label{fm15}
\end{eqnarray}
where the generalized Pauli principle $(-1)^{s+t+\lambda}=-1$ is
used for the two-nucleon part of $f_\Gamma(-\bp_\beta, \bq_\beta)$.
The isospin factors $X^{\tau (\alpha \beta)}_{\widetilde{t},t}$ are
explicitly given in Appendix B.

We assign the direct breakup amplitude to $f_\Gamma(\bp, \bq)$ in
\eq{fm15} through
\begin{eqnarray}
f_{\Gamma, S_c S_{cz}}(\bp, \bq)
= -(2\pi)^2 \left(\frac{2M}{3\hbar^2}\right)
\langle \bp \bq \Gamma|T|\phi_{\bq_0}; S_c S_{cz} \rangle_0\ ,
\label{fm16}
\end{eqnarray}
with $T=t \widehat{Q}$. The partial wave decomposition is given by
\begin{eqnarray}
f_{\Gamma, S_c S_{cz}}(\bp, \bq)
& = & (4\pi) \sum^\prime_{\gamma, \ell^\prime, J J_z}
f^{({\rm db})J}_{\gamma, (\ell^\prime S_c)}(q)
\sum_M \langle L M S S_z|J J_z\rangle
\,Y_{(\lambda \ell)LM}(\widehat{\bp}, \widehat{\bq})\nonumber \\
& & \times \sum_{m^\prime}
\langle \ell^\prime m^\prime S_c S_{cz}|J J_z\rangle
\,Y^*_{\ell^\prime m^\prime}(\widehat{\bq}_0)\ ,
\label{fm17}
\end{eqnarray}
where the prime on the sum implies that we take all the orbital
angular momentum sum for the $LS$ coupling scheme of $\gamma$; 
i.e., the sum over only $(\lambda \ell) L$ 
with $\gamma=[(\lambda \ell) L \Gamma_\sigma]J J_z;\Gamma_\tau$.
(Note the extra $(4\pi)$ factor for the scattering amplitude.)
It is convenient to define 
\begin{eqnarray}
\widehat{Q}^{(\ell S_c)J}_{i \mu \gamma}
=\sum_{(\ell^\prime S^\prime_c)}
\widetilde{Q}^{(\ell^\prime S^\prime_c)J}_{i \mu \gamma}
\,f^J_{(\ell^\prime S^\prime_c)(\ell S_c)}\ ,
\label{fm18}
\end{eqnarray}
by the solutions, $\widetilde{Q}_{i \mu \gamma}
={p_i}^2 \omega_i q_\mu \sqrt{\omega_\mu} \langle p_i, q_\mu, \gamma
|\widetilde{Q}|\psi\rangle$, of the basic AGS equation in Eq.\,I(2.59),
and the elastic scattering
amplitude $f^J_{(\ell^\prime S^\prime_c)(\ell S_c)}$ in Eq.\,I(2.92). 
The partial-wave amplitude for the direct breakup,
$f^{({\rm db})J}_{\gamma, (\ell^\prime S_c)}(q)$,\footnote{This
corresponds to the direct term
of $f^{({\rm br})J}_{\gamma, (\ell^\prime S_c)}(q)$ in Eq.\,I(2.92).} is
expressed as
\begin{eqnarray}
f^{({\rm db})J}_{\gamma, (\ell^\prime S_c)}(q)
=\sum_i \langle p_0|t_\gamma (\hbar^2 {p_0}^2/M)|p_i \rangle
\sum_\mu S_\mu(q) \frac{1}{q_\mu \sqrt{\omega_\mu}}
\widehat{Q}^{(\ell^\prime S_c)J}_{i \mu \gamma}\ ,
\label{fm19}
\end{eqnarray}
if we use the the spline interpolation for a particular value of $q$.
For the practical calculations, it is convenient
to adopt a particular coordinate system
with $\widehat{\bq}_0=\be_z$ in \eq{fm17}.
Then the basic direct breakup amplitude in the spin-isospin space
is calculated from
\begin{eqnarray}
f_{\Gamma, S_c S_{cz}}(\bp, \bq)
& = & \sqrt{4\pi} \sum^\prime_{\gamma, \ell^\prime, J}
f^{({\rm db})J}_{\gamma, (\ell^\prime S_c)}(q)
\langle L\,(S_{cz}-S_z)\, S S_z|J S_{cz}\rangle
\nonumber \\
& & \times \widehat{\ell}^\prime
\langle \ell^\prime 0 S_c S_{cz}|J S_{cz}\rangle
\,Y_{(\lambda \ell)L\,(S_{cz}-S_z)}(\widehat{\bp}, \widehat{\bq})\ ,
\label{fm20}
\end{eqnarray}
with $\widehat{\ell}^\prime=\sqrt{2\ell^\prime+1}$.
The differential cross sections in the cm system are given by
\begin{eqnarray}
& & \frac{d^5\,\sigma}{d\,\widehat{\bp}~d\,\widehat{\bq}~d\,q}
=\frac{3}{4} \frac{p_0 q^2}{q_0}\frac{1}{6} \sum_{S_c, S_{cz}}
\left[ \sum^3_{\alpha=1} \sum_{\widetilde{\Gamma}, \Gamma}
\delta_{\widetilde{S}, S}\,\delta_{\widetilde{s},s}
X^{\tau (\alpha \alpha)}_{\widetilde{t},t}
f^*_{\widetilde{\Gamma},S_c S_{cz}}(\bp_\alpha, \bq_\alpha)
f_{\Gamma,S_c S_{cz}}(\bp_\alpha, \bq_\alpha) \right. \nonumber \\
& & \left. + \sum^\prime_{(\alpha \beta \gamma)} 
\sum_{\widetilde{\Gamma}, \Gamma}
\delta_{\widetilde{S}, S}\,(-2) X^S_{\widetilde{s},s}
\,X^{\tau (\alpha \beta)}_{\widetilde{t},t}
~{\rm Re}\left\{
f^*_{\widetilde{\Gamma},S_c S_{cz}}(\bp_\alpha, \bq_\alpha)
f_{\Gamma,S_c S_{cz}}(-\bp_\beta, \bq_\beta)\right\} \right]\ .
\label{fm21}
\end{eqnarray}
The breakup differential cross sections in the lab system, 
$(d^5\,\sigma/d\,\widehat{\bk}_1\,d\,\widehat{\bk}_2\,d\,S)$,
are specified by the two directions $\widehat{\bk}_1$, $\widehat{\bk}_2$,
and the energy $S$ measured along the locus of the $E_1$-$E_2$ energy plane.
They are obtained from \eq{fm21} by a simple change
of the phase space factor \cite{PREP}
\begin{eqnarray}
\ \hspace{-10mm} 
\rho_{\rm cm} & = & \frac{3}{4} \frac{p_0 q^2}{q_0} \quad \longrightarrow
\quad \rho_{\rm lab} = \frac{M}{\hbar^2}
\frac{3k_1 k_2}{2q_0} 
\left[\left(2-\frac{k_{\rm lab}}{k_2}\cos\,\theta_2
+\frac{k_1}{k_2}\cos\,\theta_{12}\right)^2 \right. \nonumber \\
& & \ \hspace{36mm} \left. +\left(2-\frac{k_{\rm lab}}{k_1}\cos\,\theta_1
+\frac{k_2}{k_1}\cos\,\theta_{12}\right)^2\right]^{-1/2}\ .
\label{fm22}
\end{eqnarray}
The details of the three-body kinematics are summarized in Appendix A.

\subsection{Three-nucleon breakup kinematics}

Assuming that we detect two outgoing particles 1 and 2,
the breakup differential cross sections are specified
by two polar angles $\theta_1$, $\theta_2$,
and a difference of azimuthal angles $\phi_{12}=\phi_1-\phi_2$,
in addition to the energy $S$ determined from the kinematical curve
($S$-curve) in the $E_1$ and $E_2$ plane.
The starting value of the arc length $S=0$ is quite arbitrary and 
we follow the convention by the experimental setup. 
In Appendix A, we have parametrized the locus in the $k_1$-$k_2$ plane
with an angle $\theta$, and the starting point $S=0$ is uniquely
determined by specifying $\theta_{\rm st}$.
We also assume that the beam direction of the incoming particle
is the $z$ axis and set $\phi_1=\pi$, \cite{Oh65} which determines
the $x$-axis.
 
It is customary to classify the three-body breakup kinematics
into the following six categories based on the classical
(or geometrical) argument:\cite{PREP,Ku02b}

\bigskip

\begin{enumerate}
\item[1.] The quasi-free scattering (QFS): one of the nucleons
in the final state is at rest in the lab system.
\item[2.] The final-state interaction (FSI): the relative momentum
of the two outgoing nucleons is equal to zero.
\item[3.] The collinear configuration (COLL): one of the outgoing nucleons 
is at rest in the cm system, and the other two have momenta back to back.
\item[4.] The symmetric space star configuration (SST): the three nucleons
emerge from the reaction point in the cm system, keeping equal momenta
with $120^\circ$ relative to each other and perpendicular to the beam 
direction (on the $x$-$y$ plane in the cm system).
\item[5.] The coplanar star configuration (CST): the same with the
symmetric space star configuration, but with the three momenta 
lying on the reaction plane.
\item[6.] The non-standard configuration (NS): the other 
non-specific configurations.
\end{enumerate}

\medskip 

\noindent
These are mathematically distinguished by particular values
of the lab momentum $\bk_\alpha$, the relative momenta $\bp_\alpha$
and $\bq_\alpha$, etc., and provide a rough guidance to which
portion of the two-nucleon $t$-matrix is responsible at the final stage
of the reaction, according to the structure of the direct breakup
amplitudes in \eq{fm19}. For example, Ref.~\citen{PREP} argues that
the first Born term of the QFS with $\bk_\alpha=0$ is approximately 
a product of an on-shell two-nucleon $t$-matrix 
and the deuteron state at zero momentum. 
It is known that the $3N$ force effect is rather small
for the QF condition.\cite{Ku02b} 
On the other hand, the collinear configurations with $\bq_\alpha=0$ are 
expected to be sensitive to the $3N$ force intuitively, 
and the experimental study by Correll et al.\cite{Co87} was carried out 
to study the effect of the $3N$ force intensively
in the reaction $\hbox{H}(\vec{d},2p)n$ around these configurations
at the deuteron incident energy $E_d=16$ MeV.
Furthermore, the FSI is characterized by $\bp_\alpha=0$, for which
the half off-shell $t$-matrix in \eq{fm19} generates a large peak
corresponding to the $\hbox{}^1S_0$ positive-energy bound state
near the zero-energy threshold. Although the height of the peak 
is influenced by the background amplitude $\widehat{Q}_{i \mu \gamma}$,
the FSI peak is usually nicely reproduced. The disagreement with the 
$nd$ data is reported at the early stage for the SST configuration,
which is still an unsolved problem called space star anomaly.\cite{Se96}  
It should be noted, however, that the disagreement between
the theory and experiment is also seen in some other
coplanar star and non-standard configurations, for which off-shell
properties of the two-nucleon $t$-matrix is expected to play
a role in different ways. We will examine these case by case 
in the next section. 

\section{Results and discussion}

\subsection{$\hbox{H}(\vec{d},2p)n$ reaction at $E_d=16$ MeV}

It is important to take enough number of discretization points 
and partial waves to get well converged results,
especially for the breakup differential cross sections.
In this paper, we take $n_1$-$n_2$-$n_3$=6-6-5 in the notation
introduced in $\S\,3.1$ of I, unless otherwise specified.
This means that the three intervals, $[0, q_M/2], [q_M/2, (\sqrt{3}/2)q_M]$
and $[(\sqrt{3}/2)q_M, q_M]$, are discretized by the six-point Gauss
Legendre quadrature for each, and the total number of the discretization 
points for $q$ is 35 ($=6 \times 3 + 6 \times 2 + 5$).
The three-body model space is truncated
by the two-nucleon angular momentum $I_{\rm max}$, which depends on the
incident energy of the neutron. We find that $I_{\rm max}=3$ is
large enough for $E_n \leq 19$ MeV. 
The Coulomb force is entirely neglected in the present calculations.

\begin{figure}[htb]
\begin{center}
\begin{minipage}{0.48\textwidth}
\includegraphics[angle=0,width=55mm]
{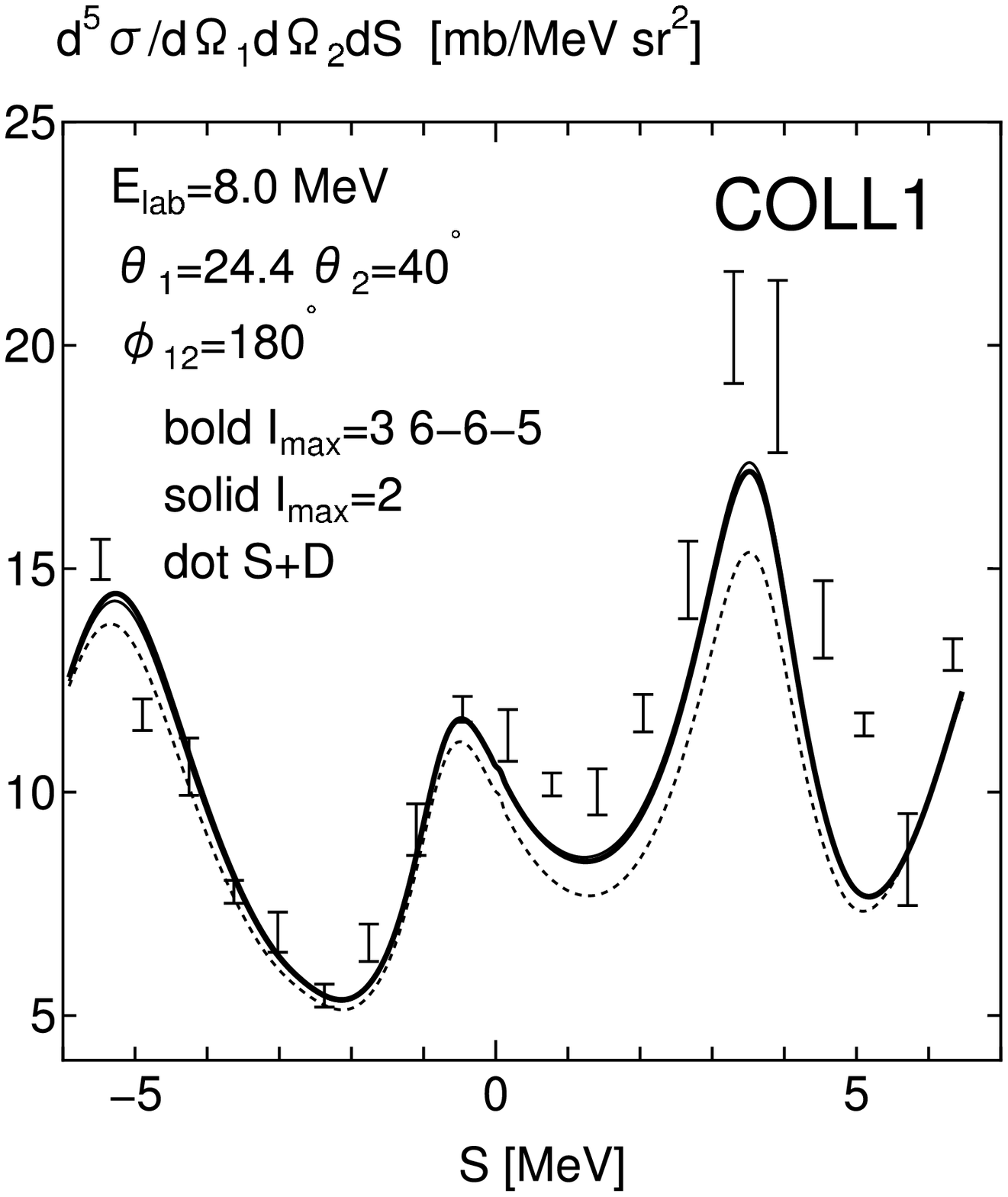}

\includegraphics[angle=0,width=55mm]
{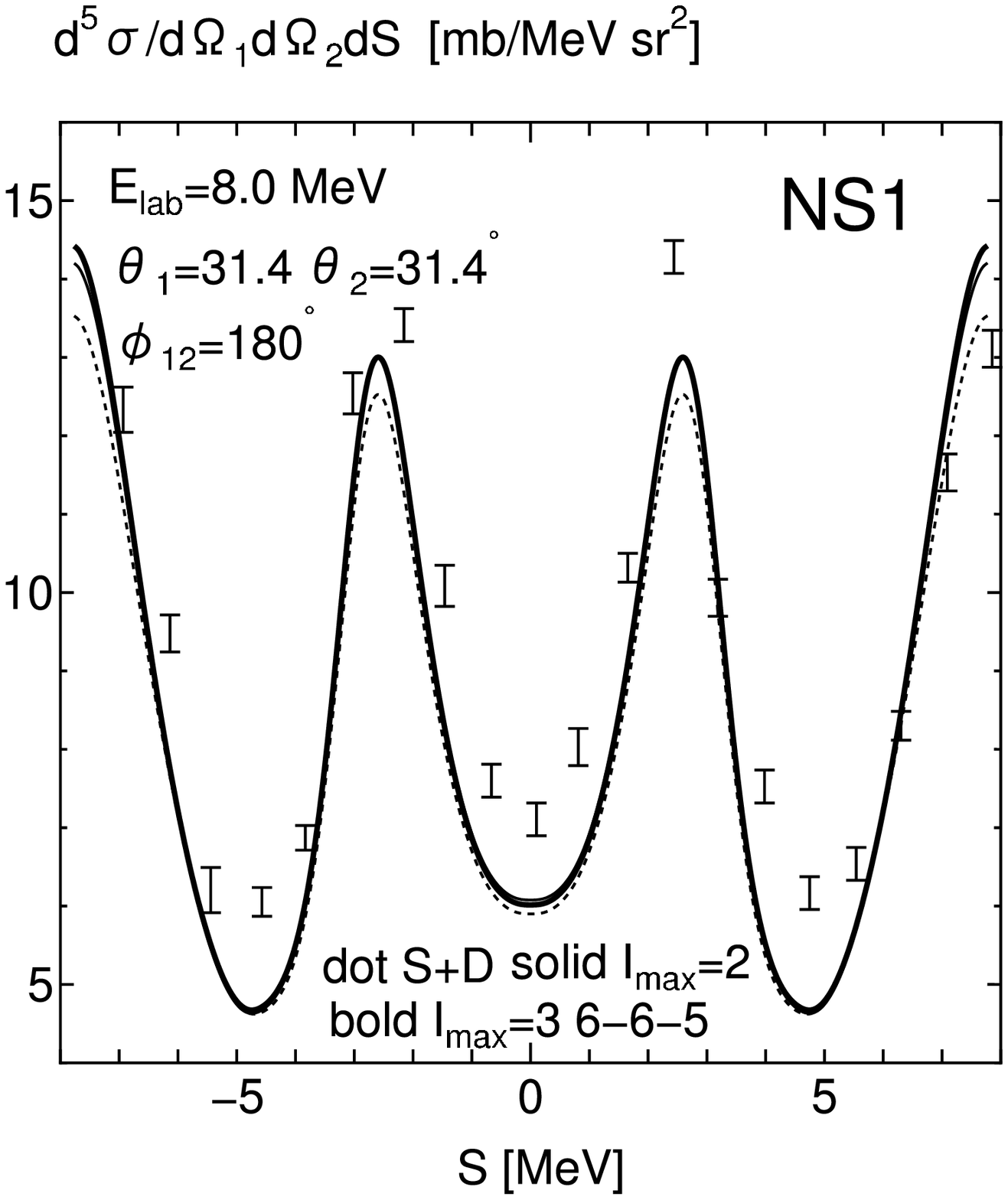}
\end{minipage}~%
\hfill~%
\begin{minipage}{0.48\textwidth}
\includegraphics[angle=0,width=55mm]
{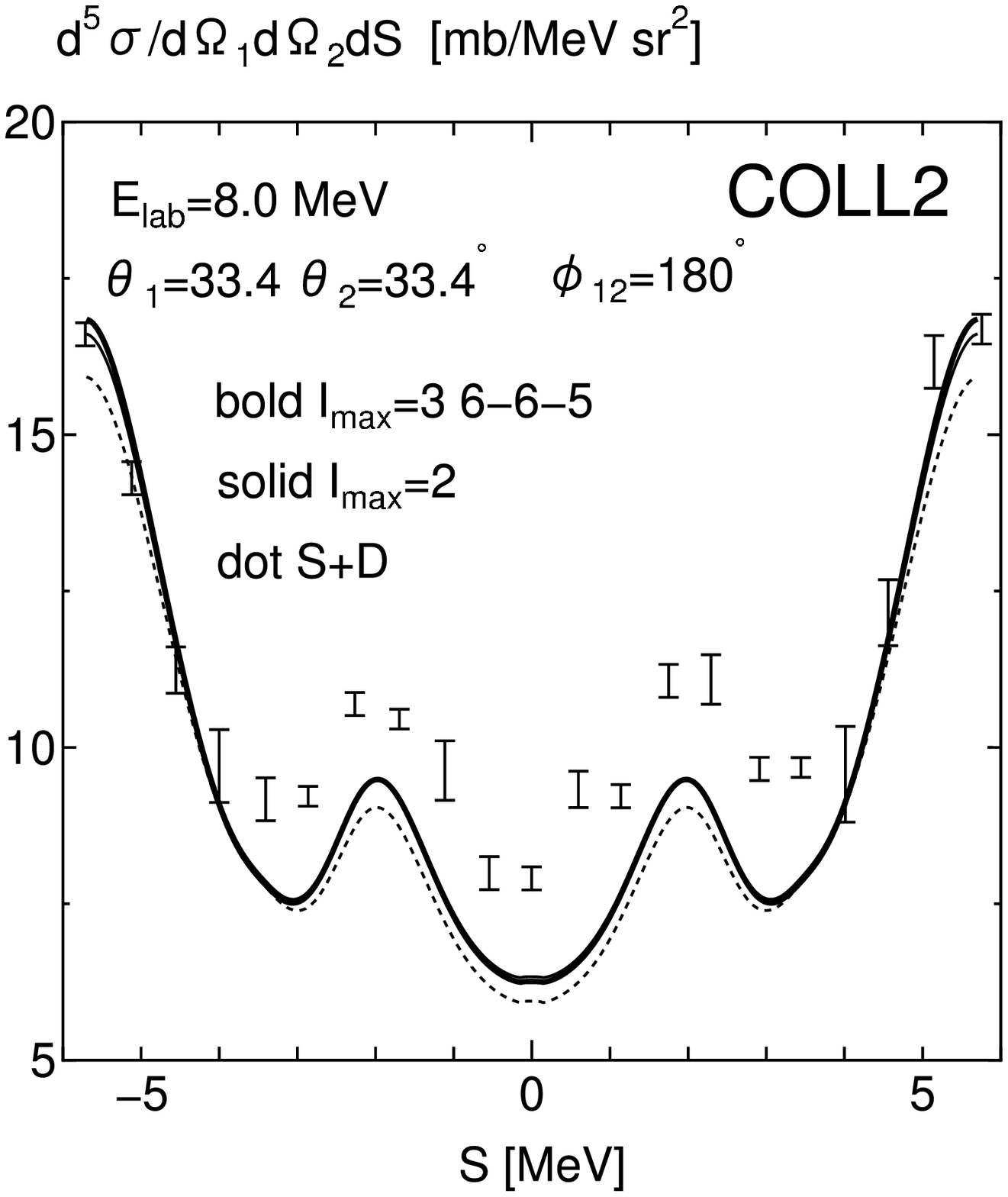}

\includegraphics[angle=0,width=55mm]
{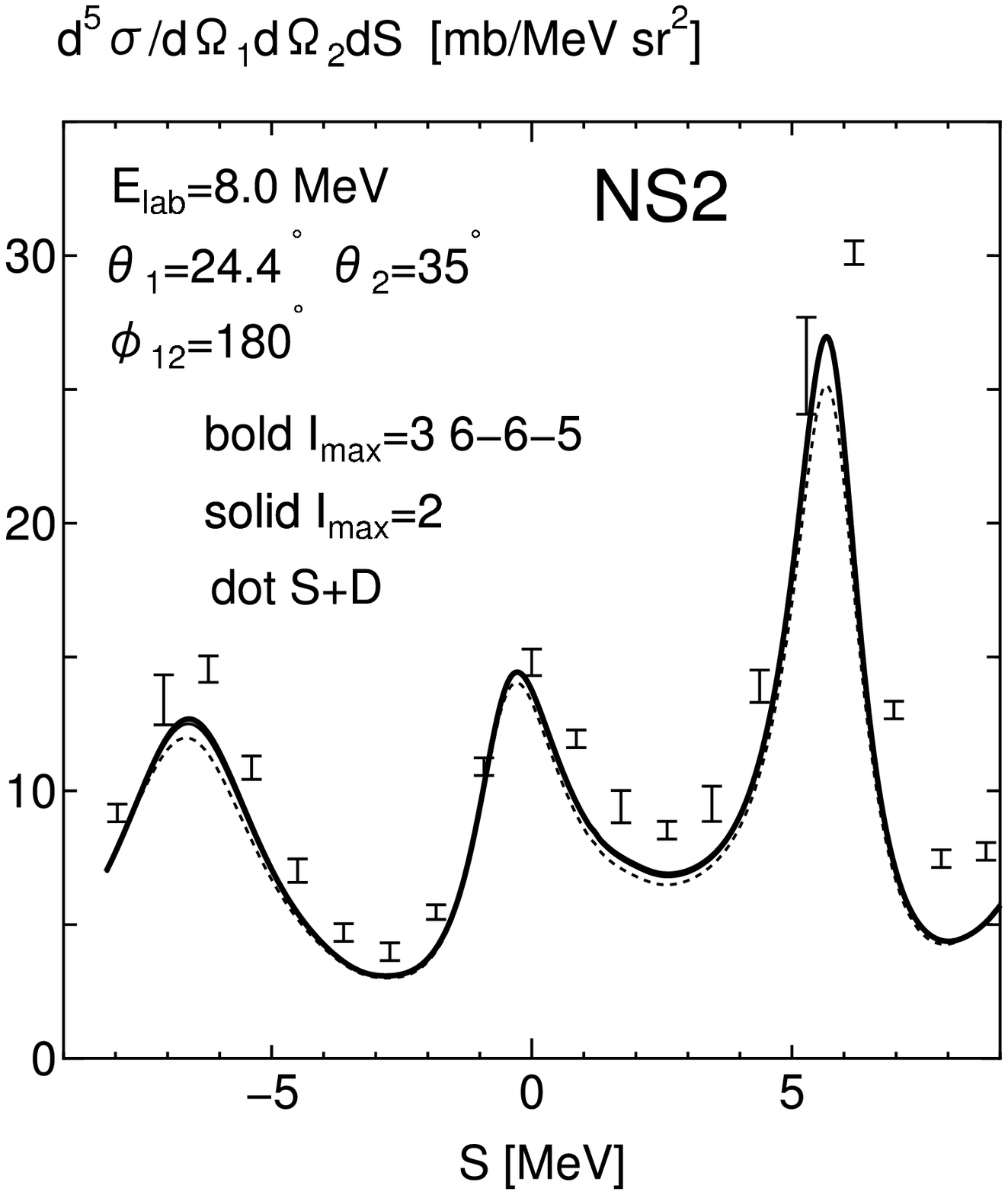}
\end{minipage}
\end{center}
\caption{
Breakup differential cross sections
for the reaction $d(n,2n)p$ with $E_n=8$ MeV.
The experimental data are the deuteron incident reaction,
$\hbox{H}(\vec{d},2p)n$, with the energy $E_d=16$ MeV.\cite{Co87} 
}
\label{fig1}
\end{figure}

We first investigate the Correll et al.'s experiment;\cite{Co87} i.e.,
$\hbox{H}(\vec{d},2p)n$ reaction with the deuteron incident 
energy $E_d=16$ MeV. 
This corresponds to the nucleon-induced breakup reaction
of the nucleon incident energy of 8 MeV.
We generate the direct breakup amplitude
for the deuteron incident reaction by adding an extra phase factor
$(-1)^{\ell^\prime}$ to each term of \eq{fm20}, corresponding to the
change from $\bq_0$ to $-\bq_0$ in \eq{fm17}.\footnote{We appreciate 
Professor H. Wita\l a for informing us about this phase change.}
The decay kinematics for the deuteron incident reaction is
discussed in Appendix A.
The breakup differential cross sections for the $d(n,2n)p$ reaction
with $E_n=8$ MeV are compared with the Correll et al.'s data
in Fig.\,\ref{fig1}, with respect to two collinear (COLL1, COLL2) and 
two non-standard (NS1, NS2) configurations.
The starting point $S=0$ is chosen as the collinear points or the
nearest point, as is discussed in Appendix A.
The dashed curve, the solid curve, and the bold solid curve 
correspond to the S+D (i.e., $\hbox{}^3S_1+\hbox{}^3D_1$ and $\hbox{}^1S_0$
only), $I_{\rm max}=2$, and $I_{\rm max}=3$ cases, respectively.
The solid curves almost overlap with bold curves and $I_{\rm max}=2$ is
actually good enough at this energy.
We also see that even the restriction to the S+D model space is not too bad.
If we compare our results with meson-exchange predictions 
in Ref.~\citen{PREP}, we find that they are very similar to each other.
The calculated values are somewhat too small especially in COLL1, COLL2
and NS1, although to less extent for the meson-exchange predictions.
There is no exact collinear point in the case of NS2, and the best agreement
with experiment is obtained in this case.

\subsection{$d(n,2n)p$ reaction at $E_n=10.3$ MeV}

\begin{figure}[htb]
\begin{center}
\begin{minipage}{0.48\textwidth}
\includegraphics[angle=0,width=55mm]
{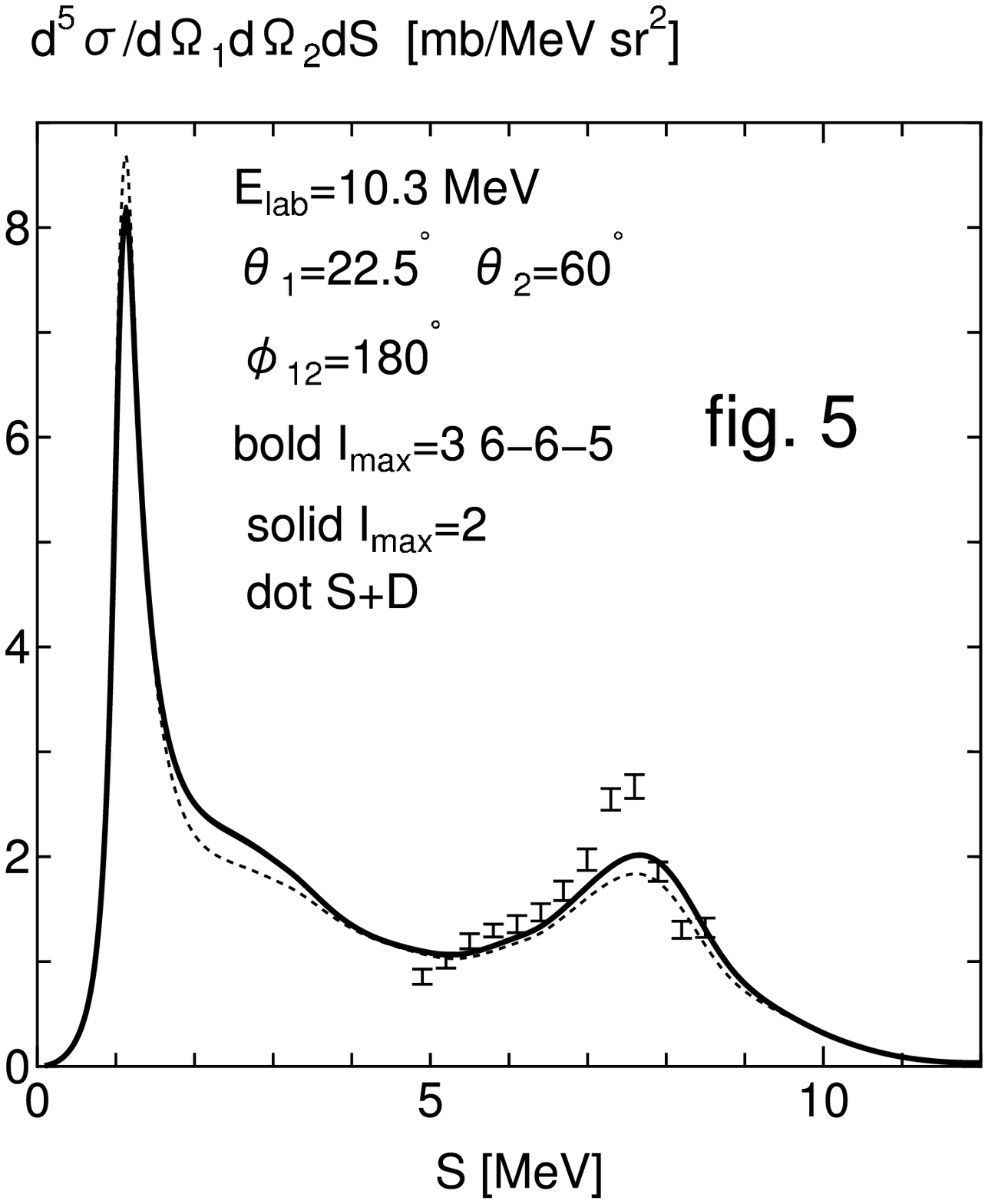}

\includegraphics[angle=0,width=55mm]
{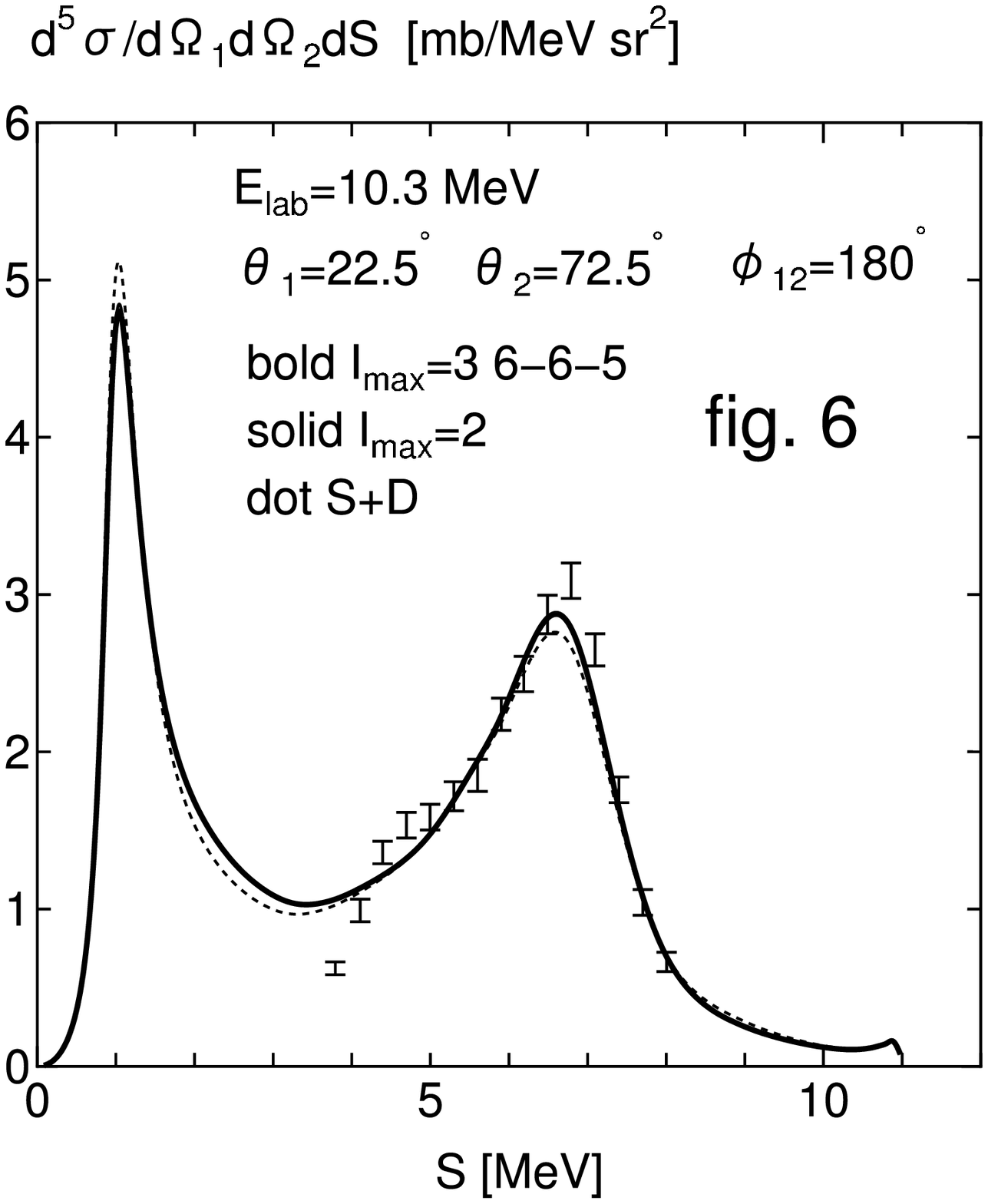}

\includegraphics[angle=0,width=55mm]
{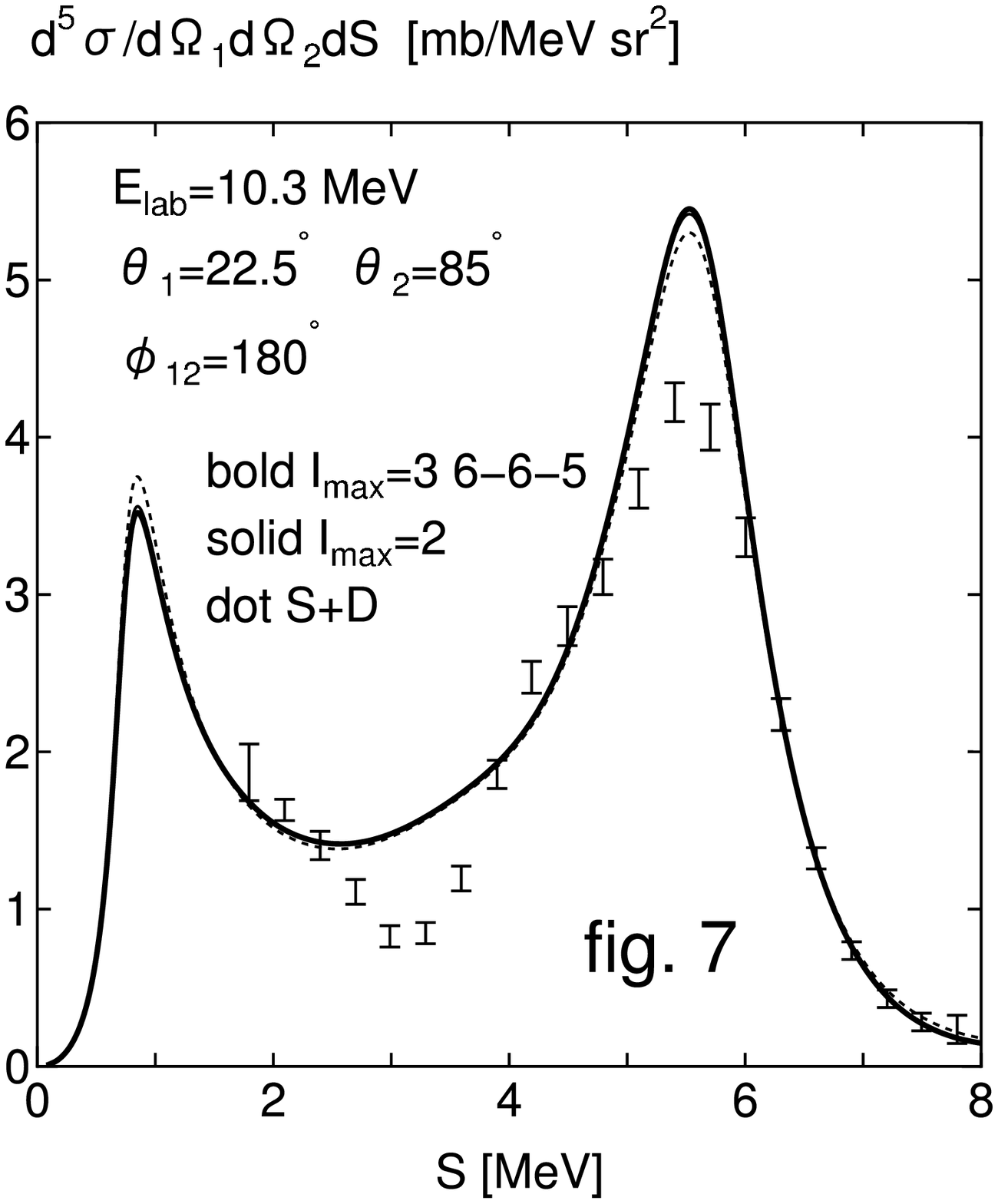}
\end{minipage}~%
\hfill~%
\begin{minipage}{0.48\textwidth}
\includegraphics[angle=0,width=55mm]
{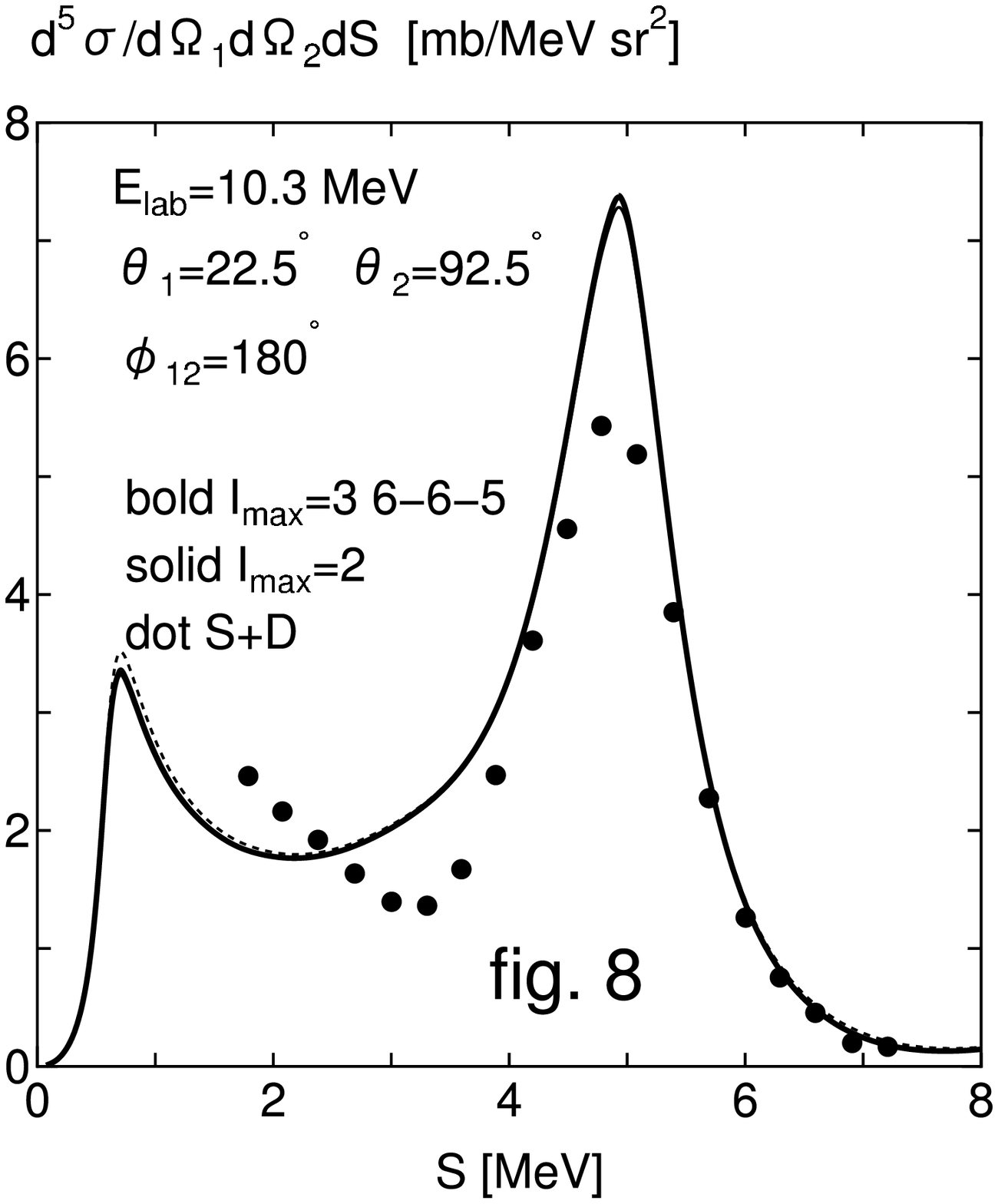}

\includegraphics[angle=0,width=55mm]
{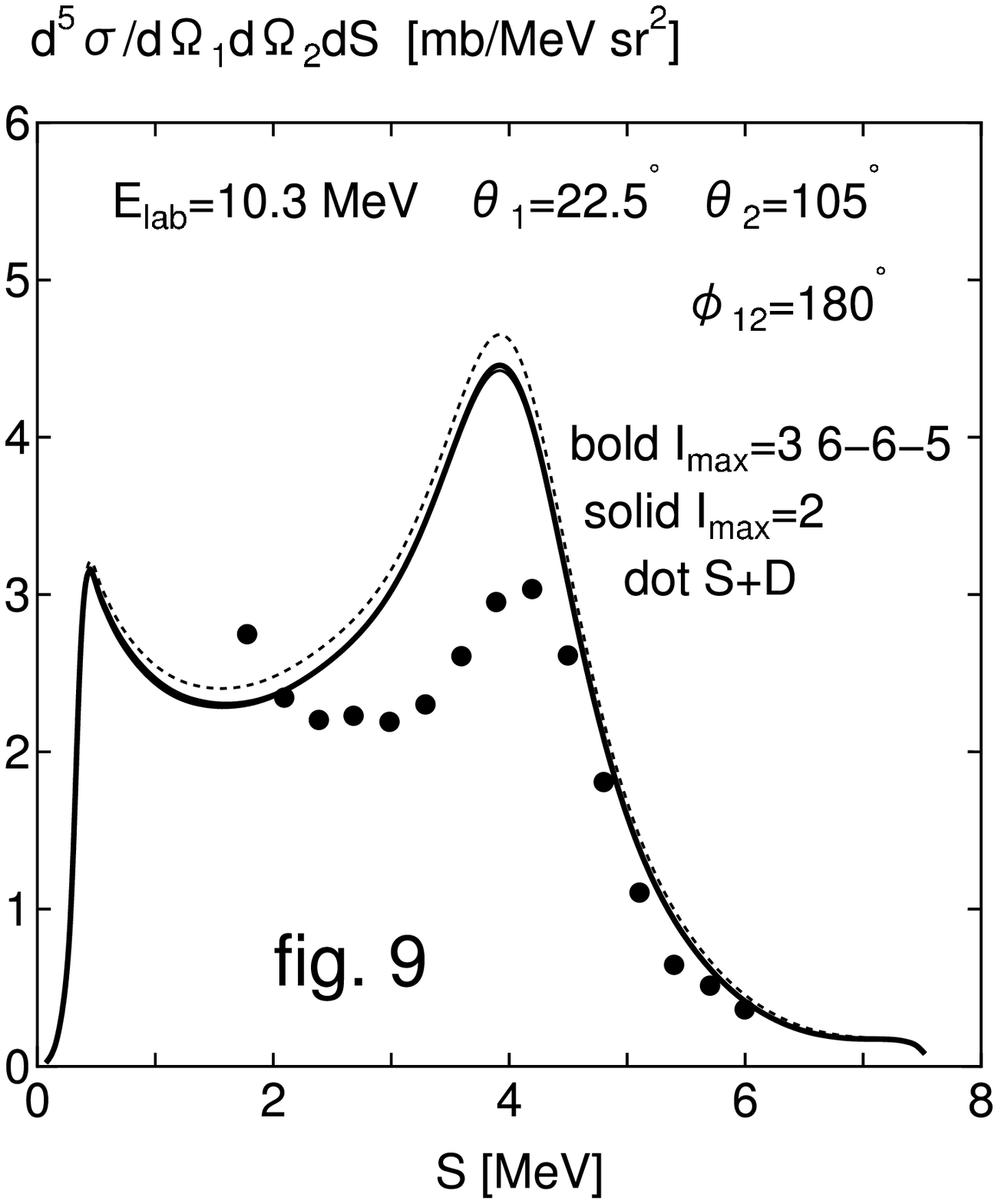}

\includegraphics[angle=0,width=55mm]
{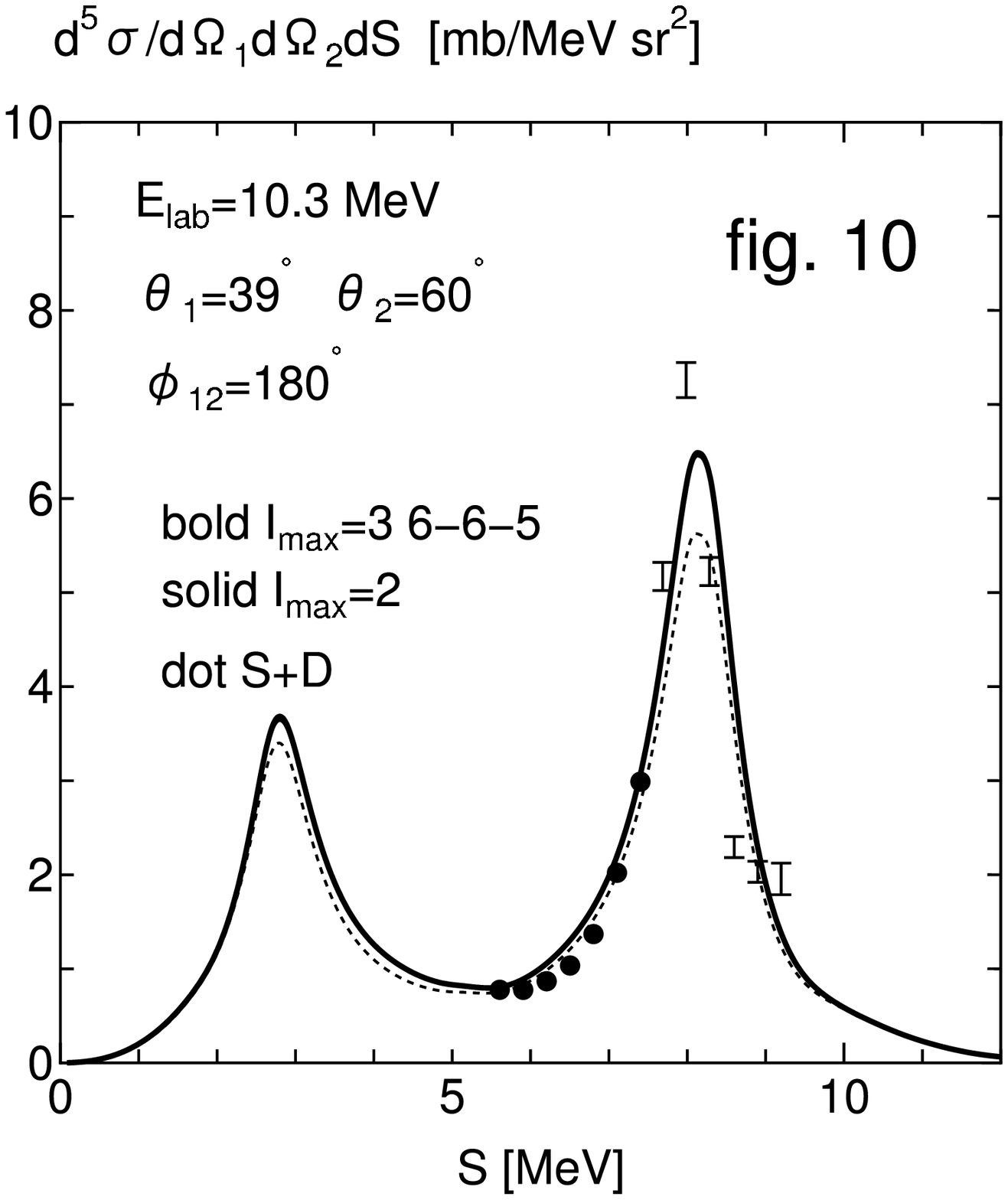}
\end{minipage}
\end{center}
\caption{
Breakup differential cross sections
for the reaction $d(n,2n)p$ with $E_n=10.3$ MeV.
The experimental data are taken from Ref.~\citen{Ge93} with
the same figure number, fig.\,5 etc., in each panel.
}
\label{fig2}
\end{figure}

\begin{figure}[htb]
\begin{center}
\begin{minipage}{0.48\textwidth}
\includegraphics[angle=0,width=55mm]
{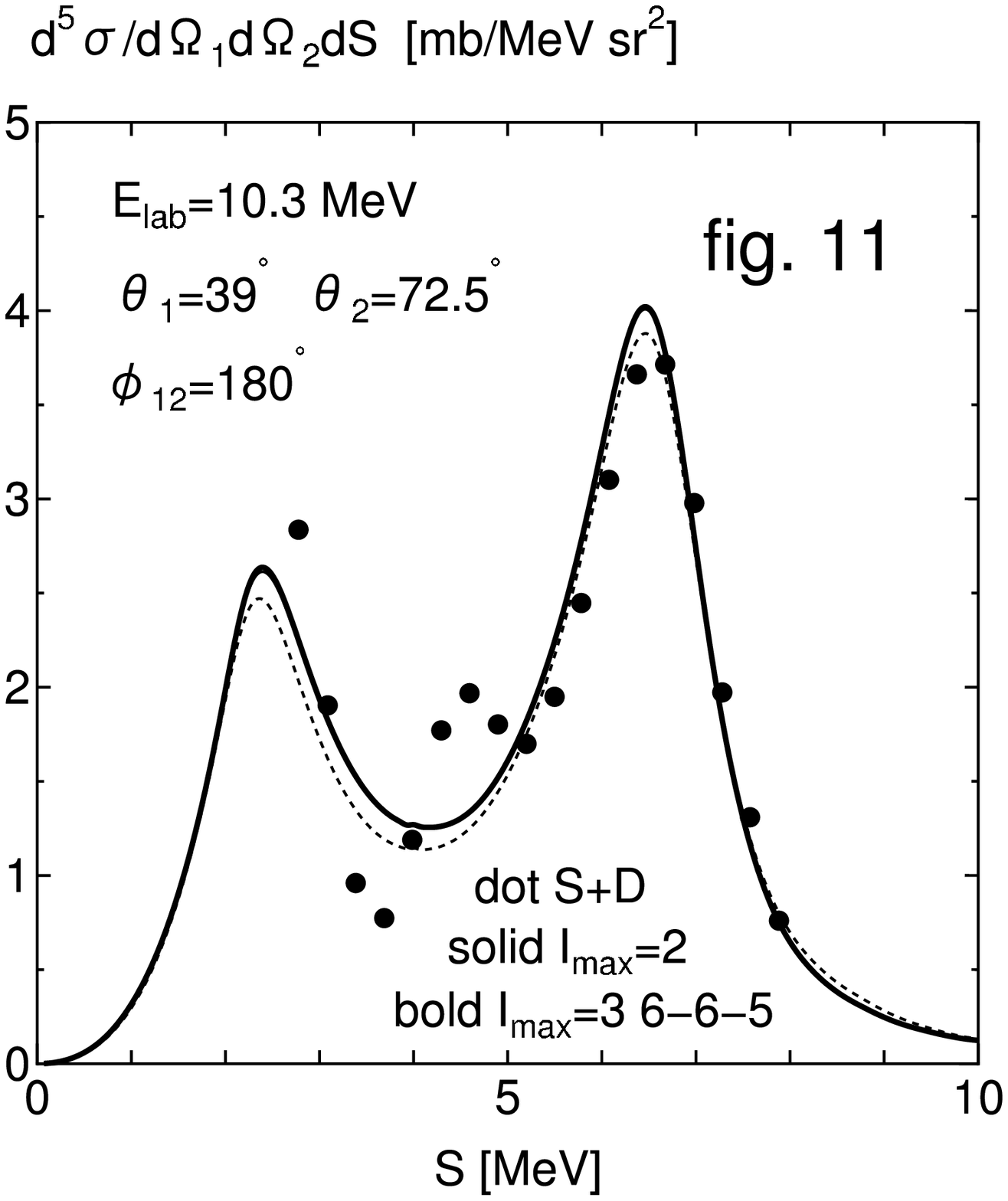}

\includegraphics[angle=0,width=55mm]
{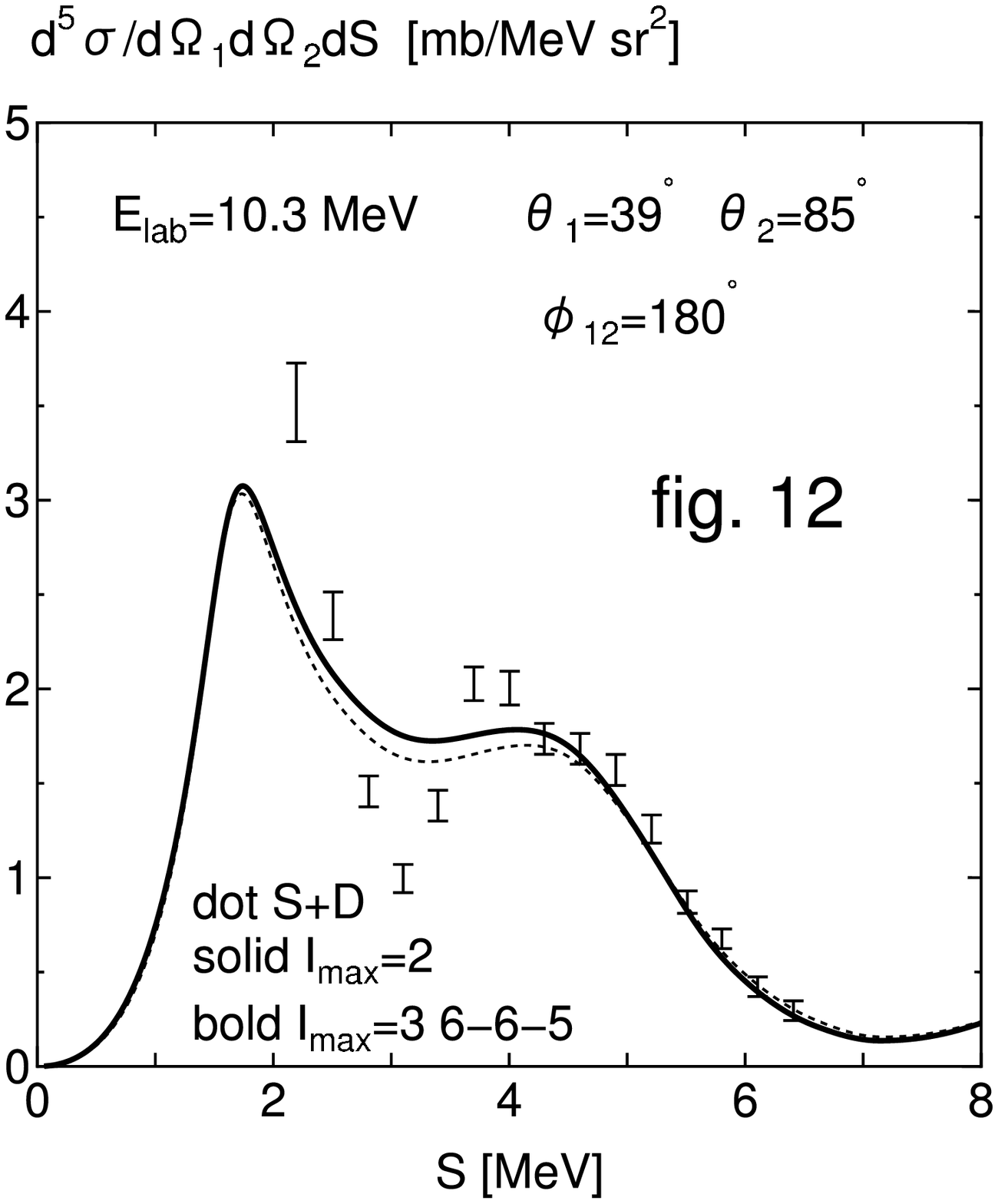}

\includegraphics[angle=0,width=55mm]
{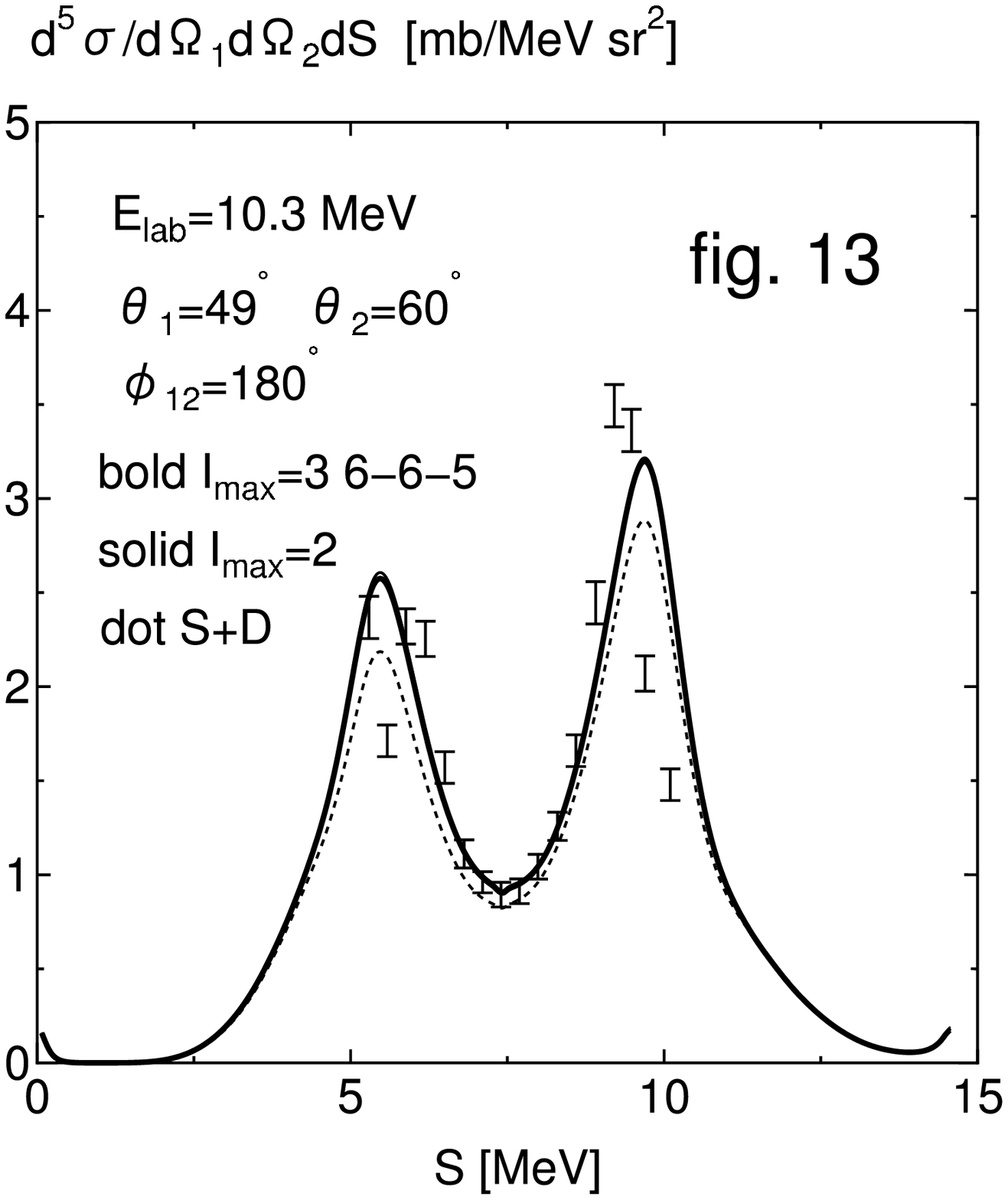}
\end{minipage}~%
\hfill~%
\begin{minipage}{0.48\textwidth}
\includegraphics[angle=0,width=55mm]
{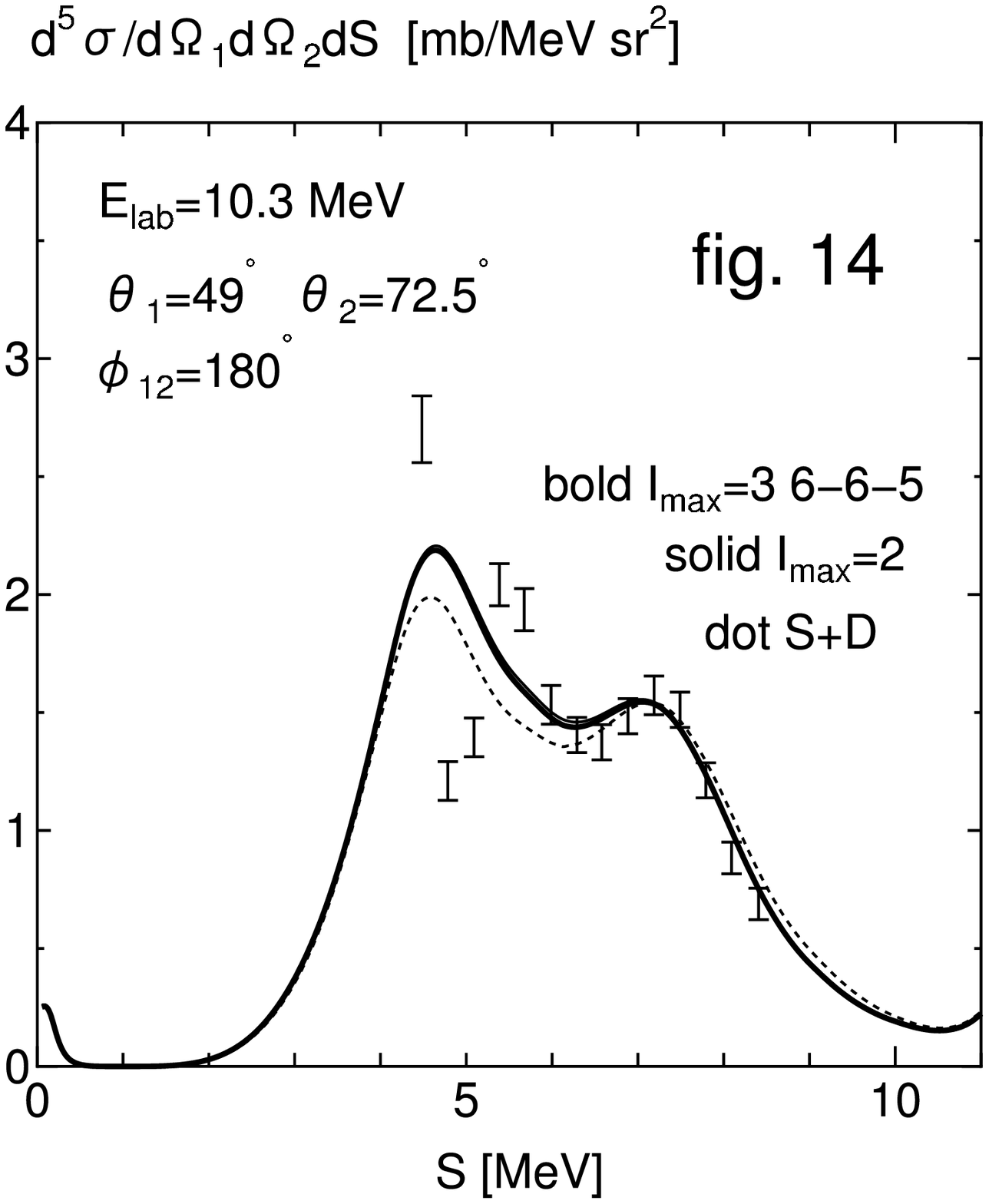}

\includegraphics[angle=0,width=55mm]
{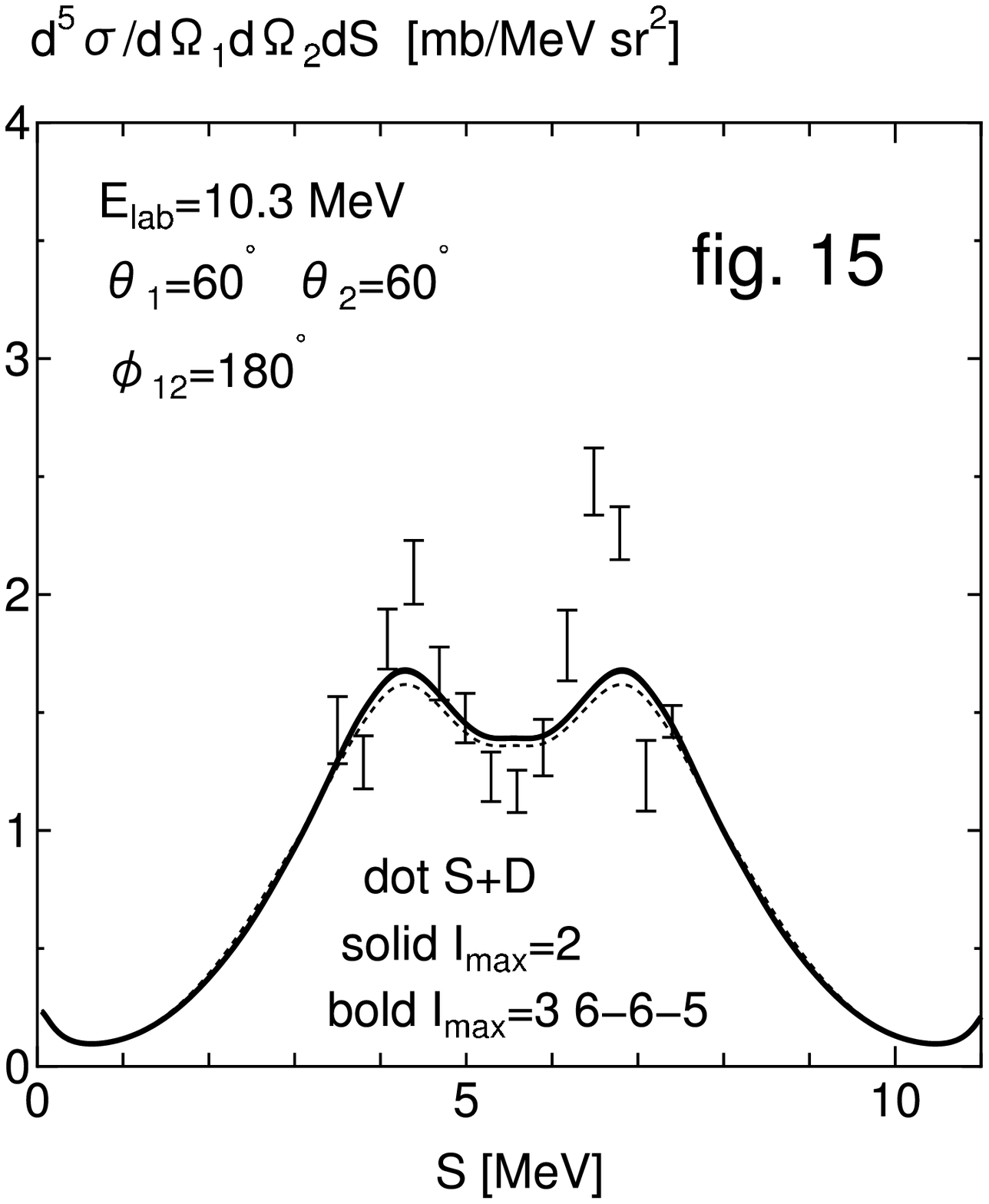}

\includegraphics[angle=0,width=55mm]
{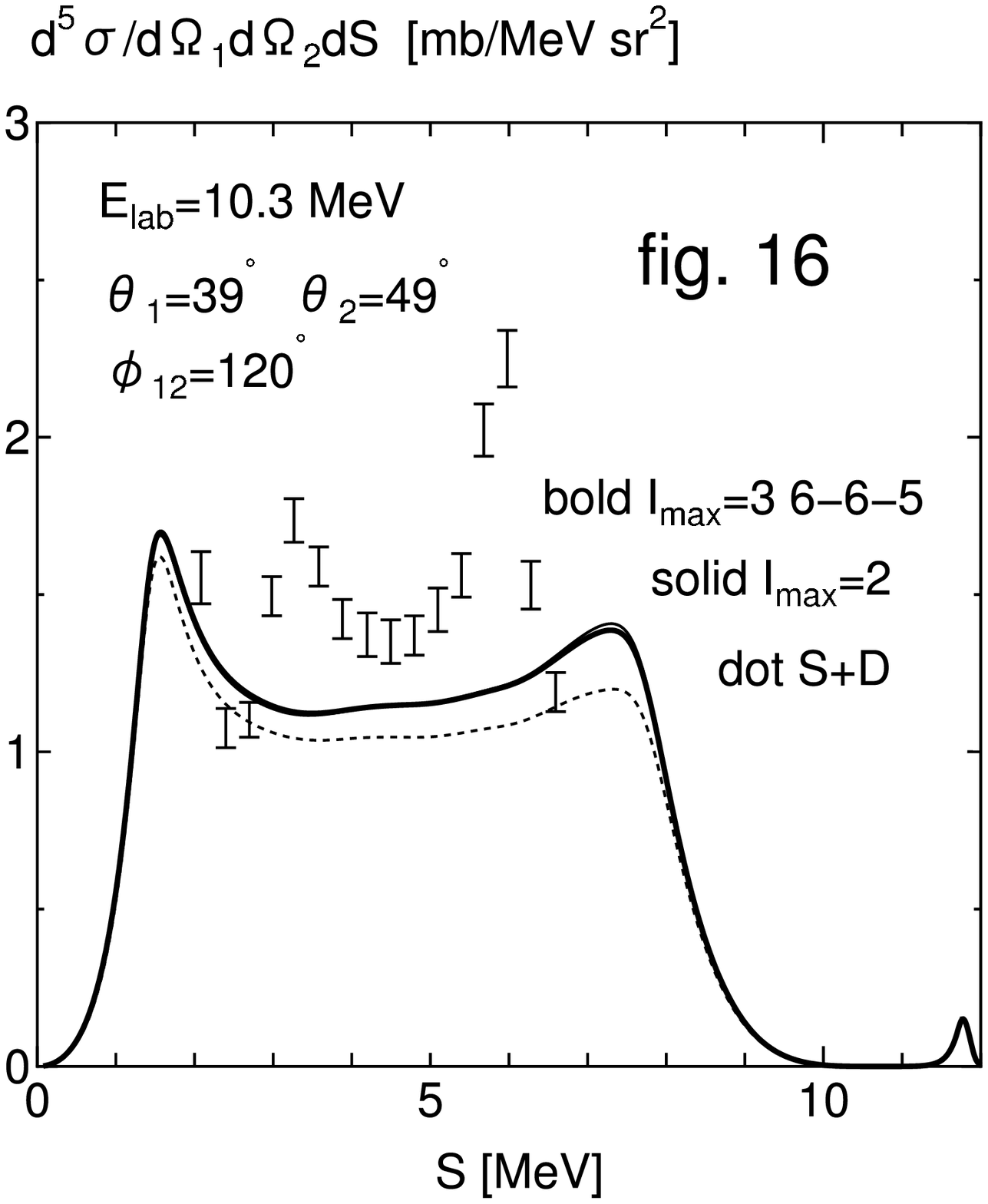}
\end{minipage}
\end{center}
\caption{
The same as Fig.\,\protect\ref{fig2}, but for other kinematical
configurations.
}
\label{fig3}
\end{figure}

The breakup differential cross sections for the $d(n,2n)p$ reaction
at $E_n=10.3$ MeV are displayed in Figs.\,\ref{fig2} and \ref{fig3},
together with the experimental data by Gebhardt et al.\cite{Ge93}
The figure number, fig.\,5 etc., in each panel
corresponds to the original one in Ref.\,\citen{Ge93}.
The large two peaks seen in fig.5 - fig.13 are the $np$ final state
interaction peaks with $\bp_1 \sim 0$ on the lower $S$ side and
those with $\bp_2 \sim 0$ on the higher $S$ side.
On the whole, the comparison with experiment gives fair agreement,
but some discrepancies found in Ref.\,\citen{Ge93} still persist.
In Ref.\,\citen{Ge93}, the experimental data are compared with
the solutions of the AGS equations in the $W$-method, using a
charge-dependent modification of the Paris potential.
Their results and ours are strikingly similar to each other,
sharing the same problems for the detailed fit to the experiment.
The peak heights for the $\bp_2 \sim 0$ final state interaction peaks
are not precisely reproduced in fig.\,5, fig.\,7, fig.\,8 and fig.\,10,
probably because we did not take into account the charge dependence
of the two-nucleon interaction. 
In fig.\,9, our result is worse than the theoretical calculation
in Ref.\,\citen{Ge93}. 
The collinear point is realized at $S=4$ MeV in fig.\,11, 
at $S=7.5$ MeV in fig.\,13, at $S=6$ MeV in fig.\,14, 
and at $S=5.6$ MeV in fig.\,15.
In fig.11, we have obtained a smooth curve around the collinear point,
just as the theoretical calculation in Ref.\,\citen{Ge93}.
The breakup differential cross sections around the collinear points
are well reproduced.
In fig.\,16, the flat structure between $S=3$ - 7 MeV is just the same as
the theoretical calculation in Ref.\,\citen{Ge93}. 
The experimental data of Ref.\,\citen{Ge93} for the symmetric space star 
configuration is plotted in the first panel of Fig.\,\ref{fig6}.
Here again, we have obtained very similar result with 
the theoretical calculation in Ref.\,\citen{Ge93}.

\subsection{Quasi-free scattering}

\begin{figure}[htb]
\begin{center}
\begin{minipage}{0.48\textwidth}
\includegraphics[angle=0,width=55mm]
{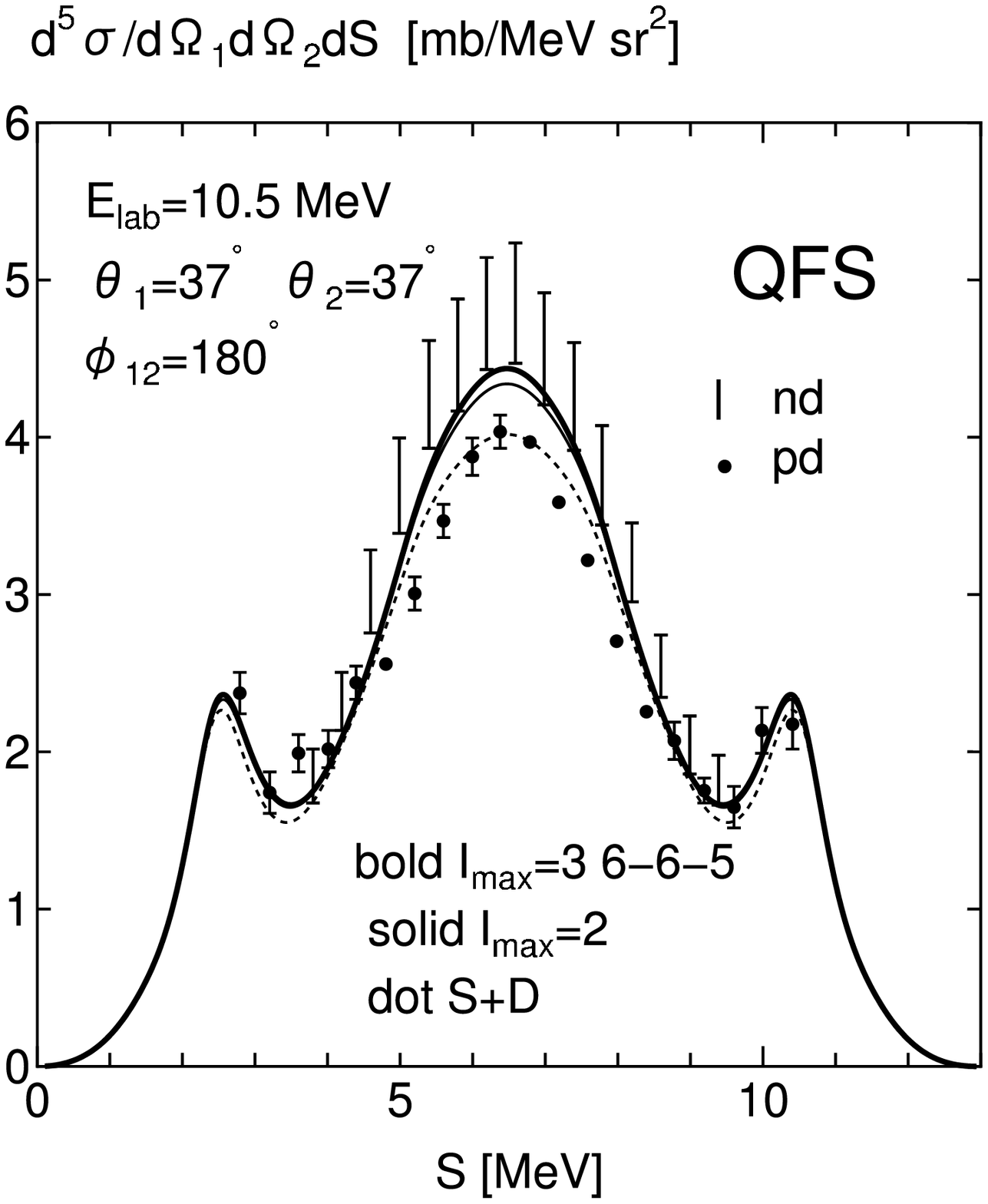}

\includegraphics[angle=0,width=55mm]
{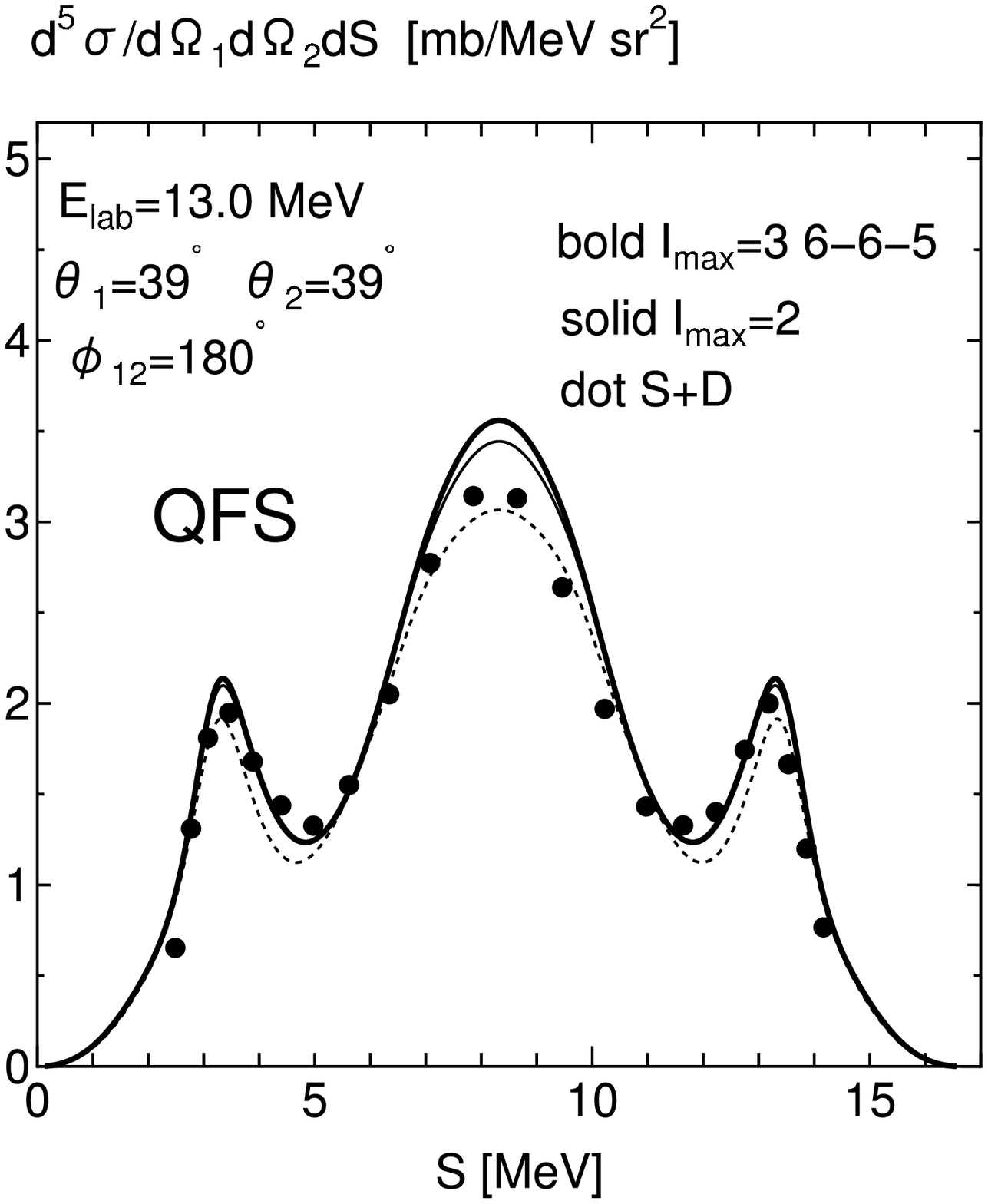}

\includegraphics[angle=0,width=55mm]
{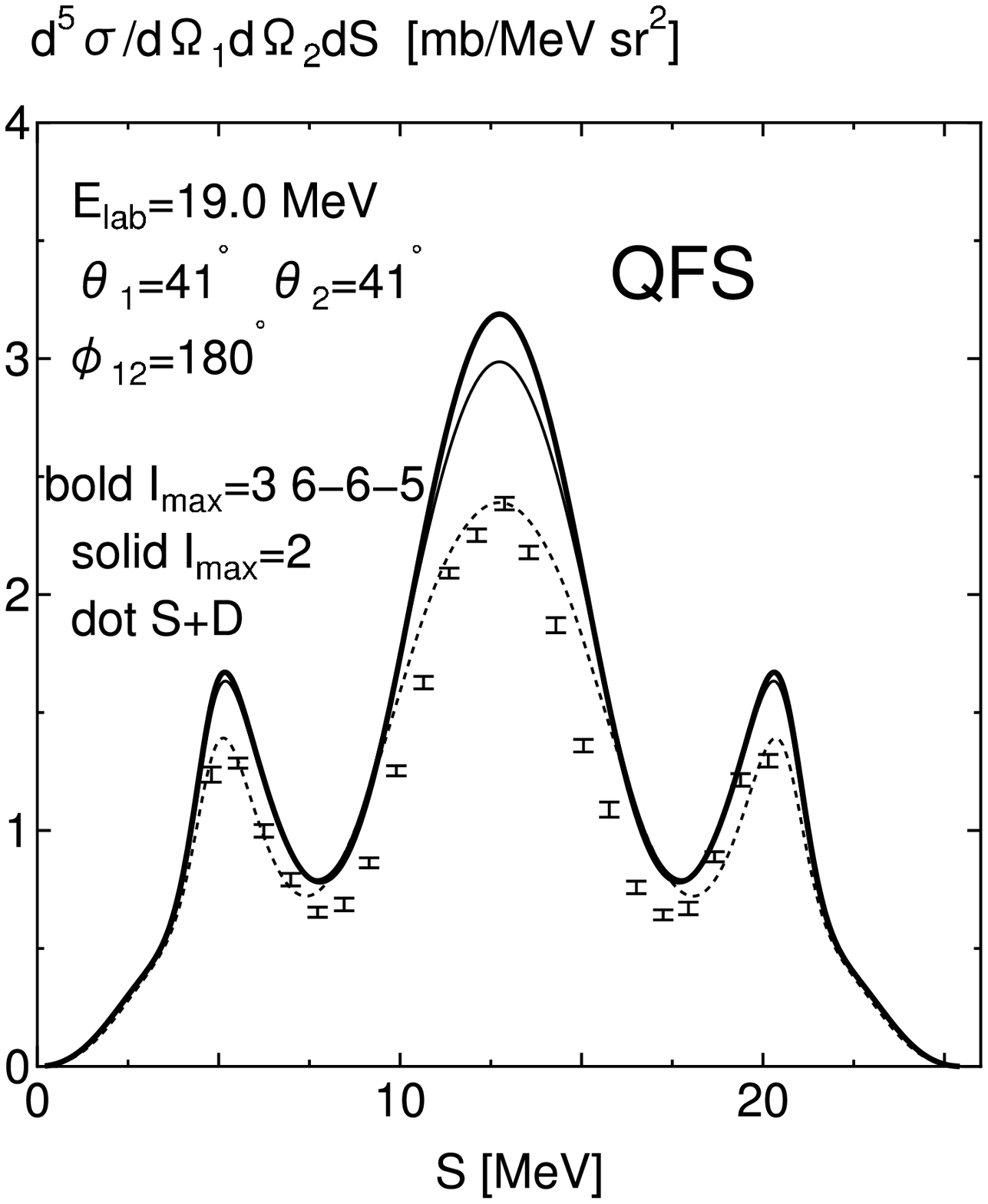}
\end{minipage}~%
\hfill~%
\begin{minipage}{0.48\textwidth}
\includegraphics[angle=0,width=55mm]
{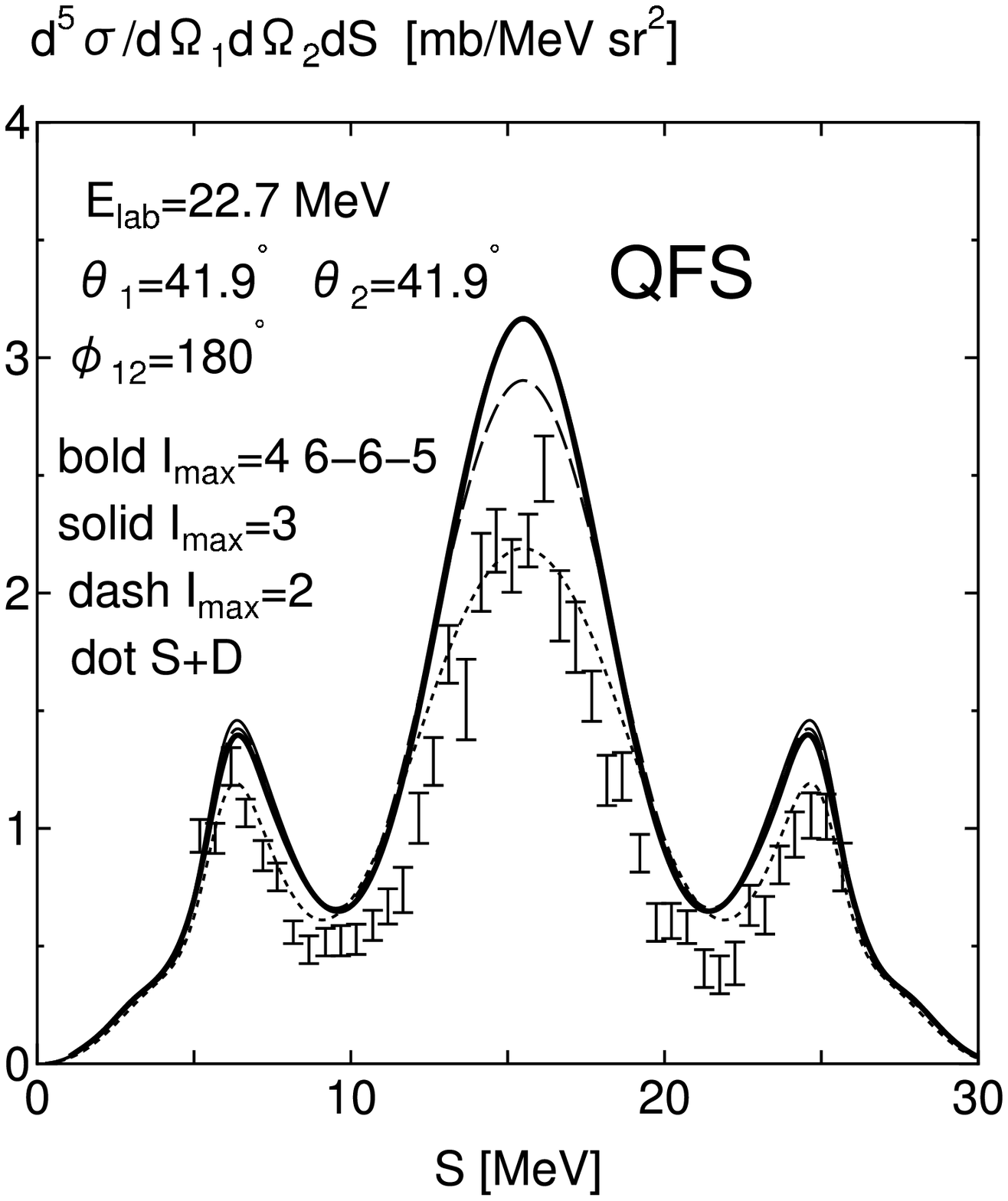}

\includegraphics[angle=0,width=55mm]
{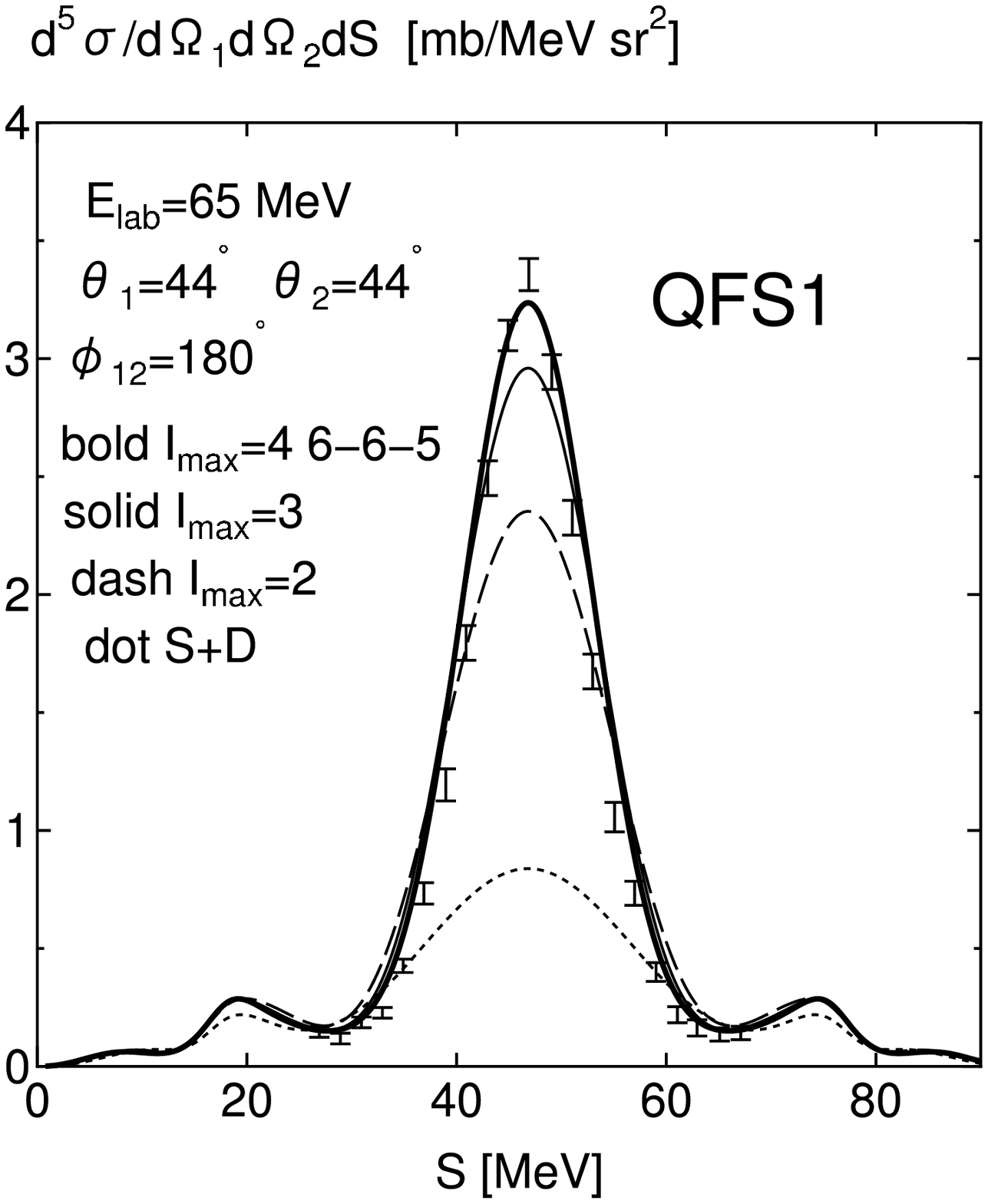}

\includegraphics[angle=0,width=55mm]
{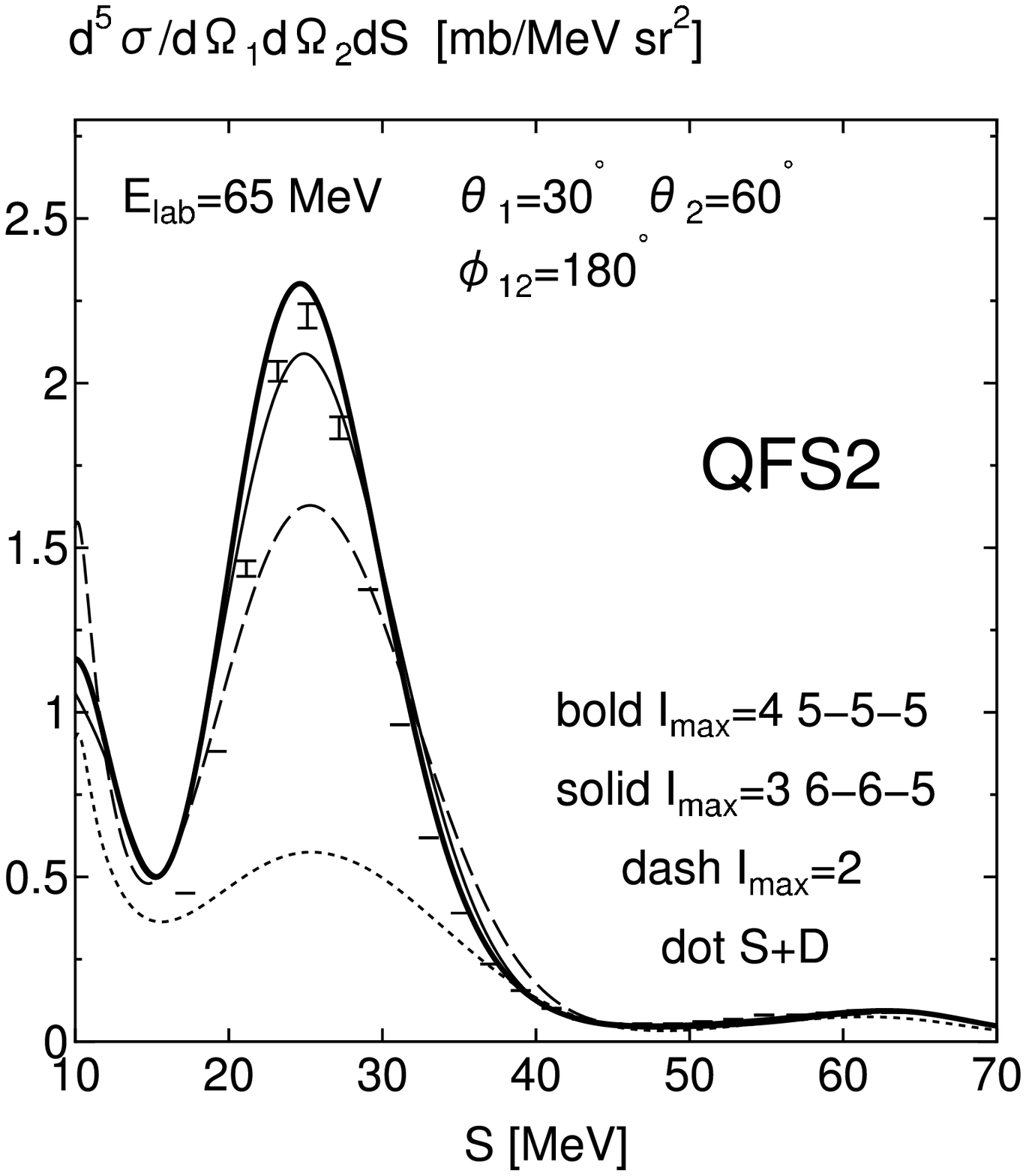}
\end{minipage}
\end{center}
\caption{
Breakup differential cross sections
for the quasi-free scattering (QFS).
The experimental data for the $d(p,pp)n$ reaction are
taken from Refs.\,\citen{Gr96} for 10.5 MeV, \citen{Ra91} for 13 MeV,
\citen{Pa96} for 19 MeV, \citen{Za94} for 22.7 MeV, 
and \citen{Al96} for 65 MeV.
For the reaction $d(n,2n)p$ with $E_n=10.5$ MeV,
the experimental data shown by bars are taken from Ref.~\citen{Gr96}.
}
\label{fig4}
\end{figure}

We show in Fig.\,\ref{fig4} the breakup differential cross sections
for the quasi-free scattering at energies $E_{\rm lab}=10.5$ - 65 MeV.
We find some deviation from the experimental data at the peak position
for all the energies. 
Detailed investigation of the Coulomb effect in Ref.\,\citen{De05b}
has revealed that this overestimation at the peak position is
reduced to some extent.
However, the reduction is probably not large enough except for 
the case of $E_{\rm lab}=13$ MeV.
Figure 8 of Ref.\,\citen{De05b} implies that this reduction  is
energy dependent. The large overestimation at $E_{\rm lab}=19$ MeV
may not be resolved only by the Coulomb effect.
The direct incorporation of the Coulomb force is necessary for our
QM interaction.
In Fig.\,\ref{fig4}, we can see that the roles of higher partial waves
are important for higher energies. The partial waves up to  $I_{\rm max}
=3$ is clearly necessary for $E_{\rm lab}=19$ MeV. For the energies,
$E_{\rm lab}=22.7$ and 65 MeV, we need more partial waves
up to $I_{\rm max}=4$. 

\subsection{Final state interaction}

\begin{figure}[b]
\begin{center}
\begin{minipage}{0.48\textwidth}
\includegraphics[angle=0,width=55mm]
{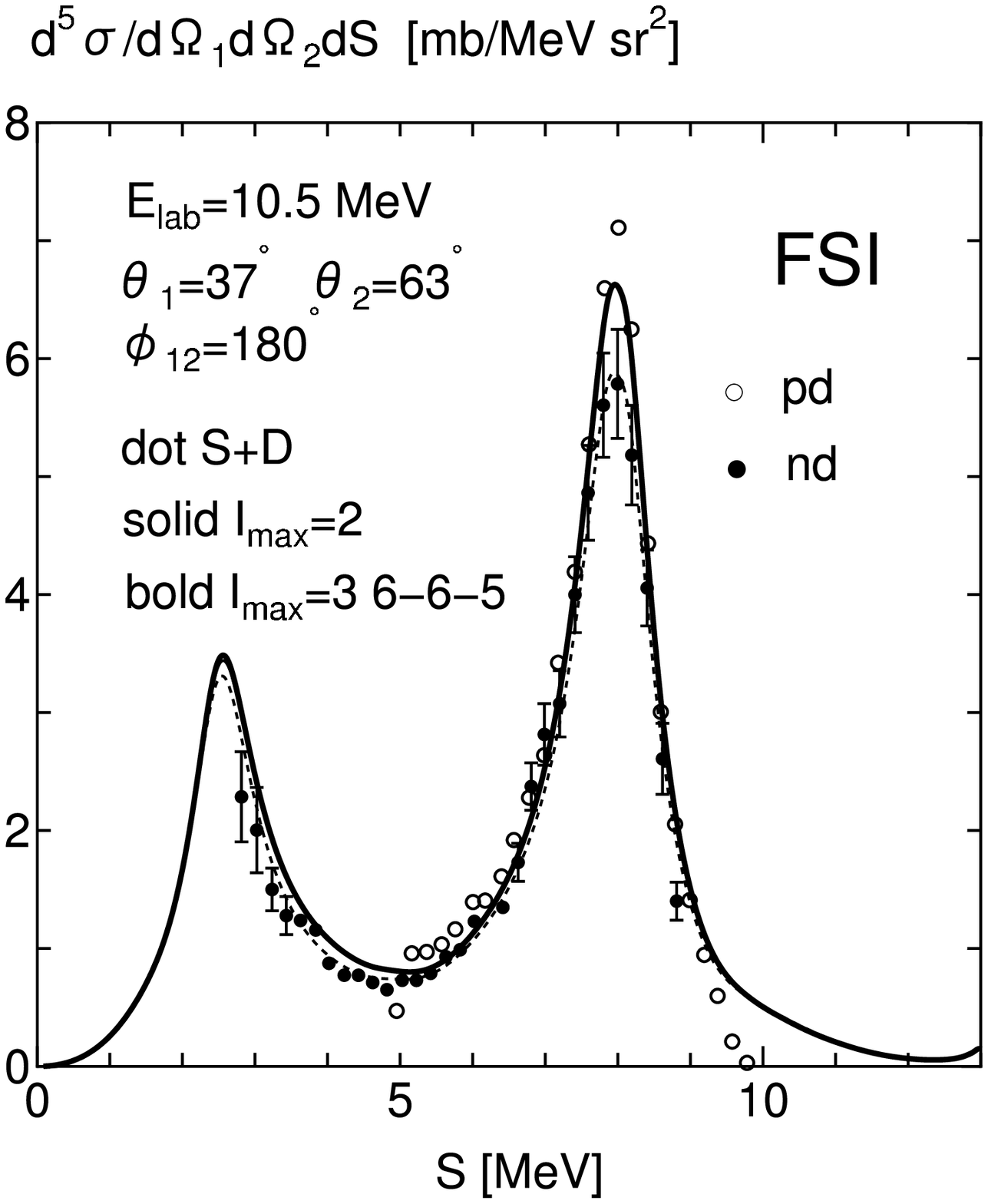}

\includegraphics[angle=0,width=55mm]
{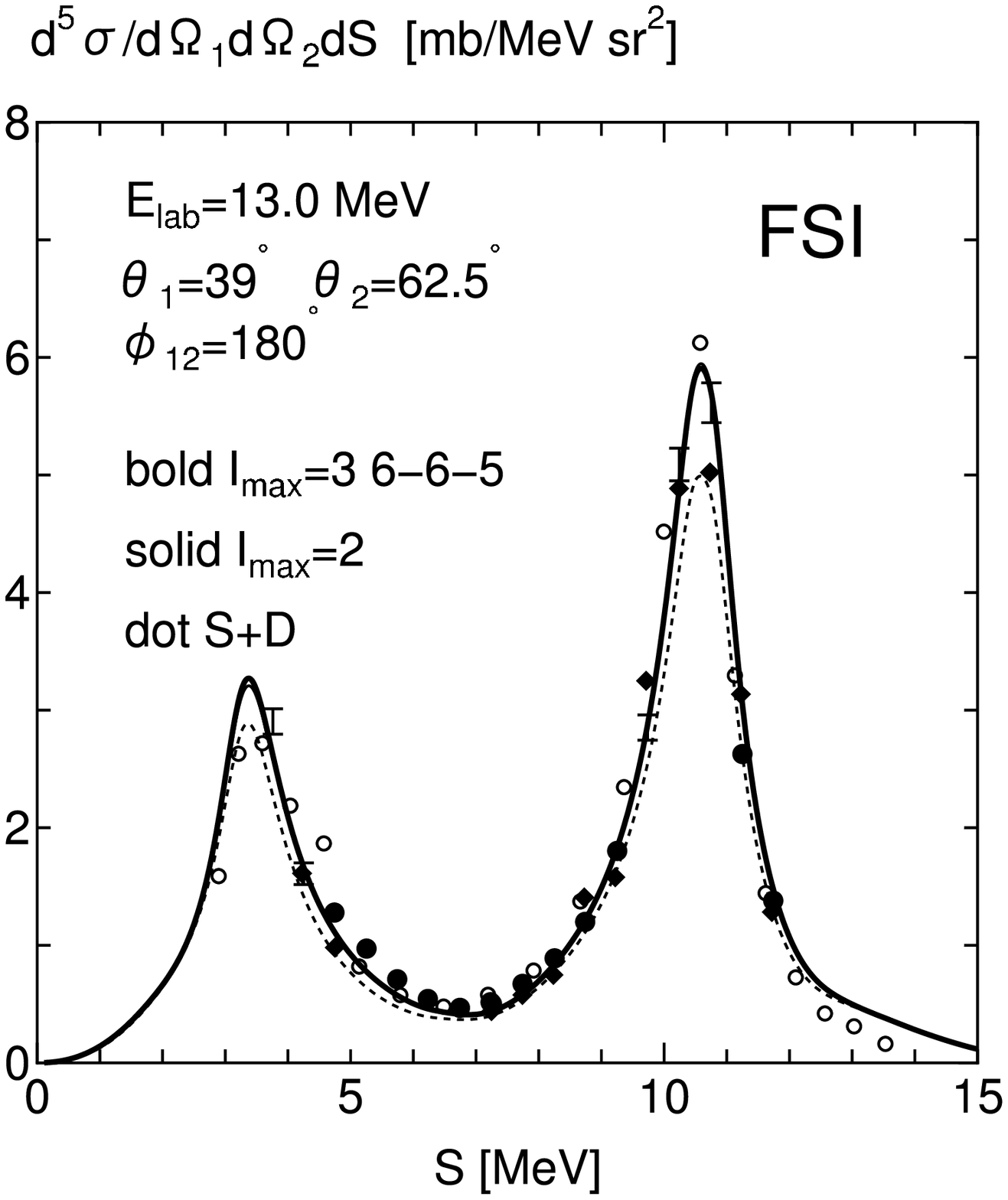}
\end{minipage}~%
\hfill~%
\begin{minipage}{0.48\textwidth}
\includegraphics[angle=0,width=55mm]
{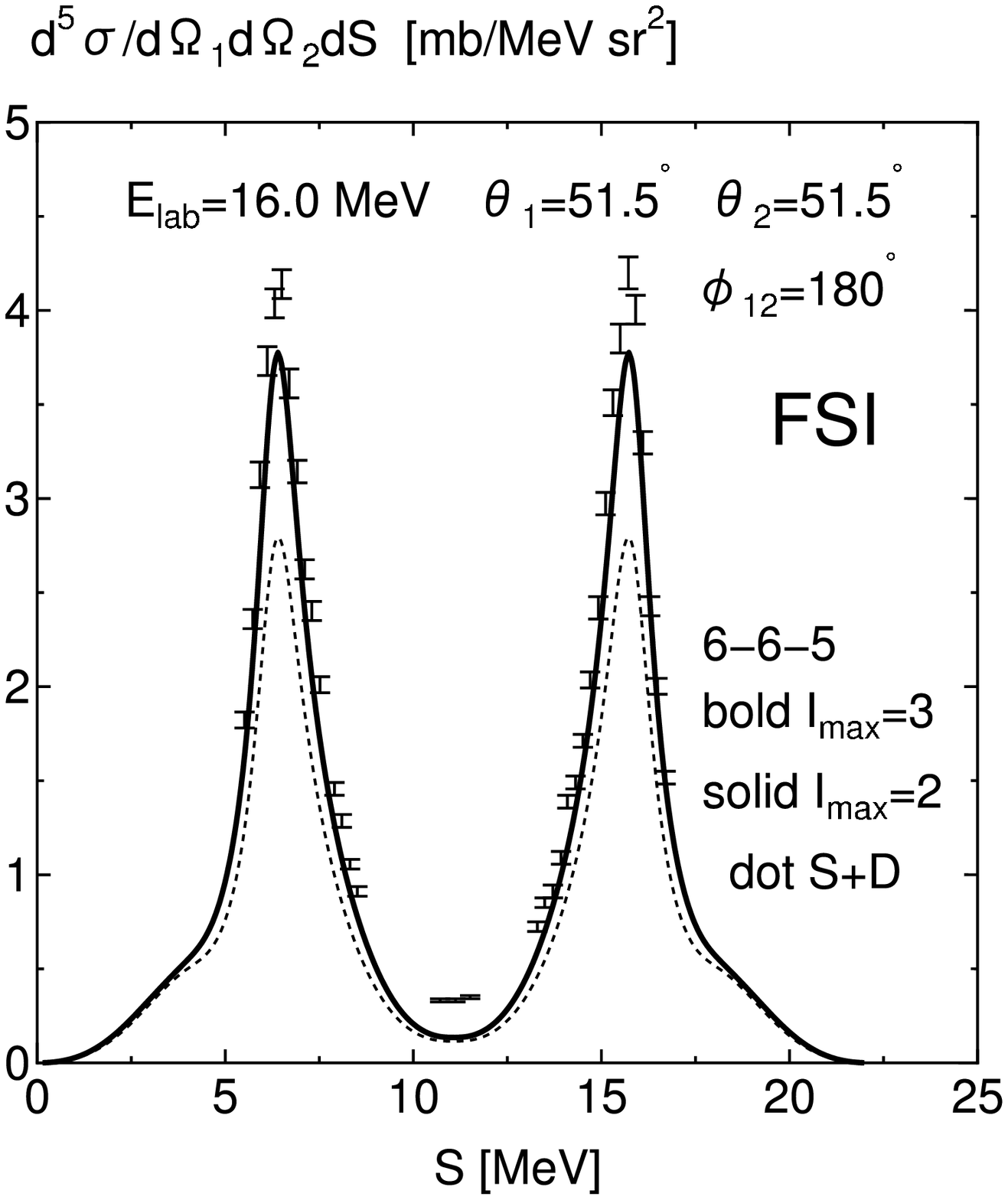}

\includegraphics[angle=0,width=55mm]
{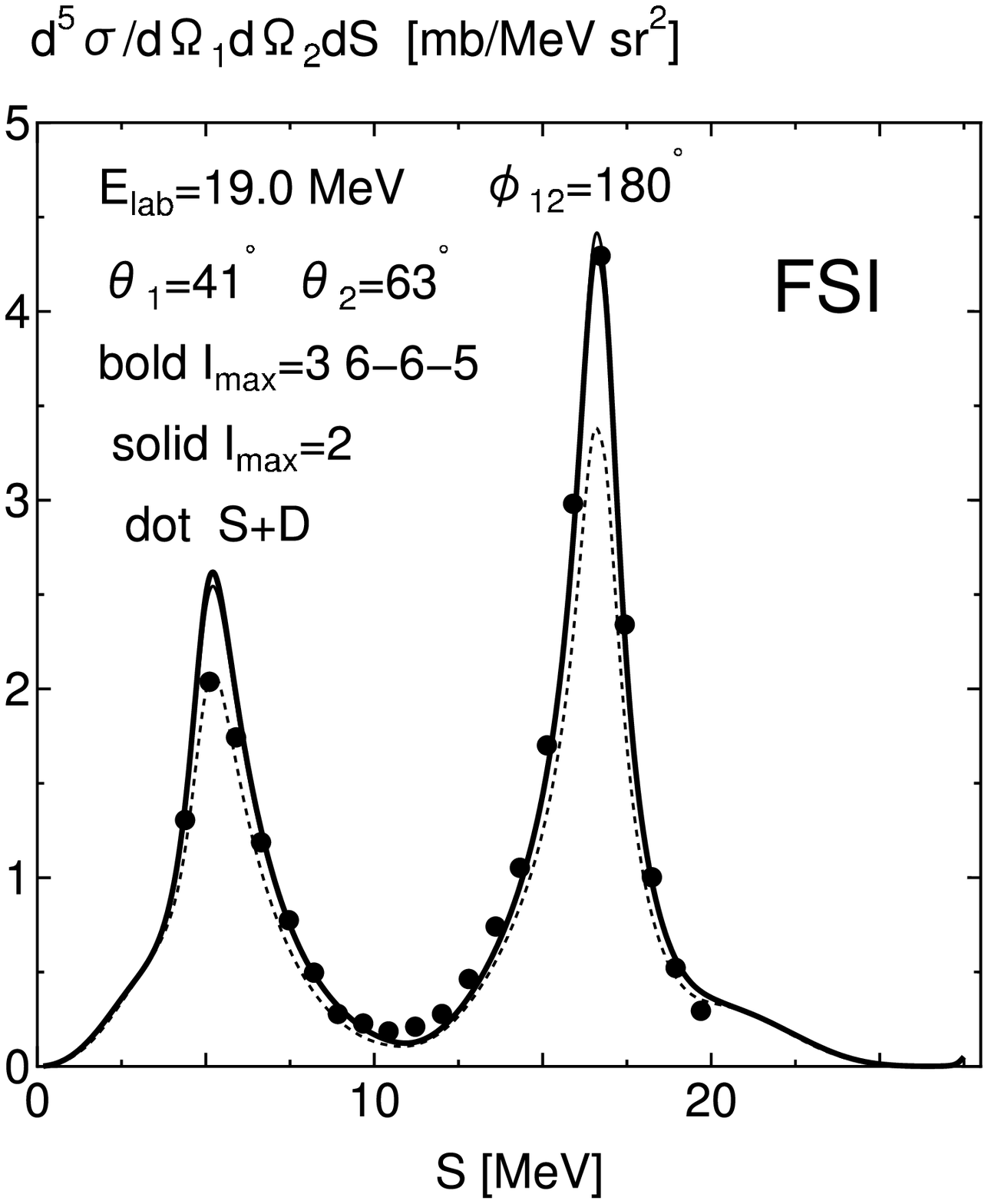}
\end{minipage}
\end{center}
\caption{
Breakup differential cross sections
for the final state interaction (FSI).
The experimental data are taken from Ref.\,\citen{Gr96} ($pd$: open circles,
$nd$: filled circles with bars) for 10.5 MeV,
Refs.\,\citen{St88,St89,Ho98} ($nd$: bars, filled circles,
filled diamonds) and \citen{Ra91} ($pd$: open circles) for 13 MeV,
\citen{Du05} ($pd$) for 16 MeV, and \citen{Pa96} ($pd$) for 19 MeV.
}
\label{fig5}
\end{figure}

Four examples of the breakup differential cross sections for
the $np$ final state interaction are displayed in Fig.\,\ref{fig5}.
Almost all the data are for the $d(p,pp)n$ reaction.
In the 10.5 MeV and 13 MeV cases, the $pd$ data are shown with
open circles, while the $nd$ data with bars, some dots and diamonds. 
The lower peaks are the $np$ final state interaction peaks 
with $\bp_1 \sim 0$, while the upper with $\bp_2 \sim 0$.
Here we find that the higher peaks are slightly too small.
The Coulomb correction increases the peak height a little, \cite{De05b}
and improves the fit to the experiment to some extent.
We probably need more careful treatment of the charge dependence
of the $NN$ interaction, just as in the previous 10.3 MeV case.    
We also see that the minimum point at $S=11$ - 12 MeV 
for the $E_{\rm lab}=16$ MeV reaction
is too low. The three-nucleon force might be necessary to increase 
the cross sections and get a good fit to the experiment.\cite{Du05}

\subsection{Symmetric space star configurations}

\begin{figure}[htb]
\begin{center}
\begin{minipage}{0.48\textwidth}
\includegraphics[angle=0,width=55mm]
{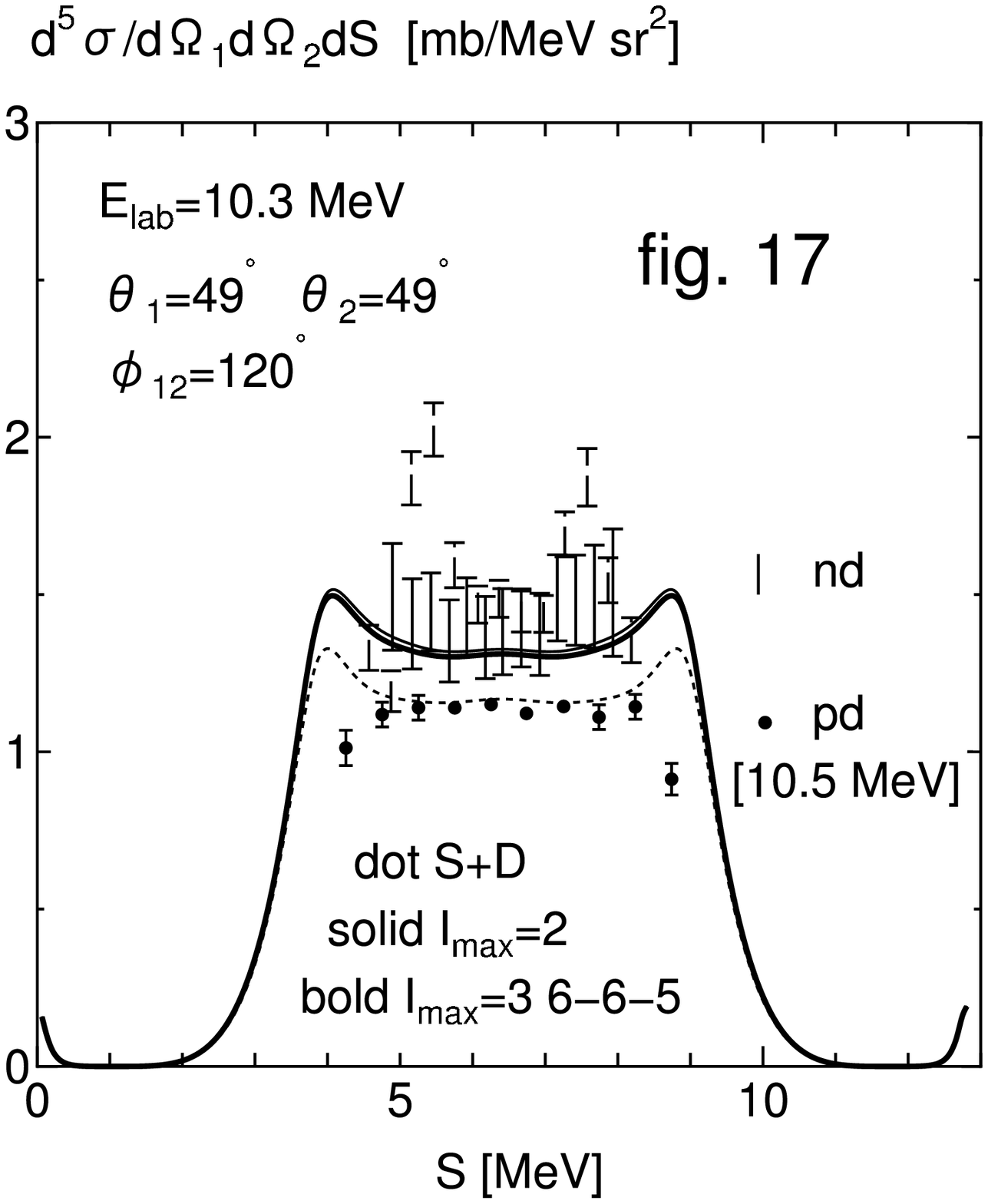}

\includegraphics[angle=0,width=55mm]
{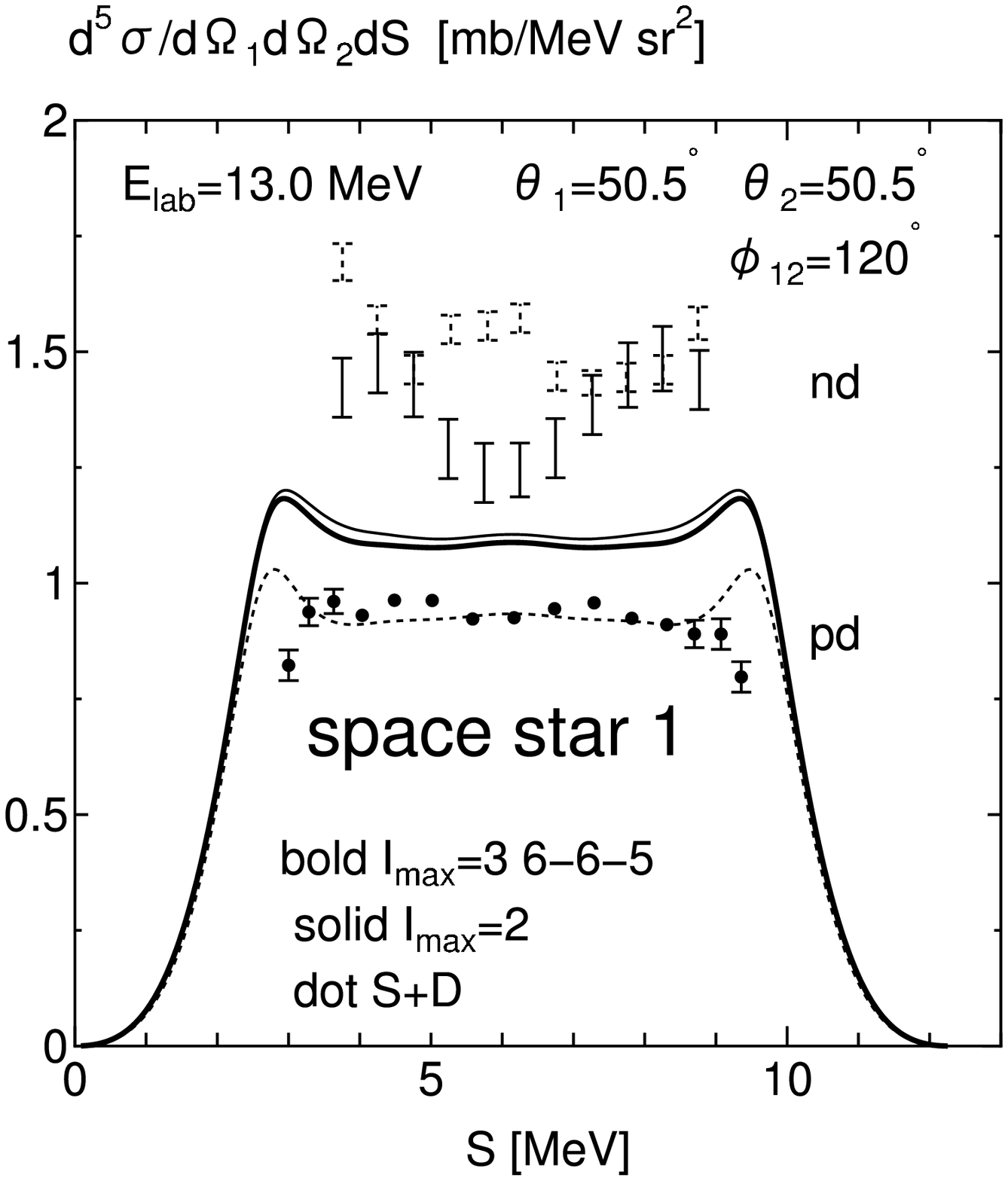}

\includegraphics[angle=0,width=55mm]
{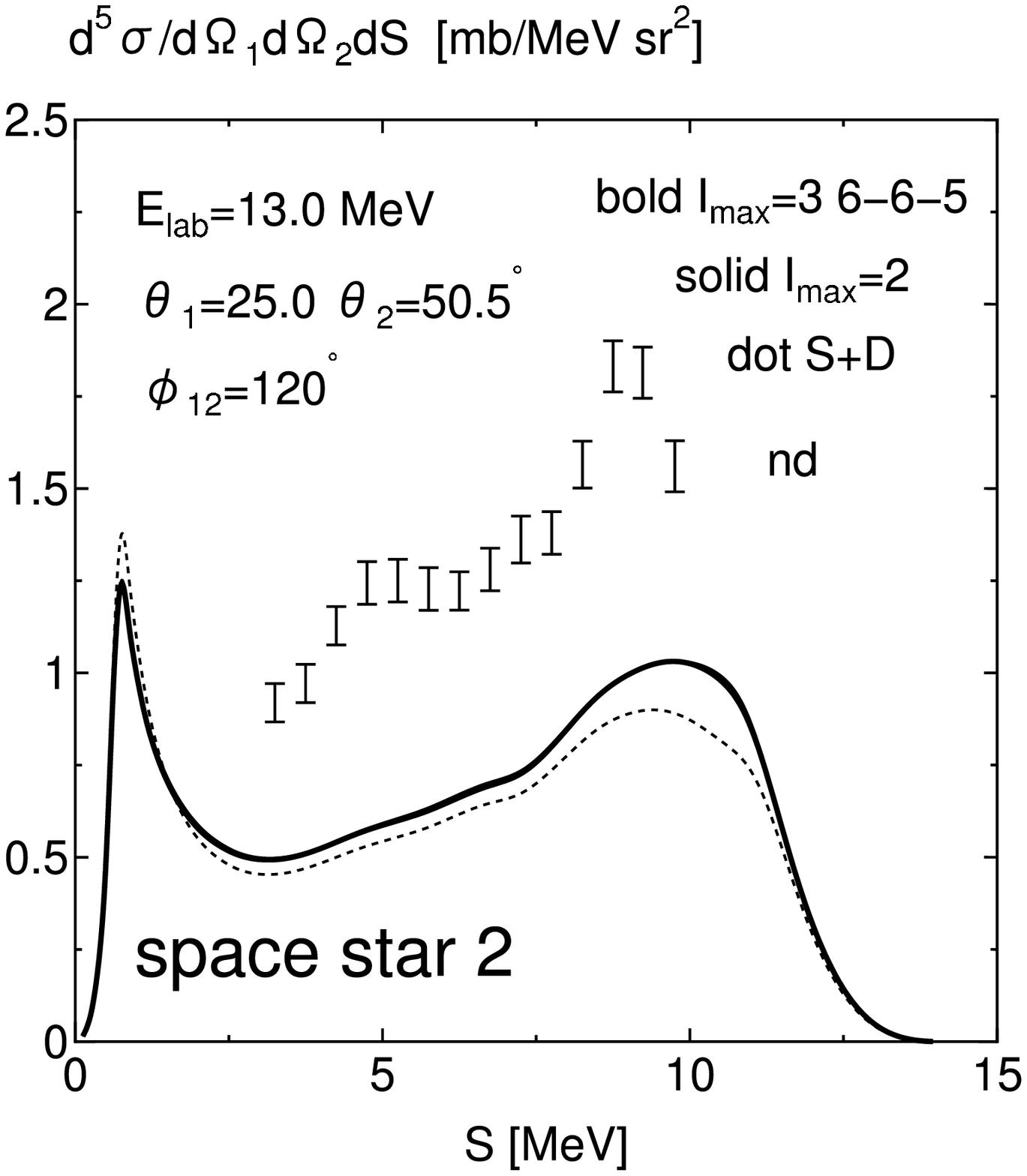}
\end{minipage}~%
\hfill~%
\begin{minipage}{0.48\textwidth}
\includegraphics[angle=0,width=55mm]
{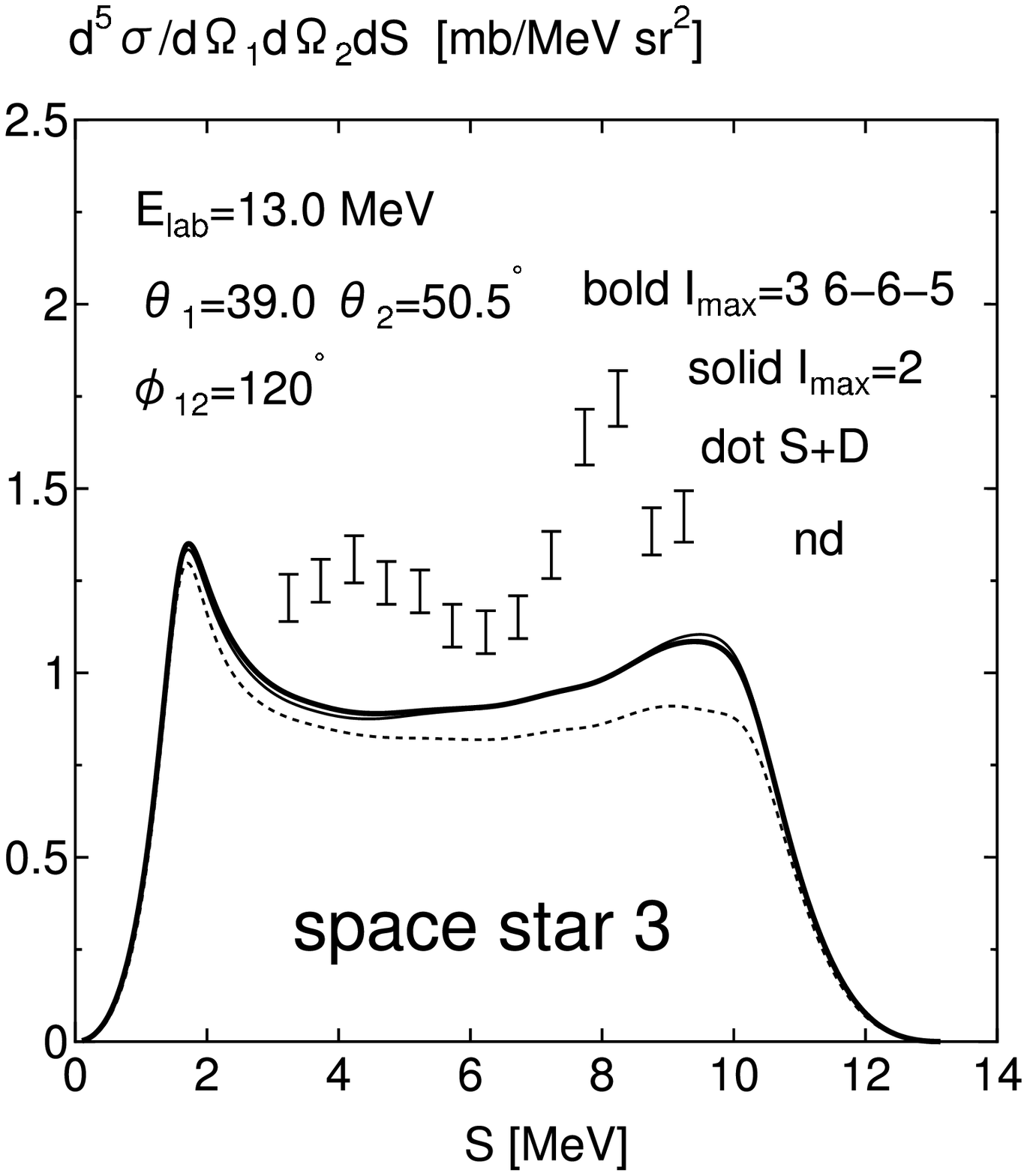}

\includegraphics[angle=0,width=55mm]
{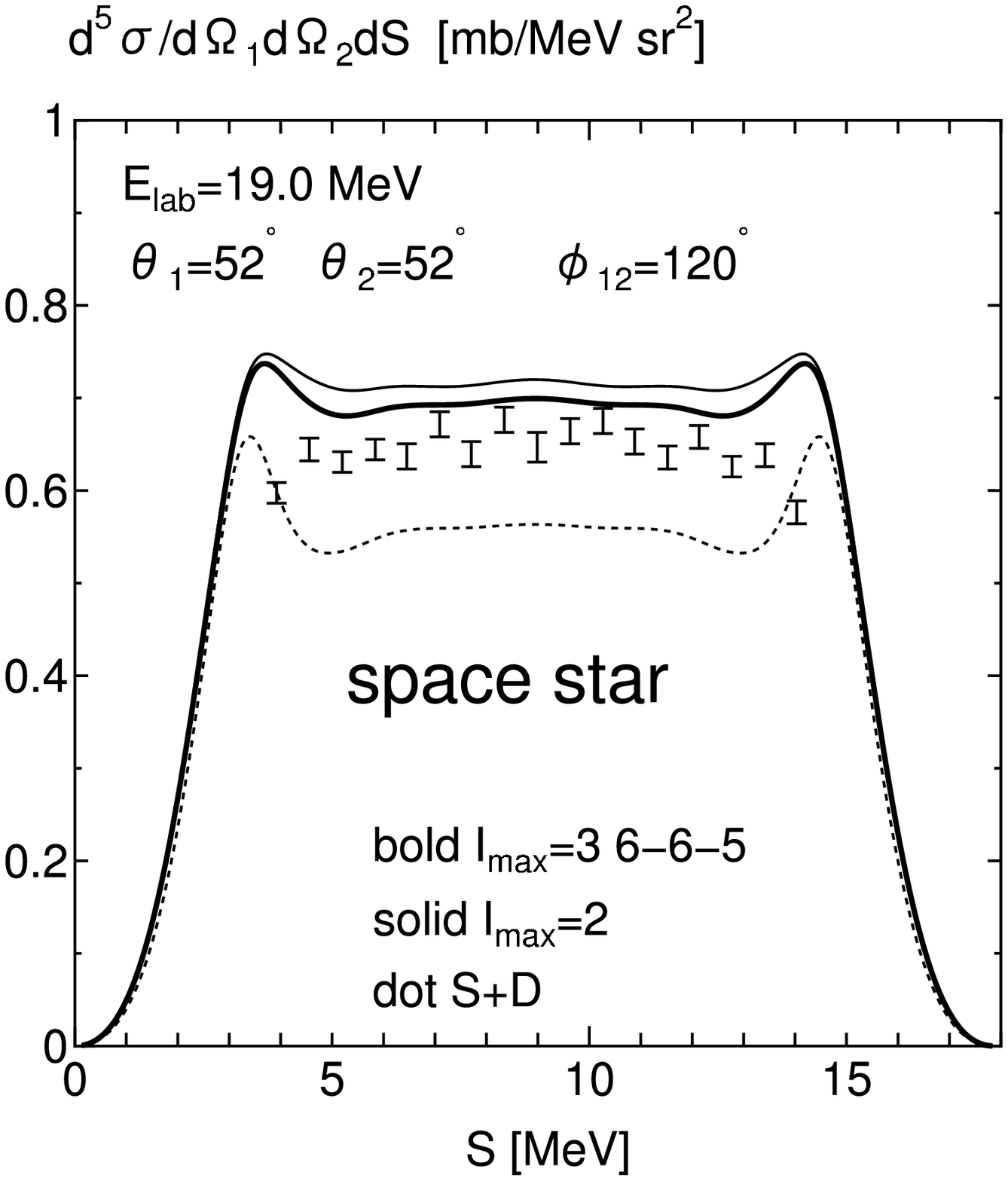}

\includegraphics[angle=0,width=54mm]
{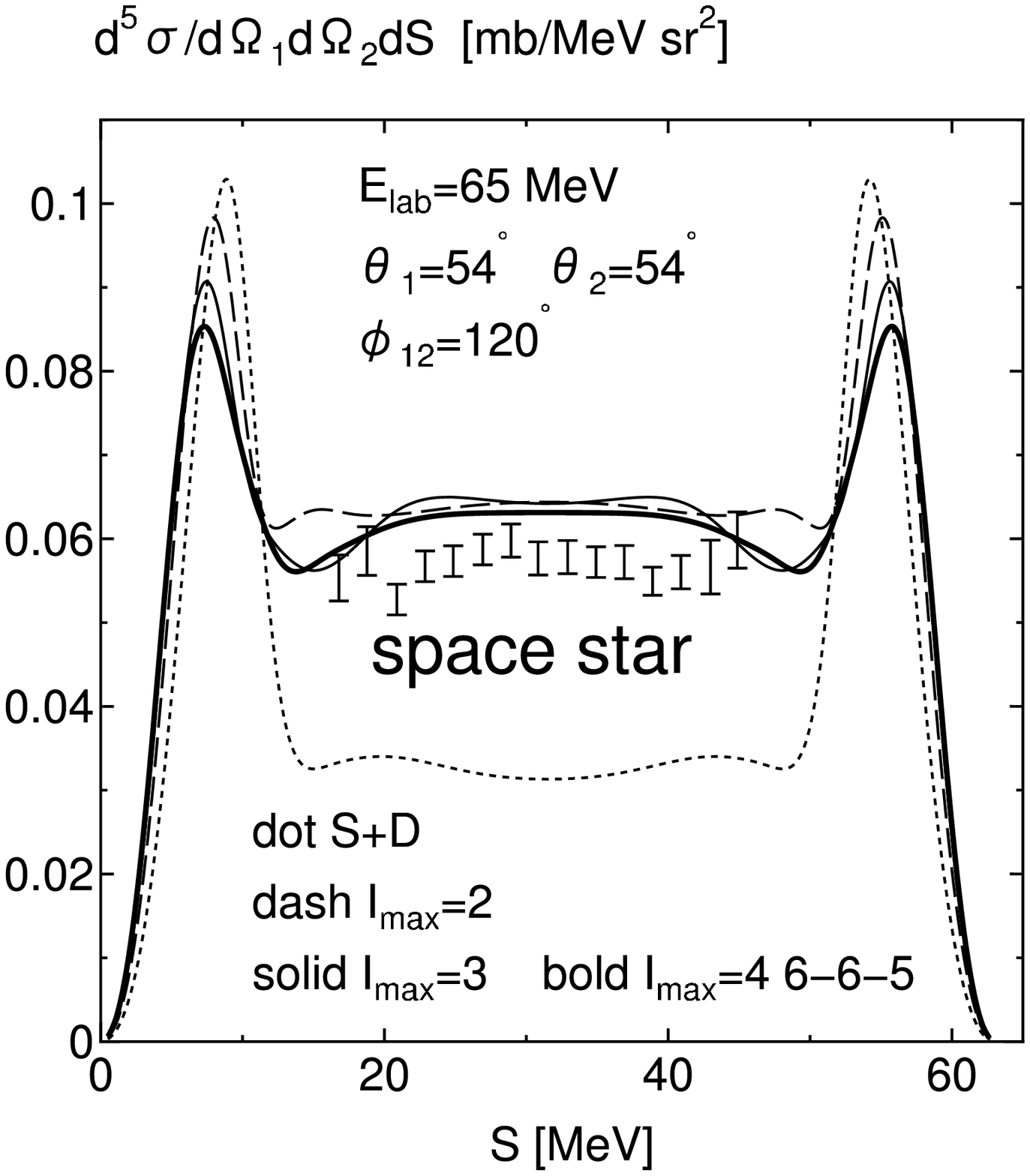}
\end{minipage}
\end{center}
\caption{
Breakup differential cross sections
for the symmetric space star (SST) configurations.
The experimental data are taken from Refs.\,\citen{Stp89,Ge93} ($nd$)
and \citen{Gr96} ($pd$: [10.5 MeV]) for 10.3 MeV, 
\citen{St88,St89,Se96} ($nd$) and \citen{Ra91} ($pd$)
for 13 MeV, \citen{Pa96} ($pd$) for 19 MeV, 
and \citen{Ze97} ($pd$) for 65 MeV.
}
\label{fig6}
\end{figure}

As is well known, a large discrepancy appears in the
breakup differential cross sections in the symmetric
space star configurations.\cite{Se96}
This is seen in Fig.\,\ref{fig6}, where our results
in the various model spaces are compared with the $nd$ and $pd$ data.
A strange thing is that any theoretical calculations of
the $d(n,nn)p$ reaction at $E_{\rm lab}=13$ MeV deviate largely
from the old and new $nd$ data \cite{St88,St89,Se96}, although
the deviation is not much for the 10.3 MeV, 19 MeV and 65 MeV data.
The $pd$ data at 13 MeV in Ref.\,\citen{Ra91} are more 
than $30\%$ smaller than the $nd$ data. A theoretical study of
the Coulomb effect for the symmetric space star configuration
in Ref.\,\citen{De05b} shows that it is generally very small irrespective
of the energy (see Fig.~6 of Ref.\,\citen{De05b}). 
Our results at $E_{\rm lab}=13$ MeV, $\theta_1=\theta_2=50.5^\circ$
and $\phi_{12}=120^\circ$ are located just between the lower $pd$ data and
the higher $nd$ data, which is very similar to other predictions
by the meson-exchange potentials. In the other geometrical configurations
at 13 MeV, the cross sections in the space star 2 case ($\theta_1=25^\circ,
~\theta_2=50.5^\circ,~\phi_{12}=120^\circ$) are about half of the 
experimental values and those in the space star 3 case ($\theta_1=39^\circ,
~\theta_2=50.5^\circ,~\phi_{12}=120^\circ$) are almost $30\%$ smaller 
than the experiment. The same situation happens in the
Faddeev calculations by the Paris potential.\cite{St88}
(See Figs.\,29 and 30 of Ref.\,\citen{St89}.) 
Note that we need enough partial waves for the convergence
of the symmetric space star configurations in particular,
which was already pointed out in Ref.\,\citen{PREP}.

\subsection{Coplanar star configurations}

\begin{figure}[htb]
\begin{center}
\begin{minipage}{0.48\textwidth}
\includegraphics[angle=0,width=55mm]
{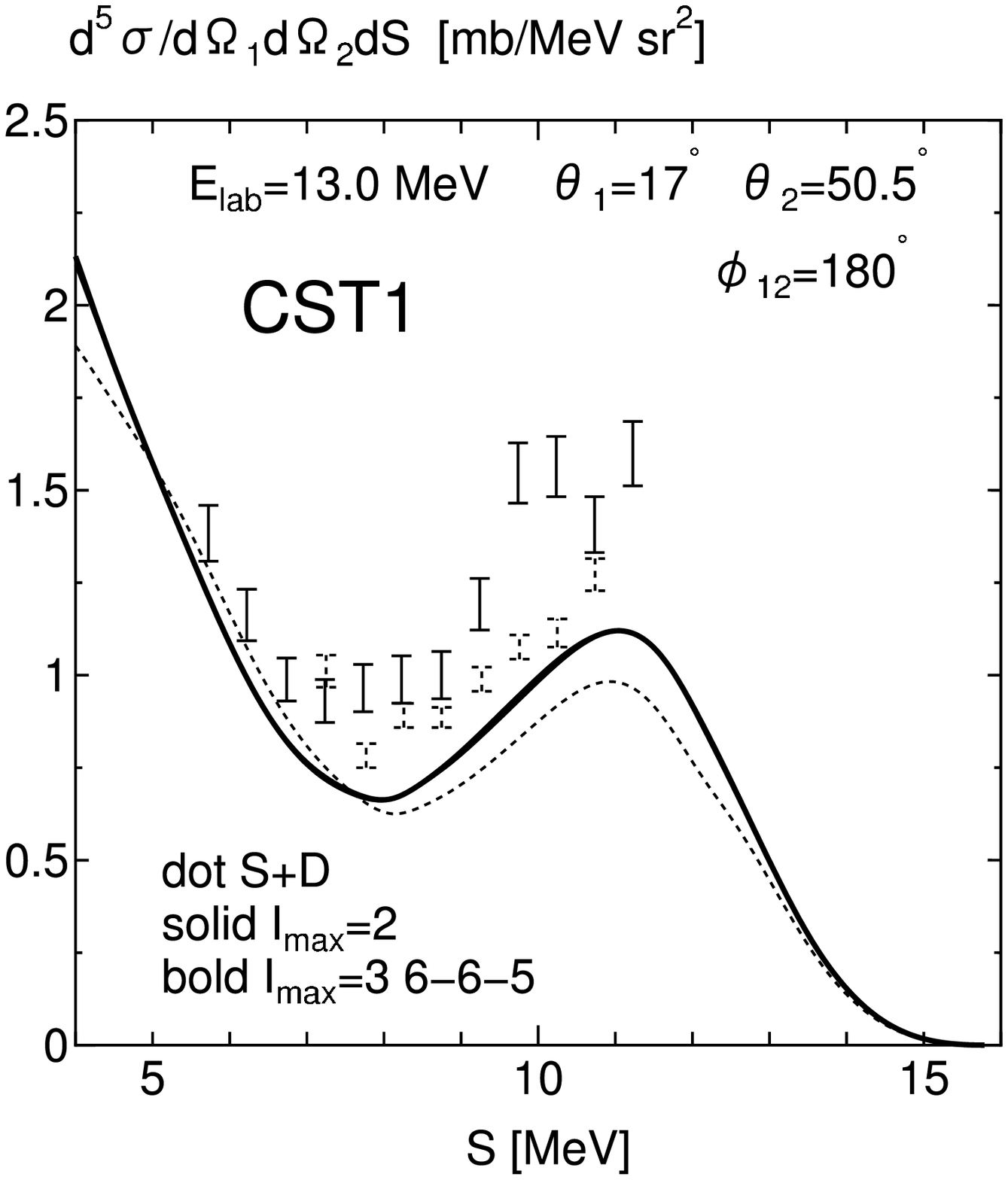}

\includegraphics[angle=0,width=55mm]
{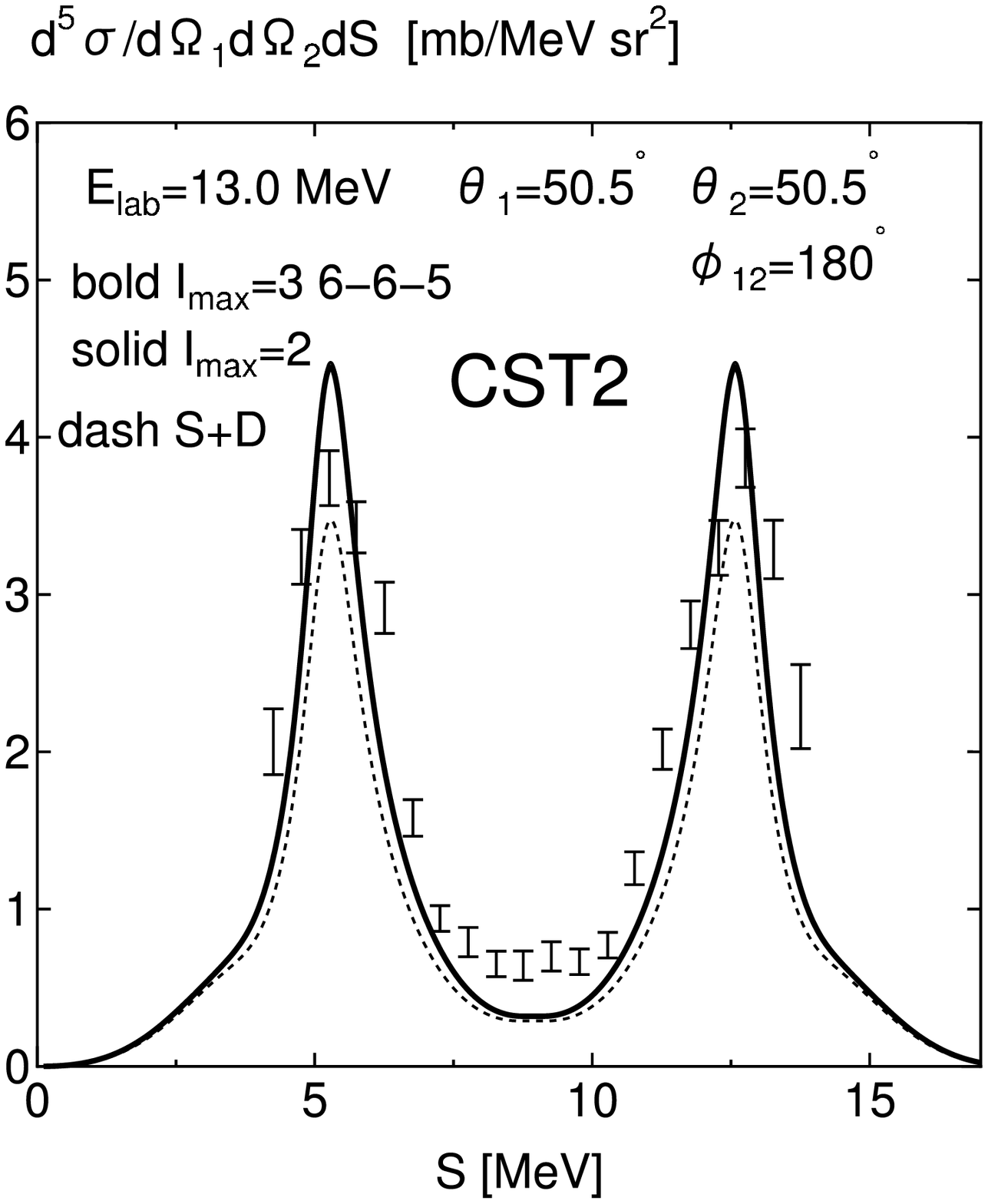}

\includegraphics[angle=0,width=55mm]
{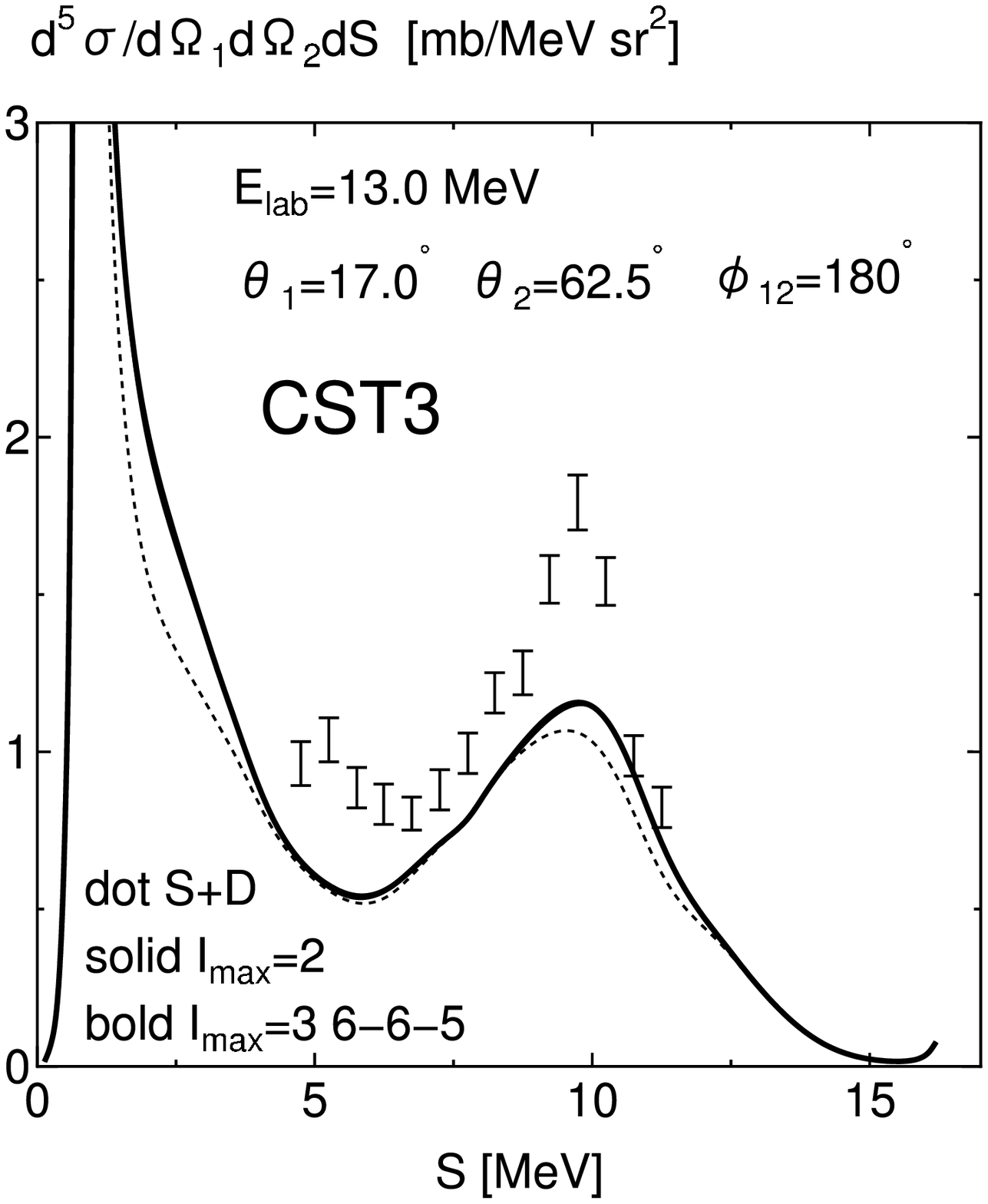}
\end{minipage}~%
\hfill~%
\begin{minipage}{0.48\textwidth}
\includegraphics[angle=0,width=55mm]
{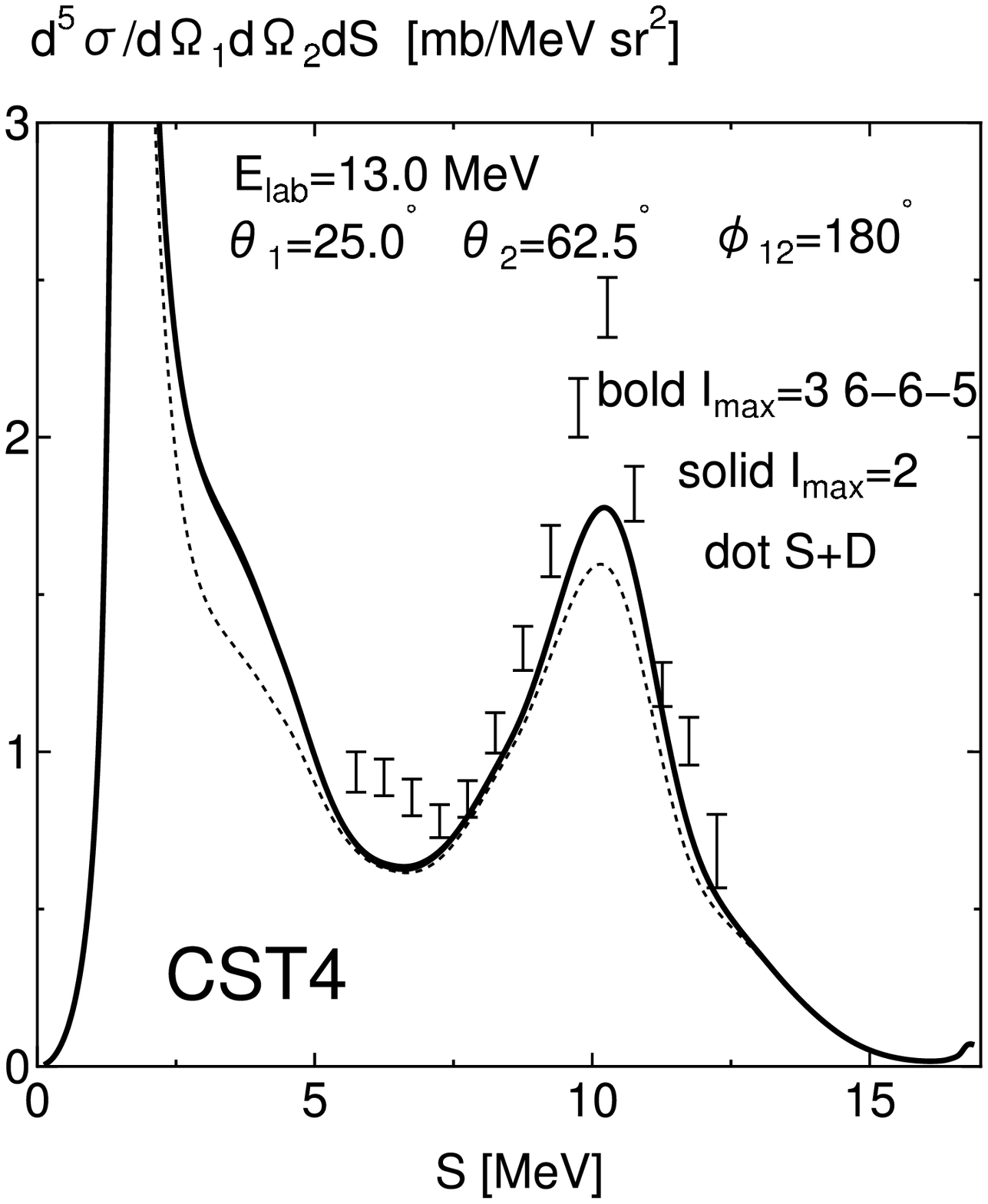}

\includegraphics[angle=0,width=55mm]
{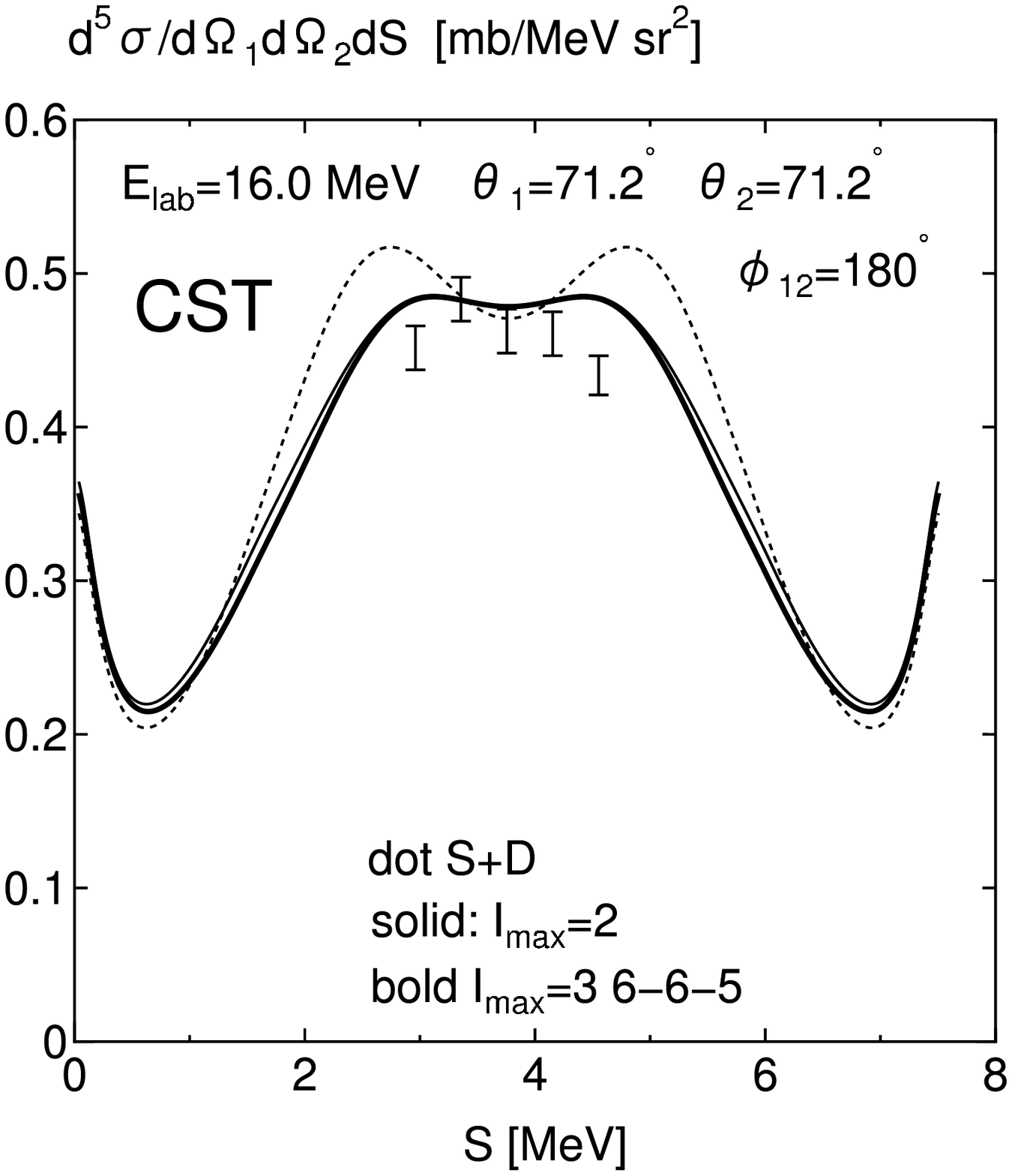}

\includegraphics[angle=0,width=55mm]
{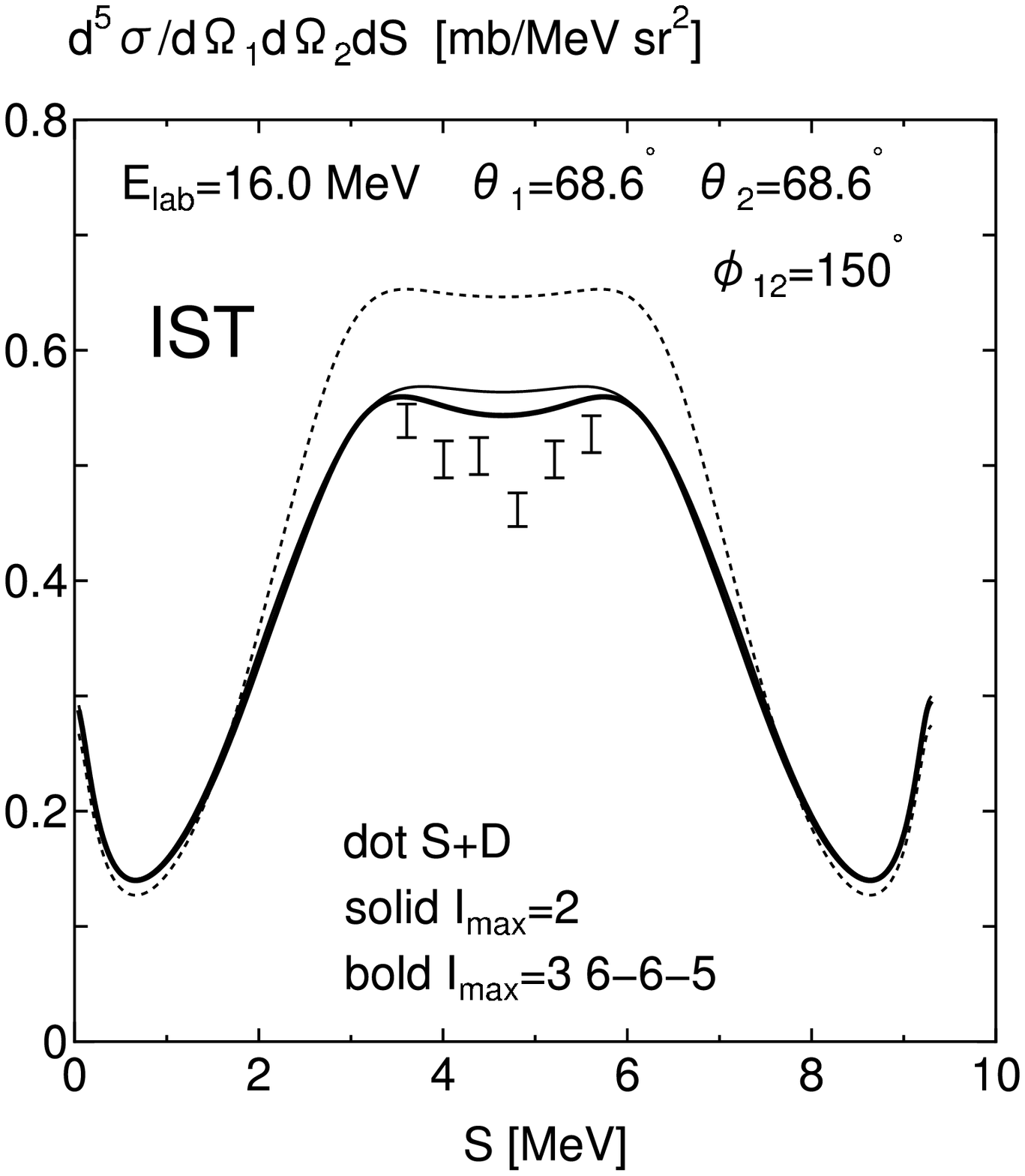}
\end{minipage}
\end{center}
\caption{
Breakup differential cross sections for the coplanar
star (CST) configurations at the energies $E_{\rm lab}=13$ and 16 MeV.
The result for the intermediate star (IST) configuration is also
shown at $E_{\rm lab}=16$ MeV.
The experimental data are taken from Refs.\,\citen{Se96,St89} ($nd$)
for 13 MeV, \citen{Du05} ($pd$) for 16 MeV.
}
\label{fig7}
\end{figure}

\begin{figure}[htb]
\begin{center}
\begin{minipage}{0.48\textwidth}
\includegraphics[angle=0,width=55mm]
{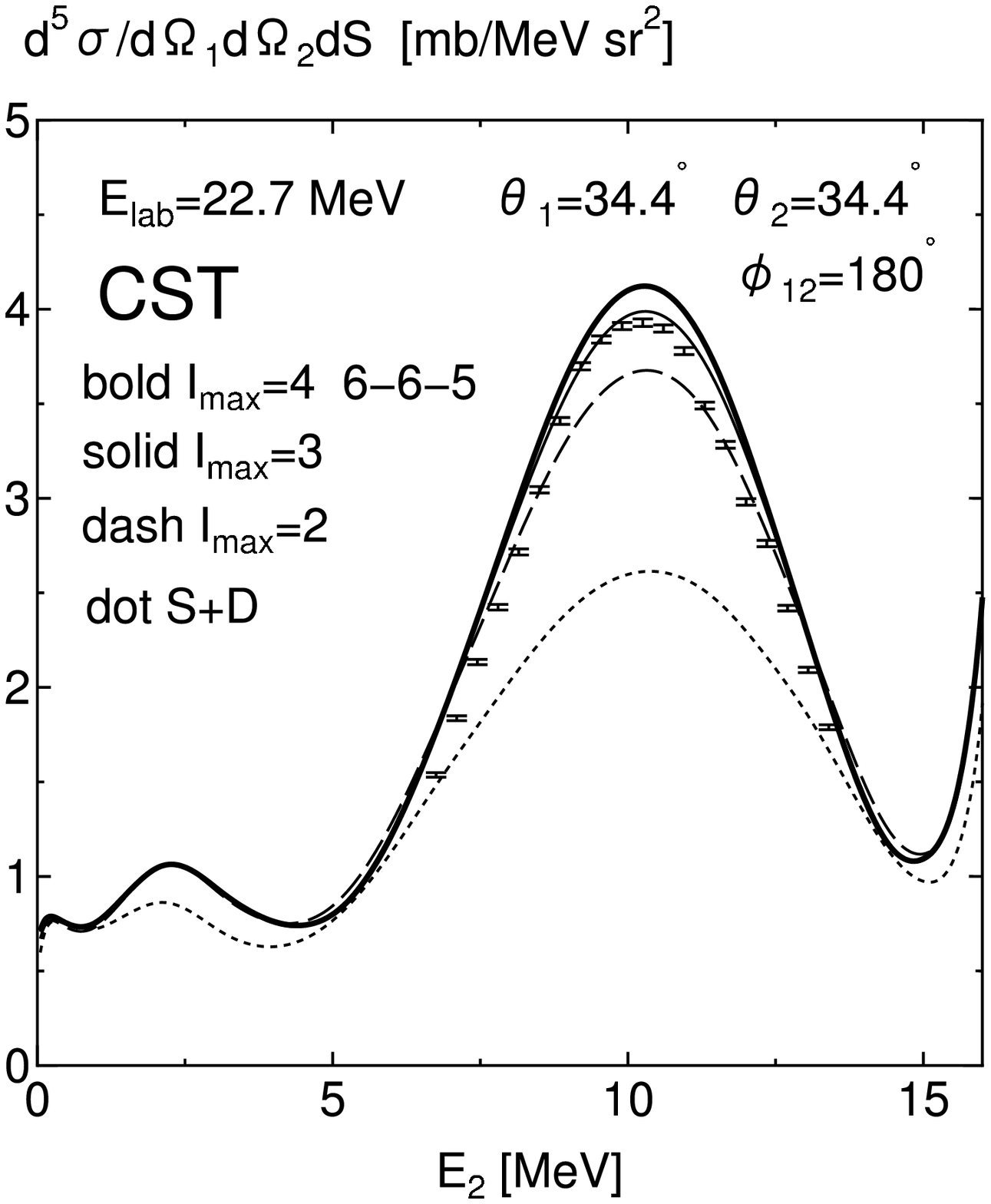}

\includegraphics[angle=0,width=55mm]
{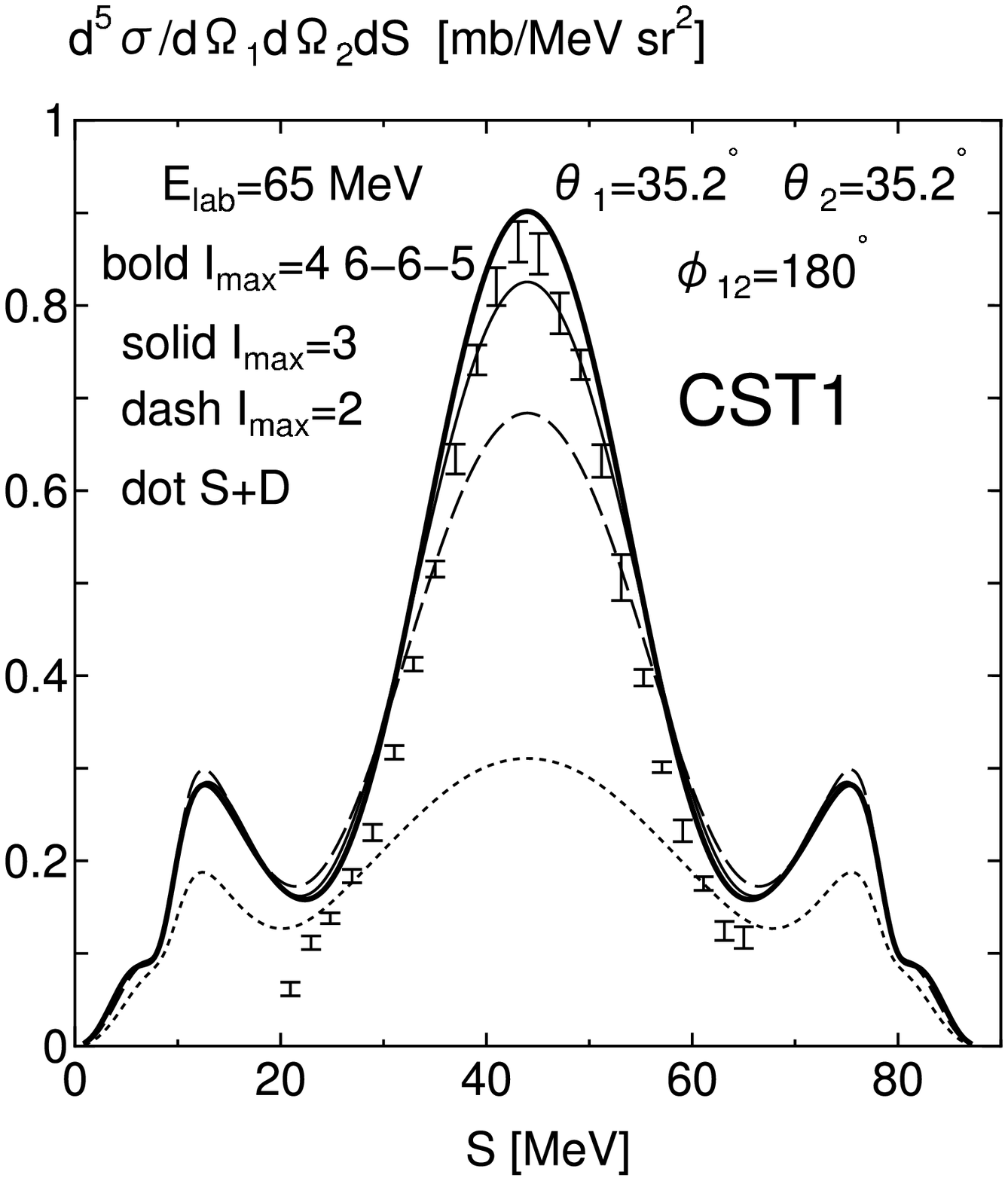}
\end{minipage}~%
\hfill~%
\begin{minipage}{0.48\textwidth}
\includegraphics[angle=0,width=55mm]
{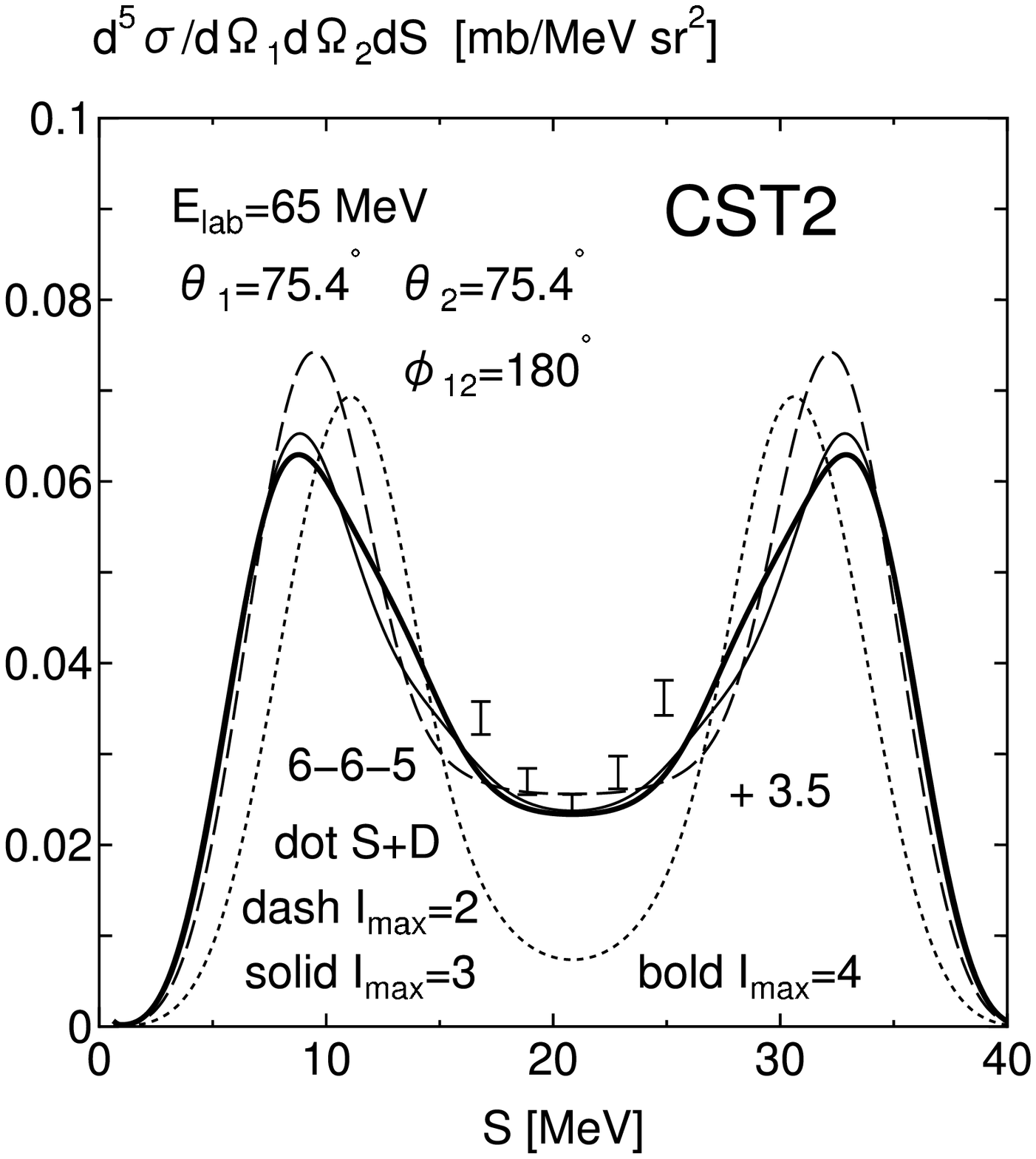}

\vspace{65mm}
\end{minipage}
\end{center}
\caption{
The same as Fig.\,\protect\ref{fig7}, but for the energies
$E_{\rm lab}=22.7$ and 65 MeV.
The experimental data are taken from Refs.\,\citen{Fo85} ($pd$) for 22.7 MeV
and \citen{Ze97} ($pd$) for 65 MeV.
}
\label{fig8}
\end{figure}

Let us move to the coplanar star configurations in Figs.\,\ref{fig7}
and \ref{fig8}.
The agreement between our calculation and the data is satisfactory
in general, but some deviations still exist. 
For example, in the first panel of 13 MeV, the new $nd$ data \cite{Se96}
are much closer to the calculation than the old data \cite{St89}, but
some underestimation still exists in the calculation.
The underestimation of the cross sections at the minimum point
$S=9$ MeV of CST2, and also 
at the $np$ final state interaction peaks around $S=10$ MeV 
in CST3 and in CST4 are the common feature with the meson-exchange
potentials. See Figs.~11,~12 and 13 of Ref.\,\citen{St89}.
For 16 MeV reactions, we have given a comparison not only for the 
coplanar star configuration, but also for the intermediate star (IST)
configuration, both of which are very similar to the predictions
by other models given in Ref.\,\citen{Du05}.
For $E_{\rm lab}=22.7$ MeV data, the curves are not plotted as a function
of $S$ but by $E_2$ for the second particle.
For this and $E_{\rm lab}=65$ MeV reactions,
we find a large contribution of higher partial waves up to $I_{\rm max}=4$.
In the symmetric backward plane star configuration 
of 65 MeV, denoted by CST2, the original experimental data 
are shifted to the larger side of $S$ by 3.5 MeV, since the
starting position of $S=0$ does not seem to be the same between our
calculation and the experiment. 

\subsection{Collinear configurations}

\begin{figure}[htb]
\begin{center}
\begin{minipage}{0.48\textwidth}
\includegraphics[angle=0,width=55mm]
{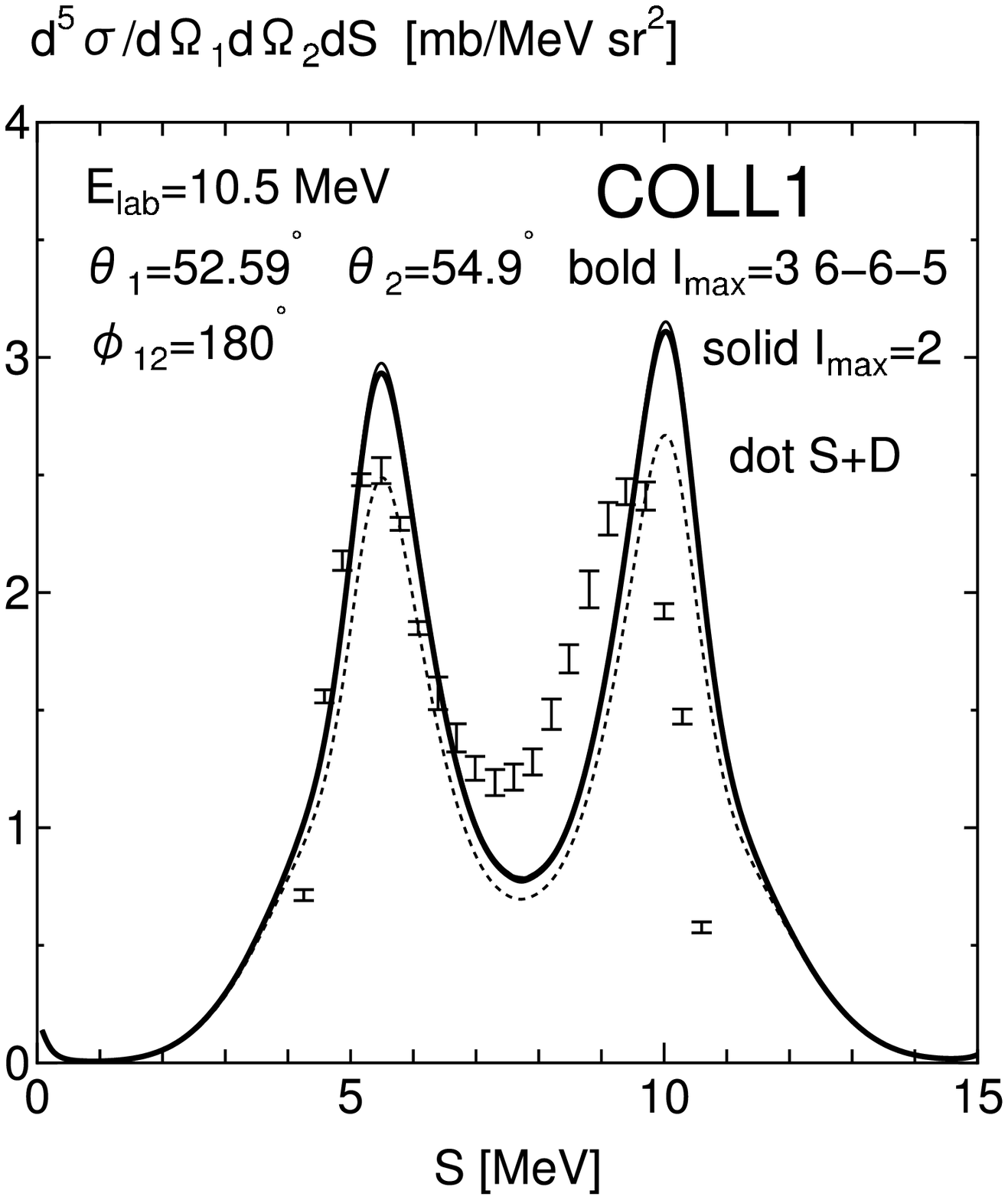}

\includegraphics[angle=0,width=55mm]
{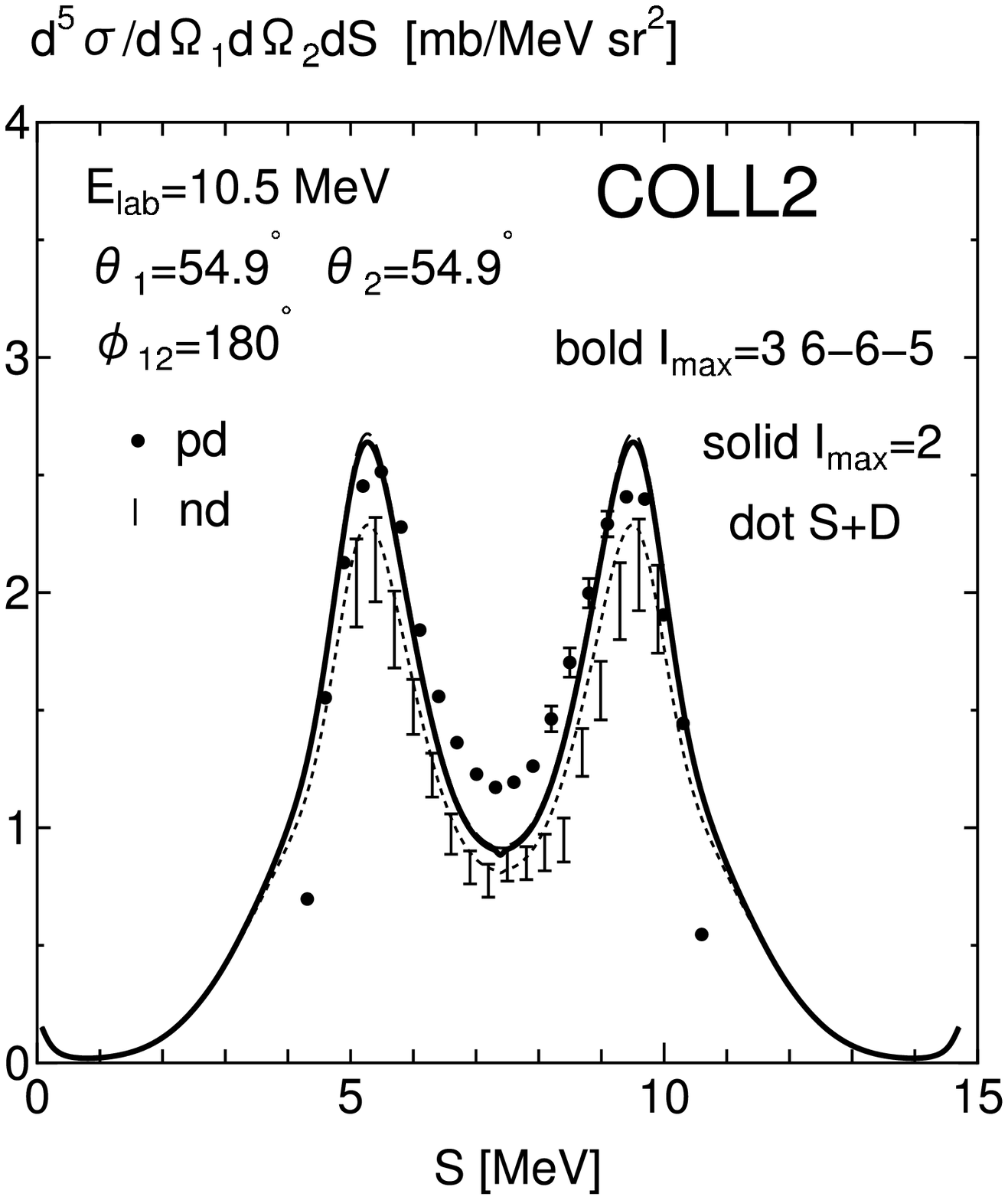}

\includegraphics[angle=0,width=55mm]
{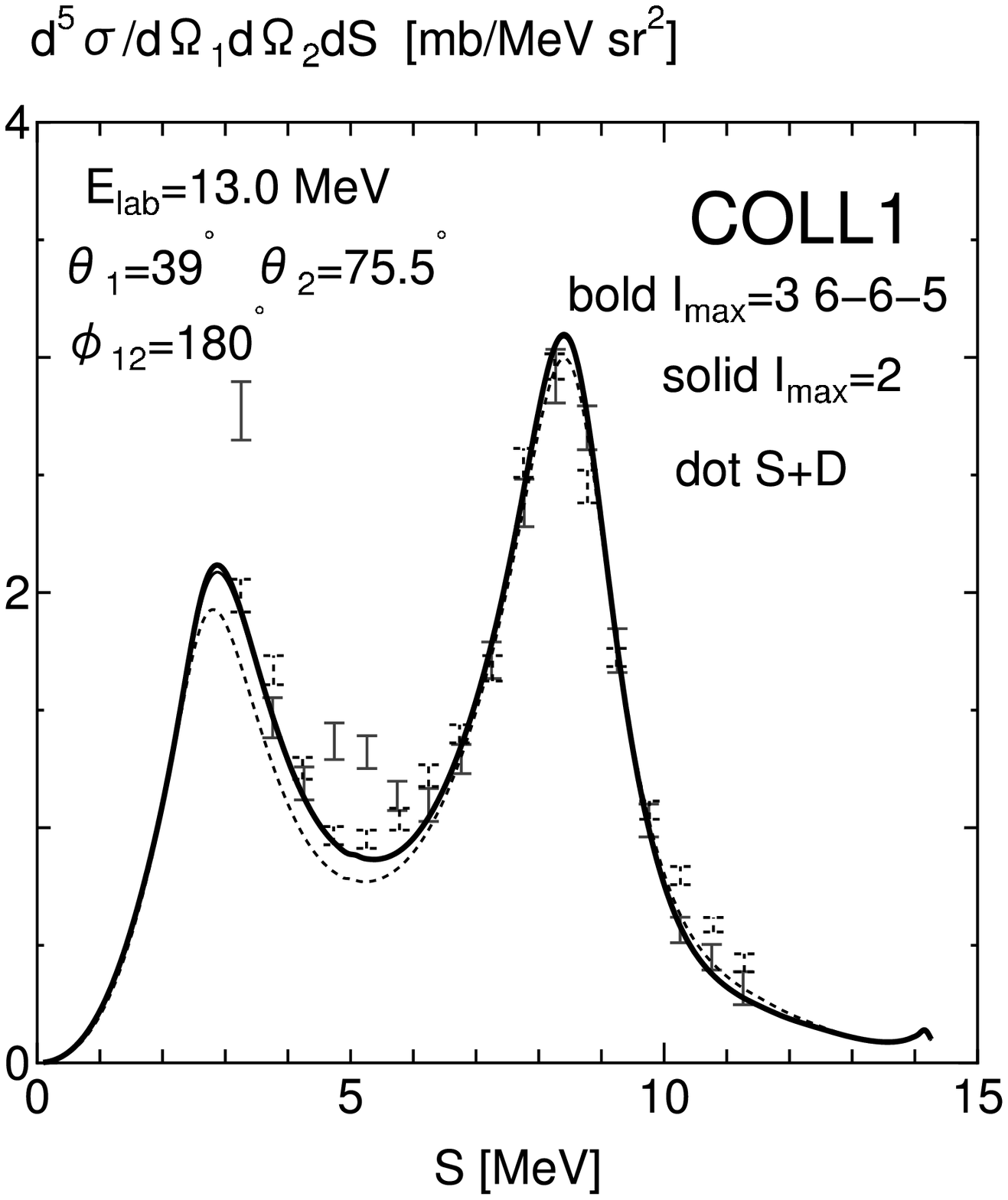}
\end{minipage}~%
\hfill~%
\begin{minipage}{0.48\textwidth}
\includegraphics[angle=0,width=55mm]
{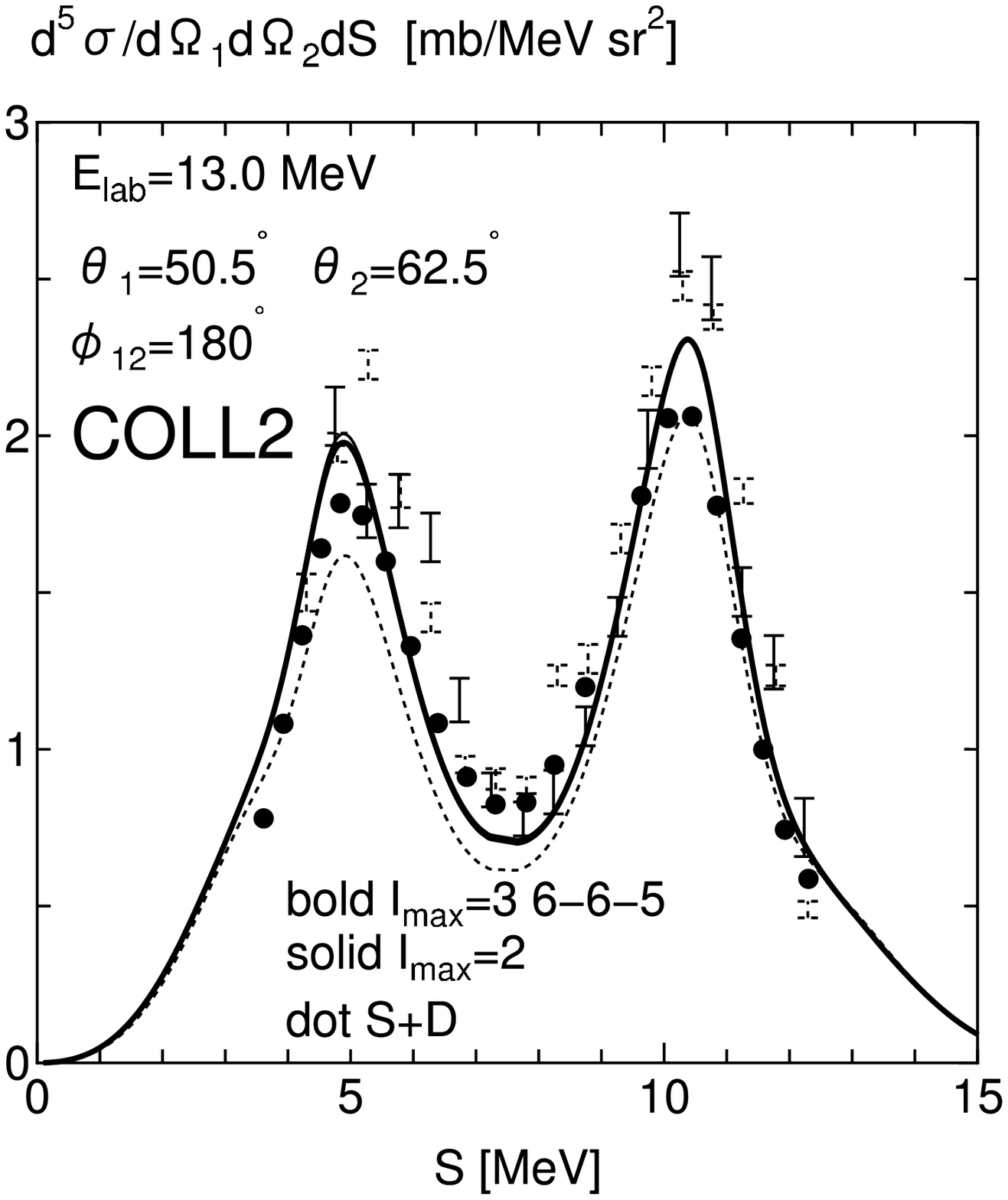}

\includegraphics[angle=0,width=55mm]
{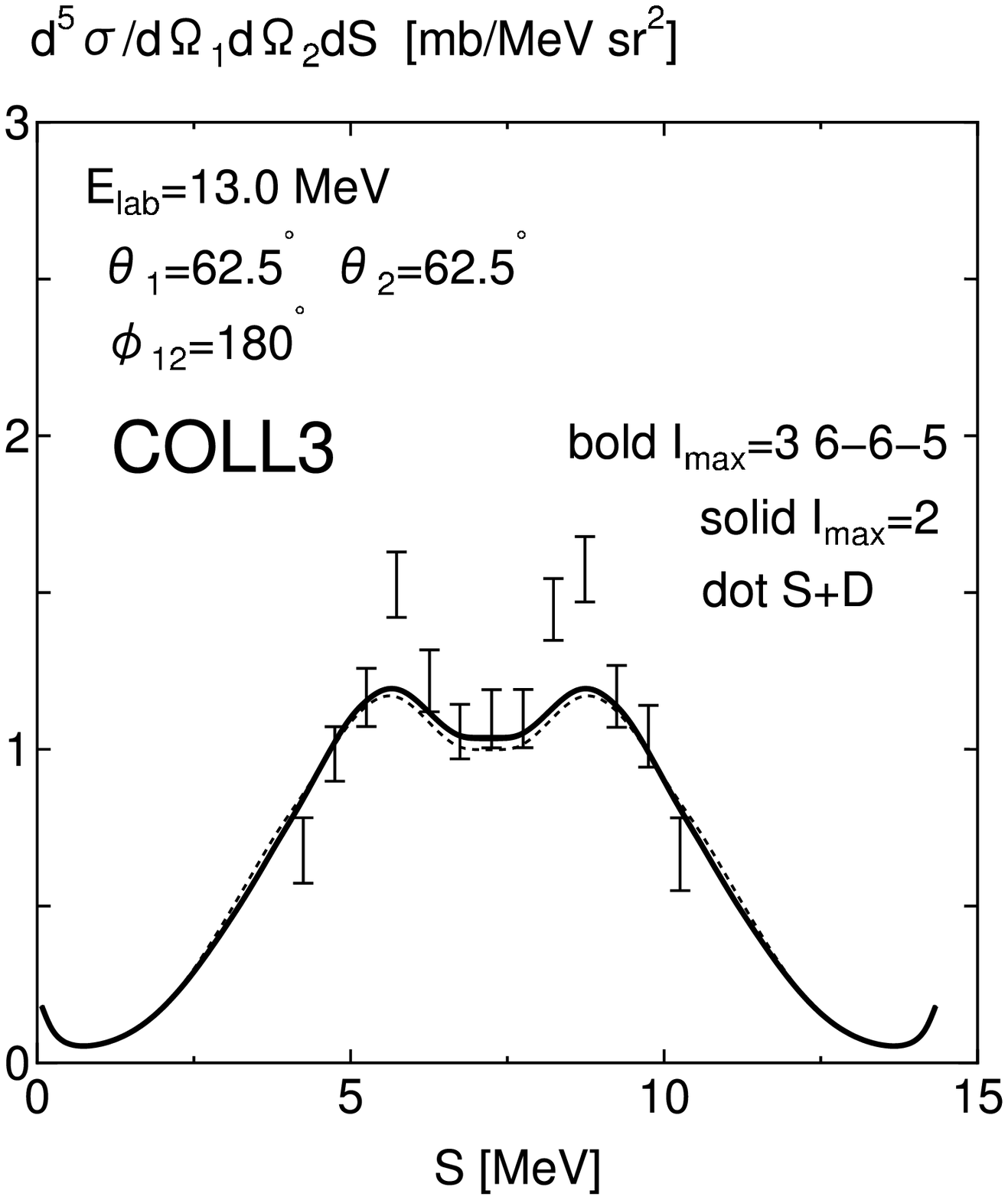}

\includegraphics[angle=0,width=55mm]
{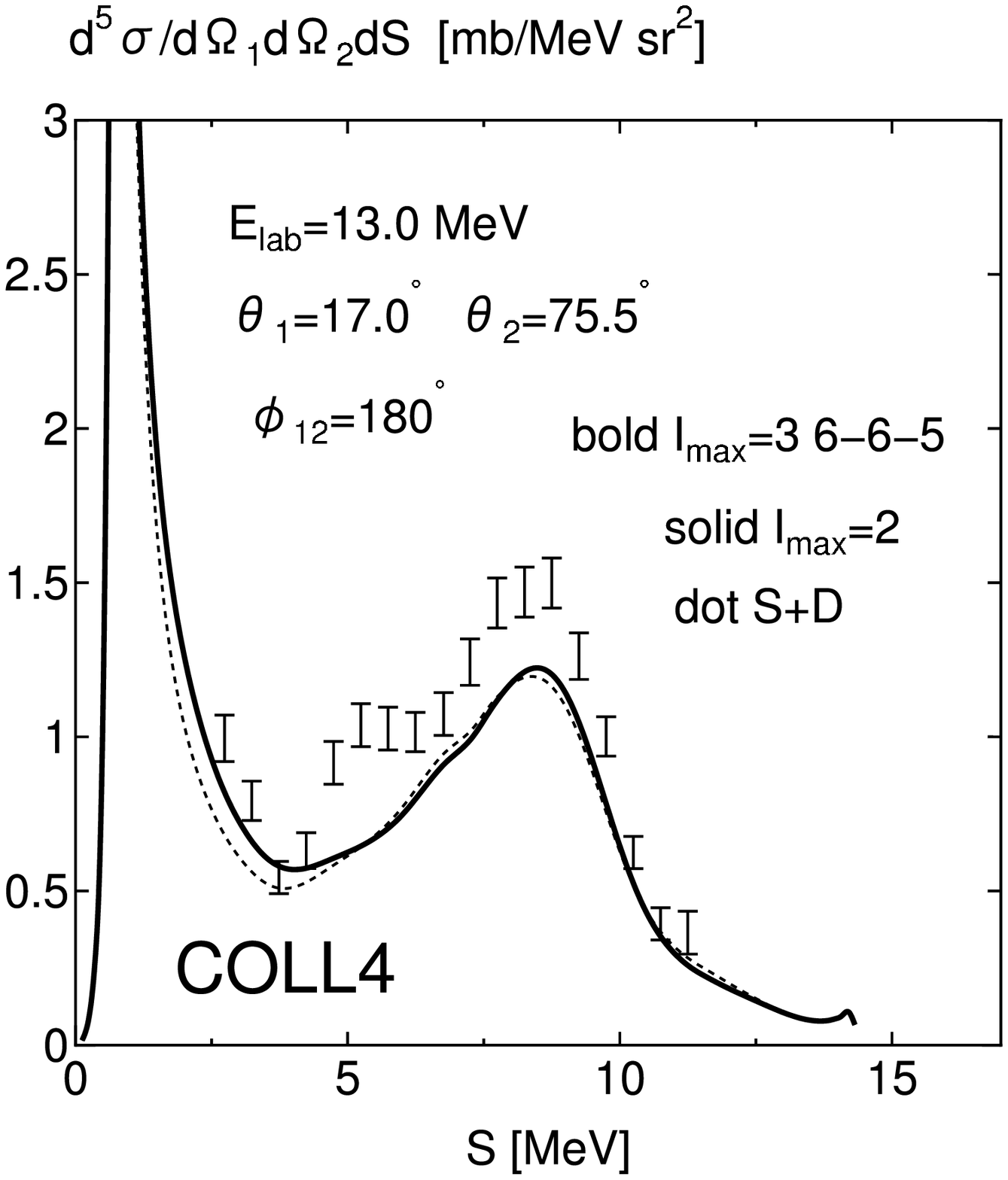}
\end{minipage}
\end{center}
\caption{
Breakup differential cross sections
for the collinear (COLL) configurations.
The experimental data are taken from Ref.\,\citen{Gr96} ($pd$ and $nd$)
for 10.5 MeV and Refs.\,\citen{Se96,St89} ($nd$) for 13 MeV.
}
\label{fig9}
\end{figure}

\begin{figure}[htb]
\begin{center}
\begin{minipage}{0.48\textwidth}
\includegraphics[angle=0,width=55mm]
{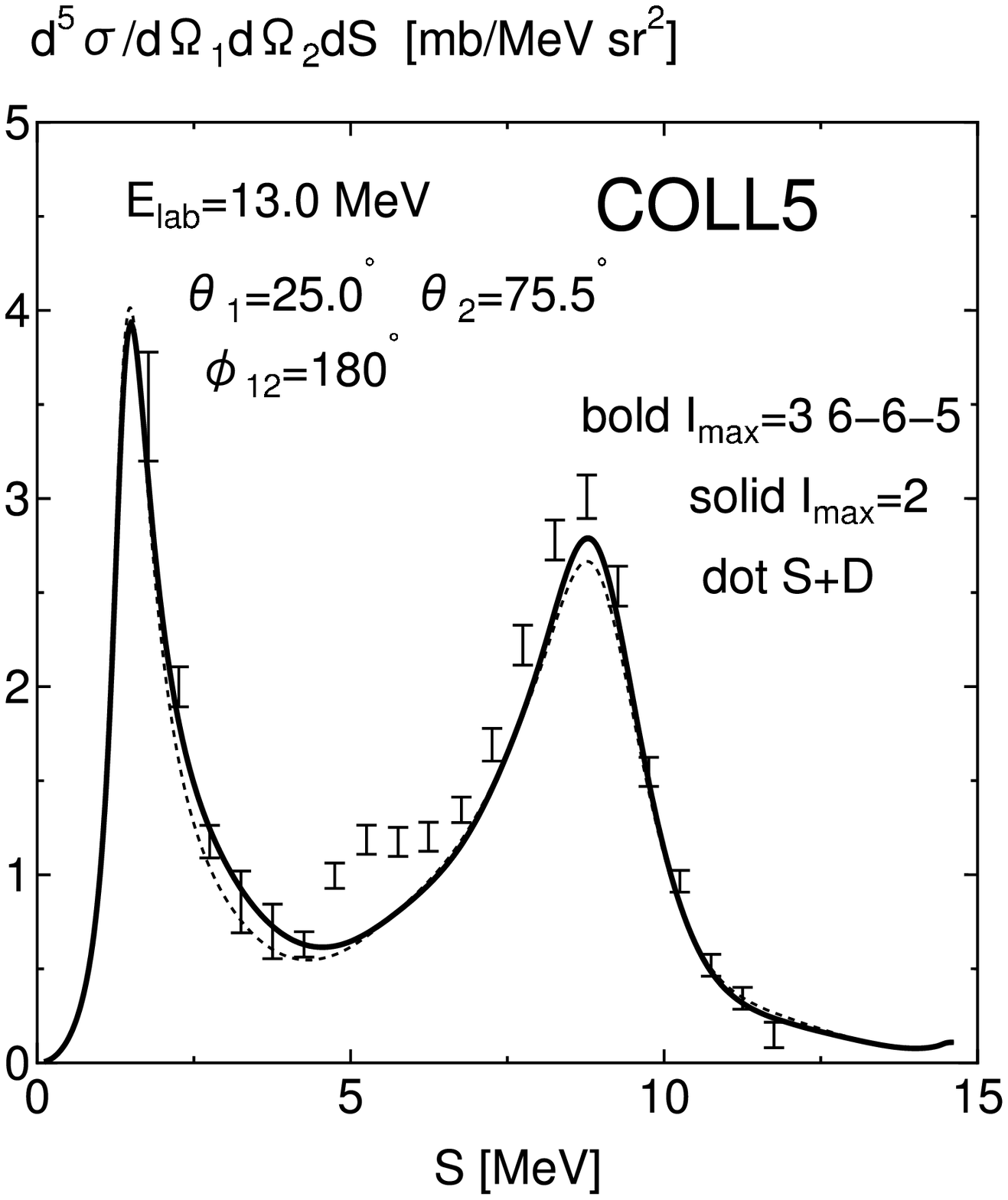}

\includegraphics[angle=0,width=55mm]
{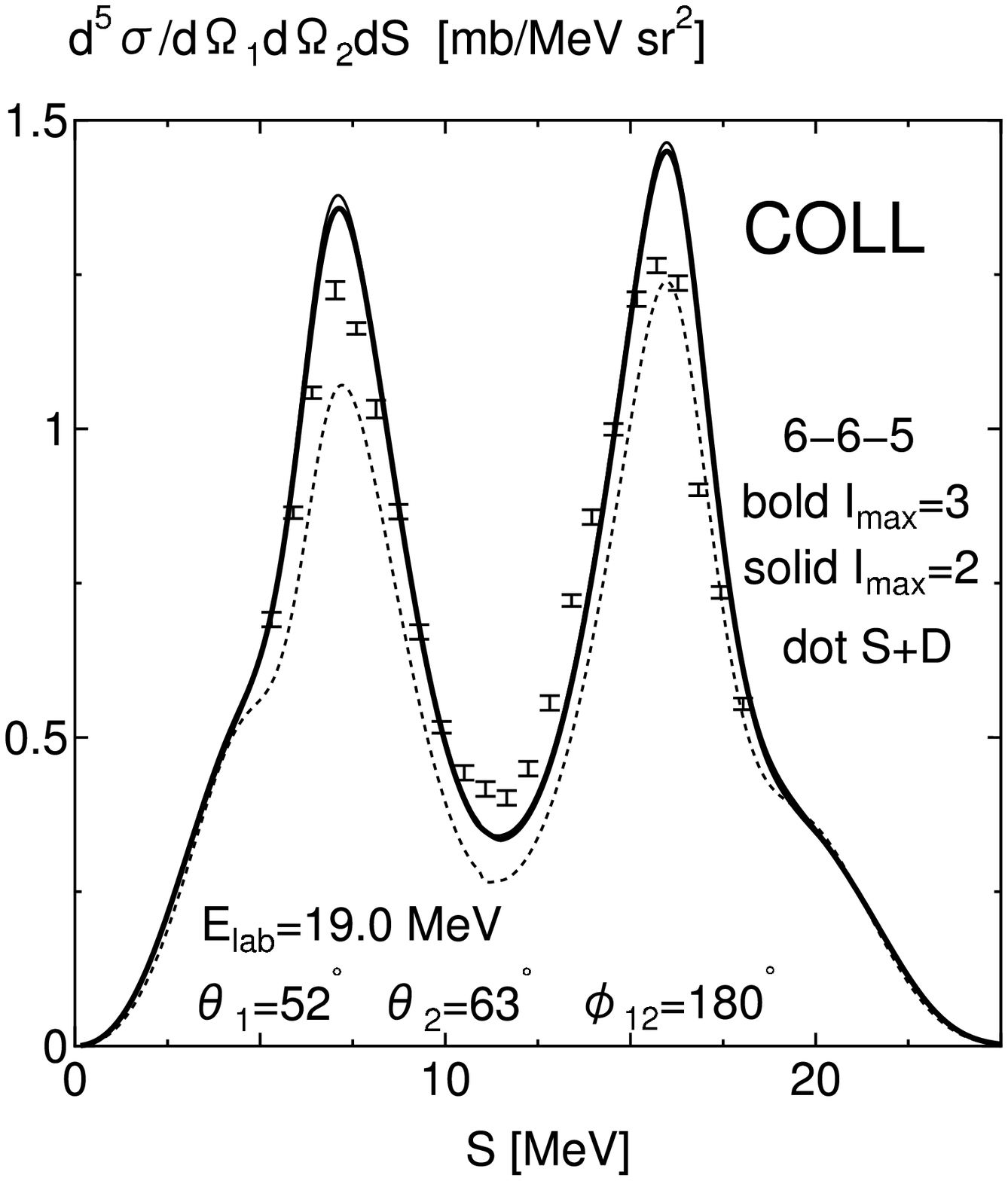}

\includegraphics[angle=0,width=55mm]
{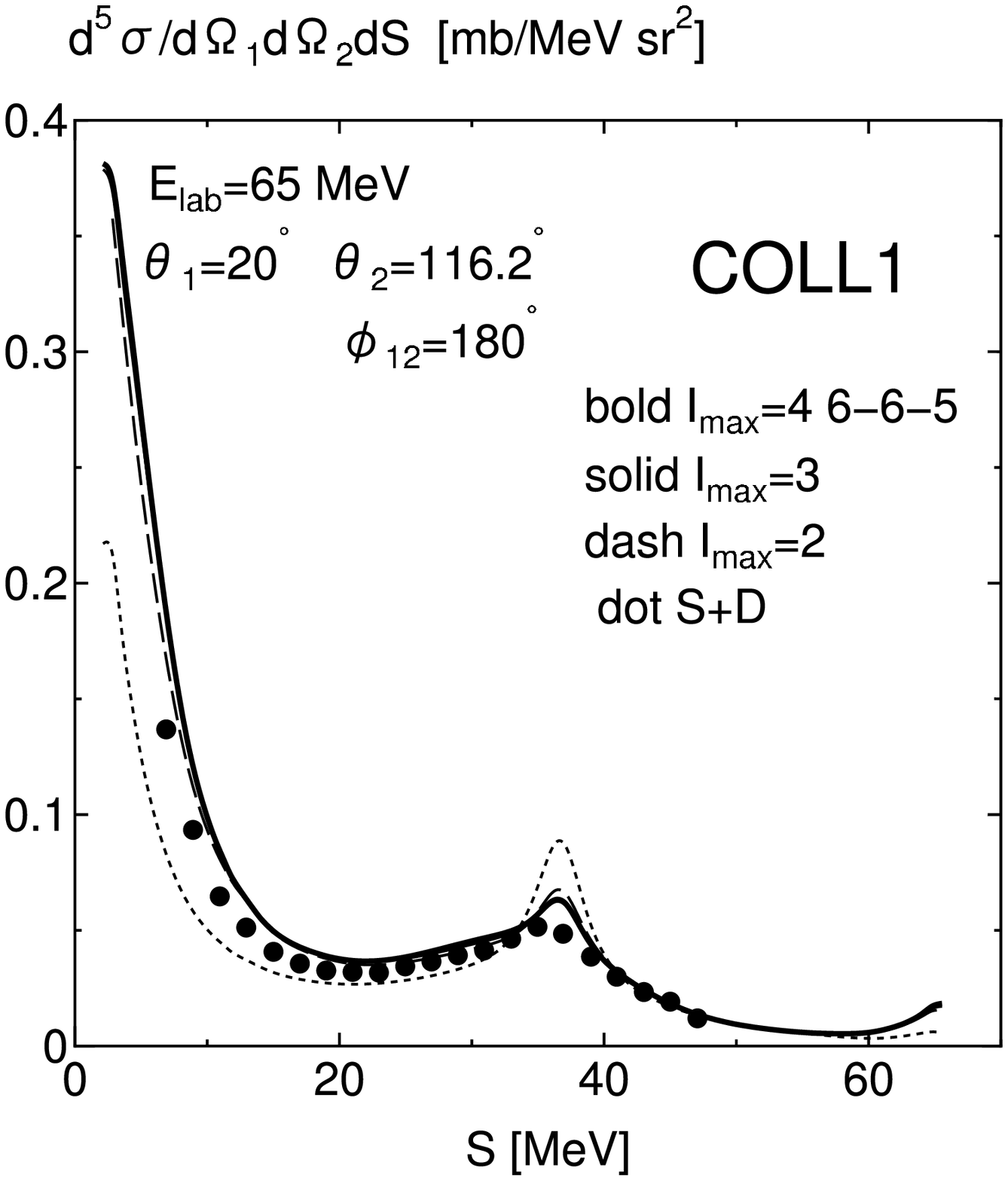}
\end{minipage}~%
\hfill~%
\begin{minipage}{0.48\textwidth}
\includegraphics[angle=0,width=55mm]
{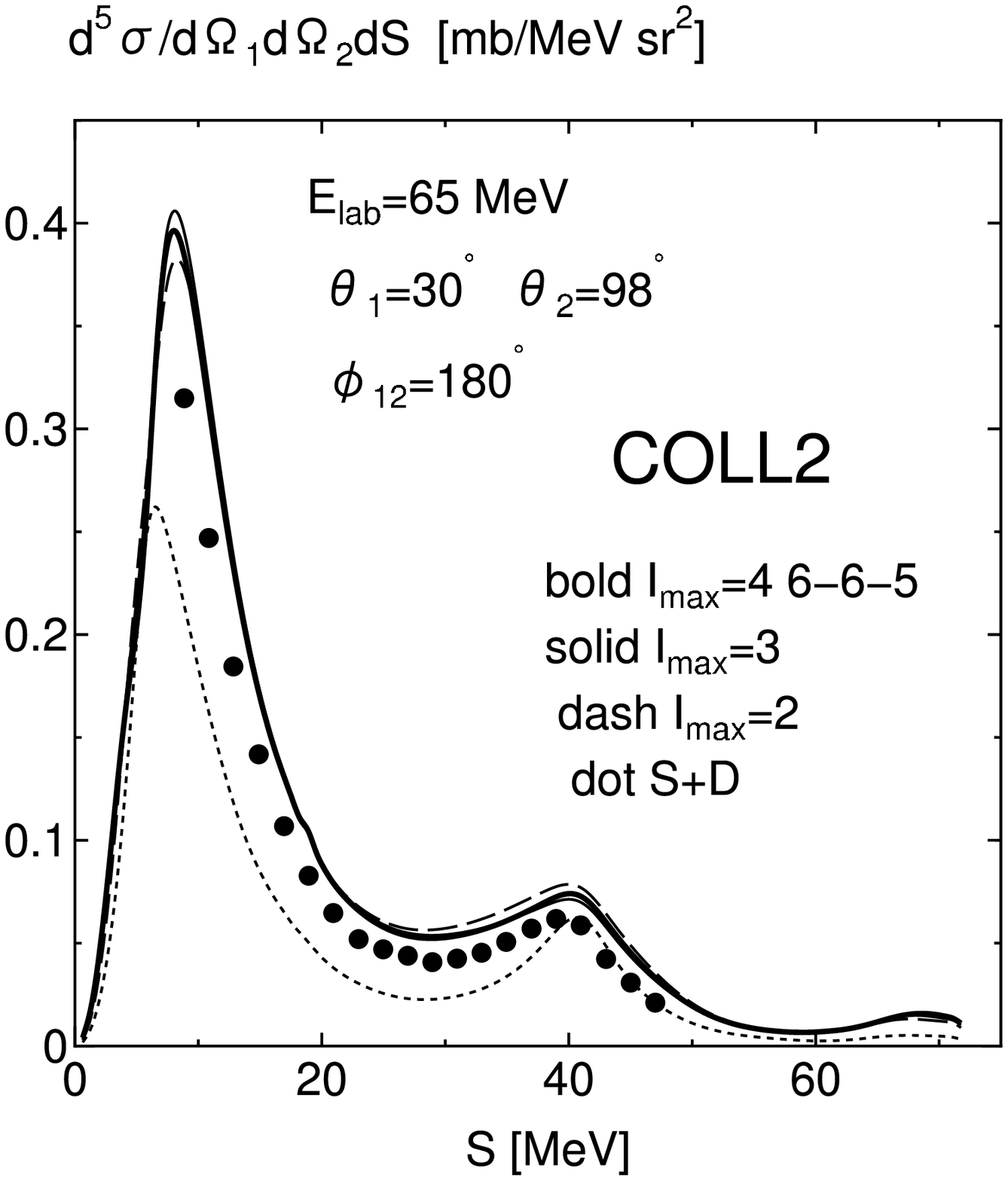}

\includegraphics[angle=0,width=55mm]
{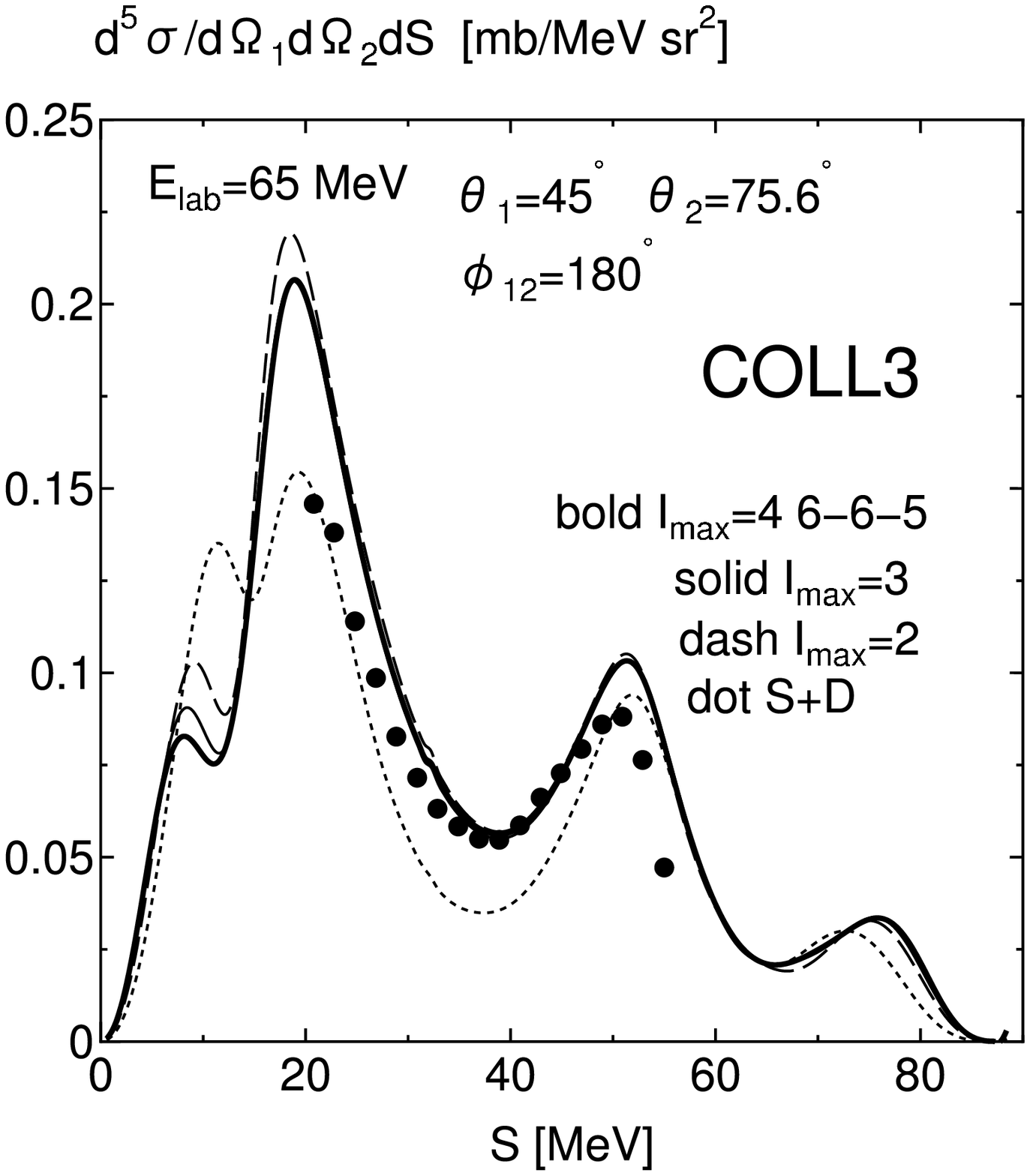}

\includegraphics[angle=0,width=54mm]
{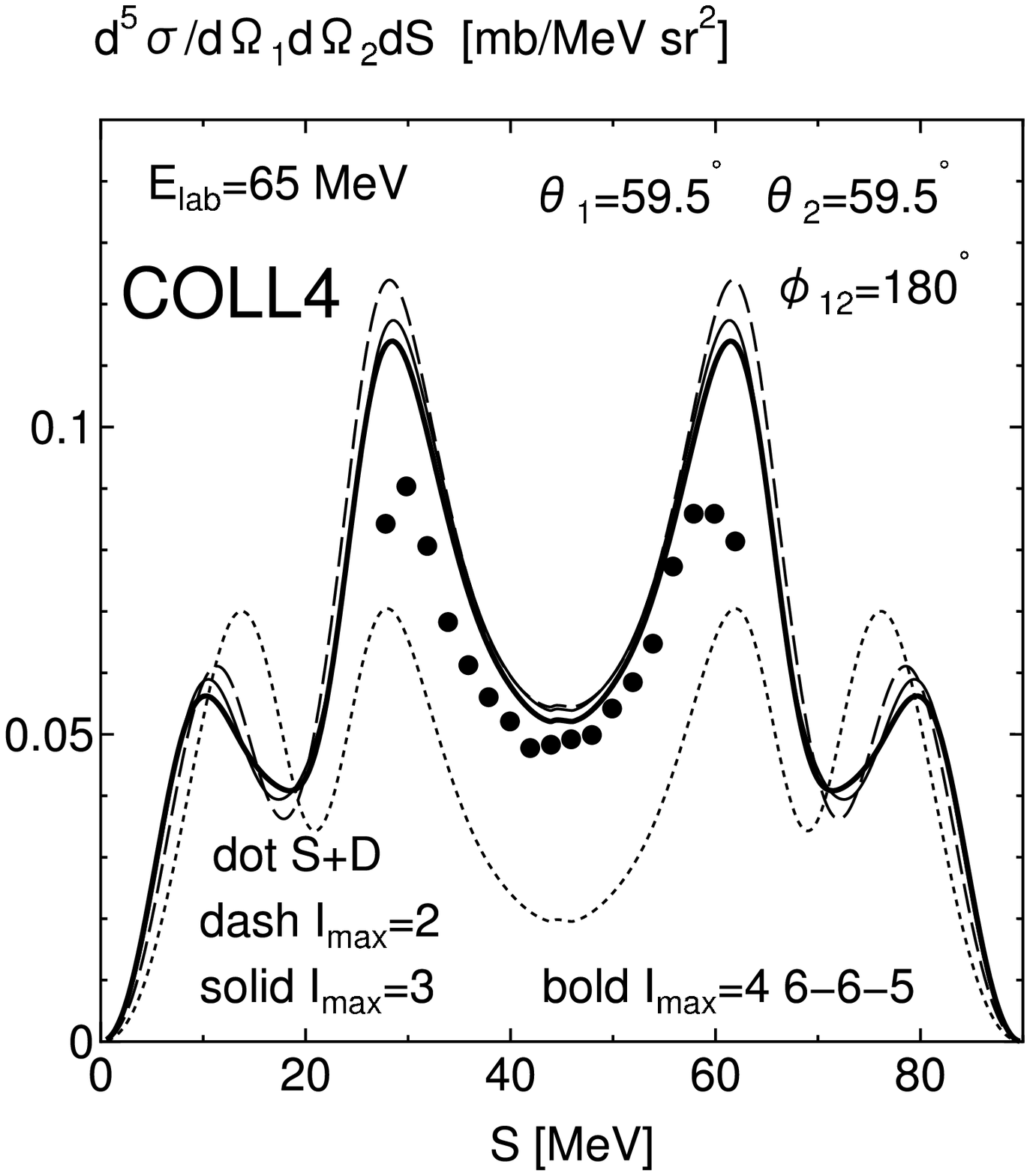}
\end{minipage}
\end{center}
\caption{
The same as Fig.\,\protect\ref{fig9}, but for other geometrical
configurations and energies.
The experimental data are taken from Refs.\,\citen{St89} ($nd$) for 13 MeV,
\citen{Pa96} ($pd$) for 19 MeV, and \citen{Al94} ($pd$) for 65 MeV.
}
\label{fig10}
\end{figure}

The comparison for the collinear configurations are displayed
in Figs.\,\ref{fig9} and \ref{fig10}.
For these configurations, the comparison with the experiment
is generally good. The Coulomb force has an appreciable effect 
to increase the breakup cross sections at the collinear point,
especially on the low-energy side, \cite{De05b} which is much more important
than the $3N$ force effect. 
In the first panel with $E_{\rm lab}=10.5$ MeV (COLL1),
we find some kinematical mismatch of the final state interaction peak
at $S \sim 10$ MeV. For COLL2 - COLL5 with $E_{\rm lab}=13$ MeV,
the small breakup cross sections around the collinear
points (minimum points) move to better 
direction to fit the experimental data by the expected Coulomb effect. 
This would also be true for the minimum point for $E_{\rm lab}=19$ MeV.
On the other hand, the breakup cross sections in COLL1 - COLL4 for 
$E_{\rm lab}=65$ MeV seem to be slightly overestimated.

\subsection{Non-standard configurations}

\begin{figure}[htb]
\begin{center}
\begin{minipage}{0.48\textwidth}
\includegraphics[angle=0,width=55mm]
{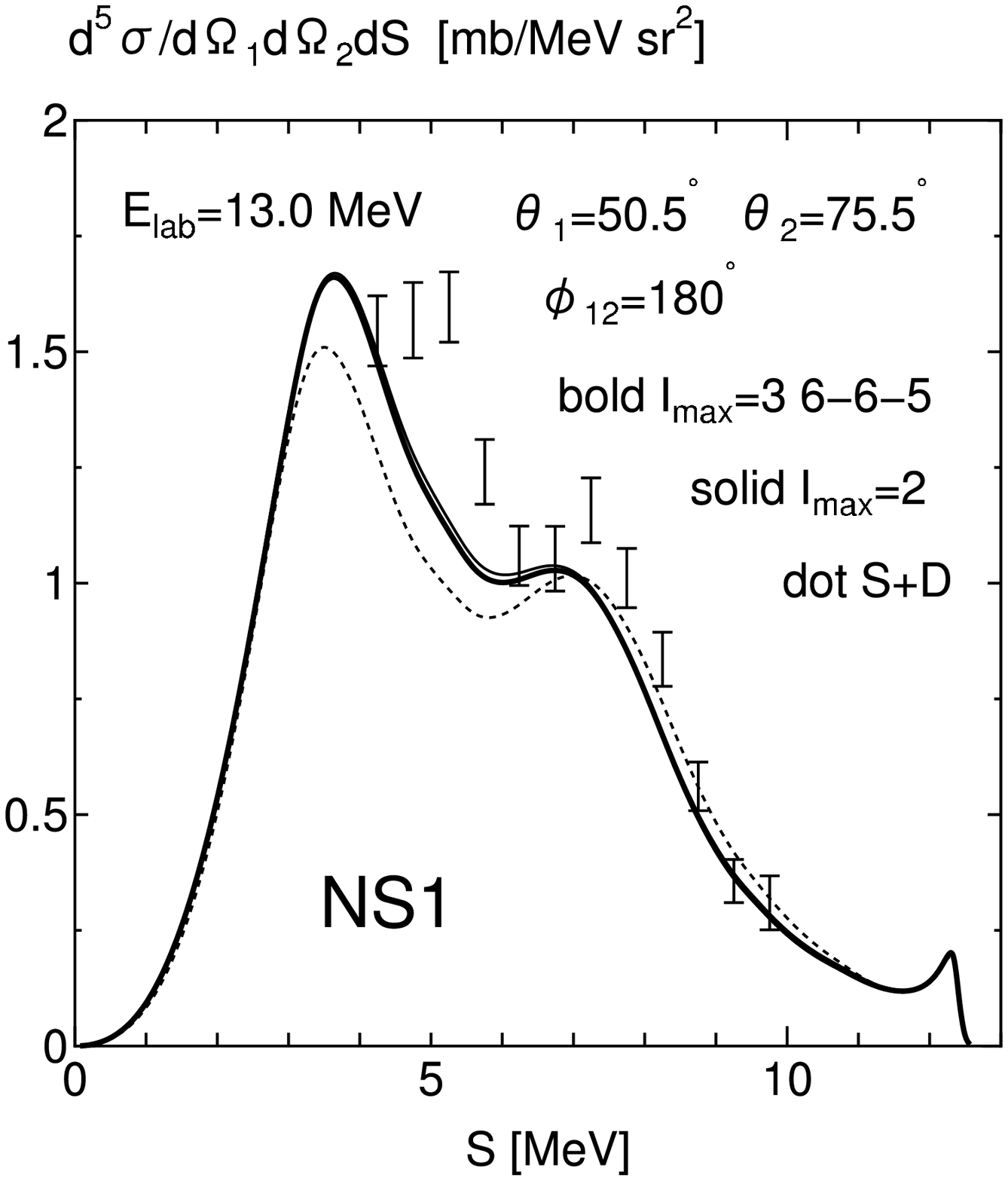}

\includegraphics[angle=0,width=55mm]
{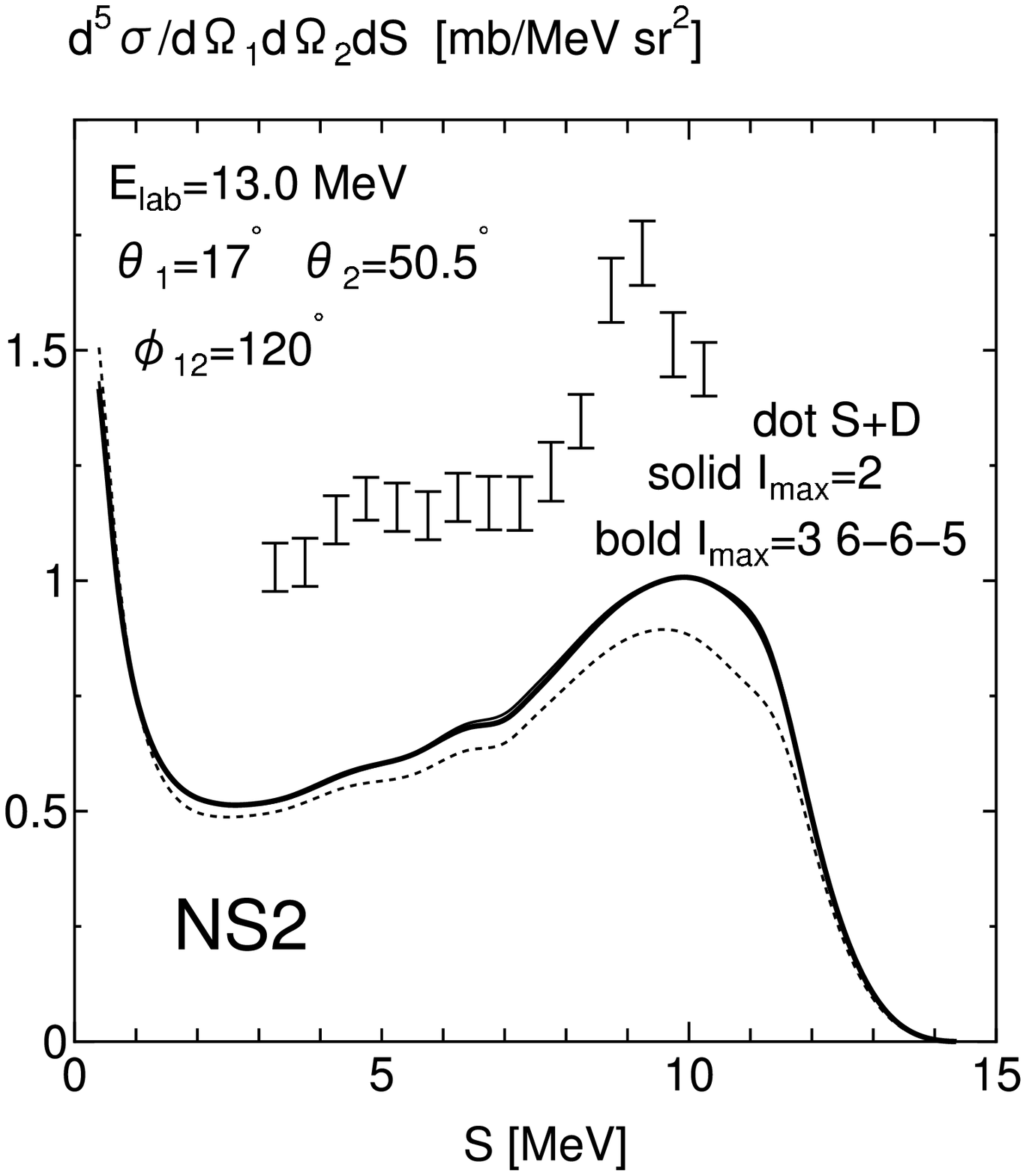}

\includegraphics[angle=0,width=55mm]
{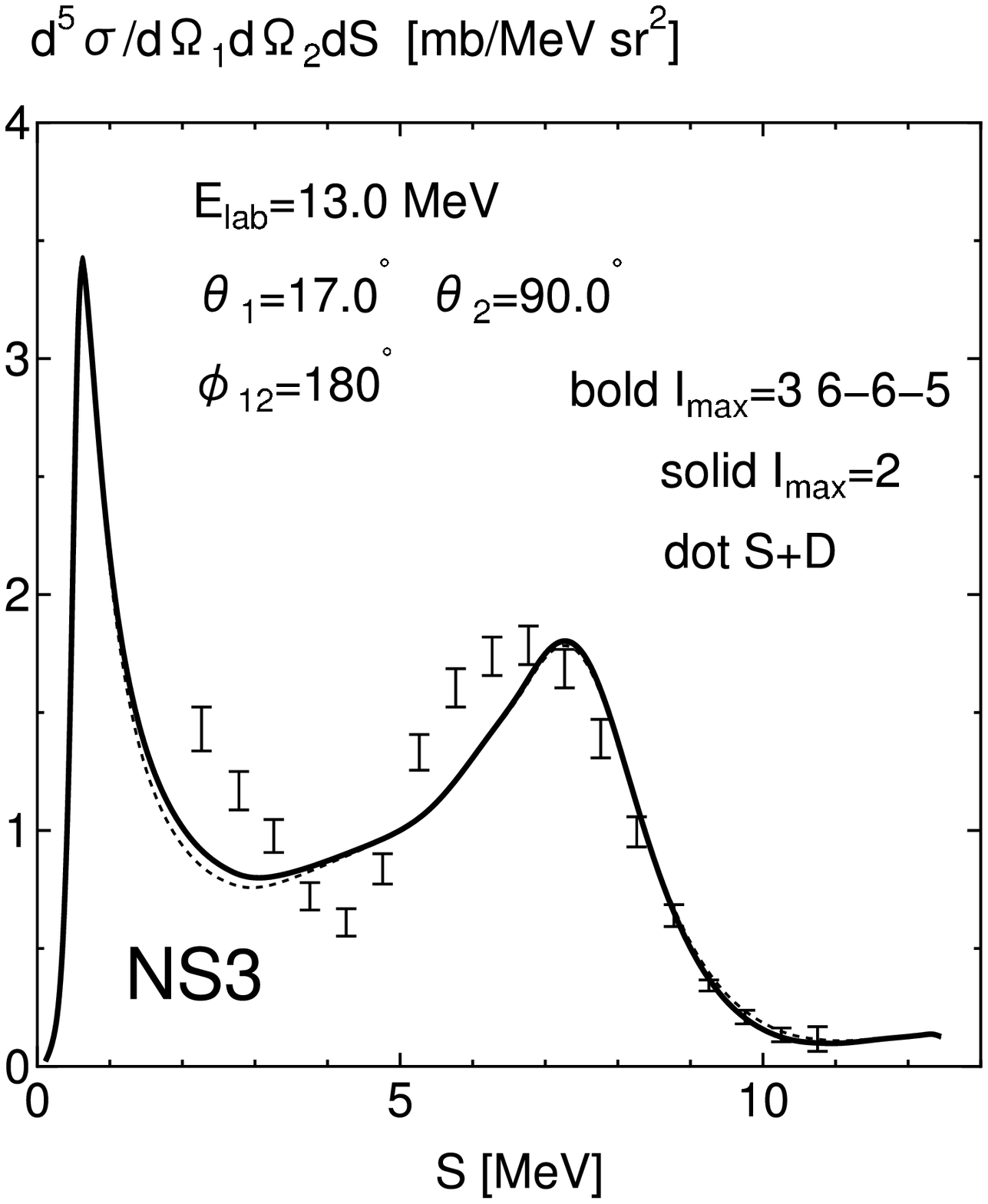}
\end{minipage}~%
\hfill~%
\begin{minipage}{0.48\textwidth}
\includegraphics[angle=0,width=55mm]
{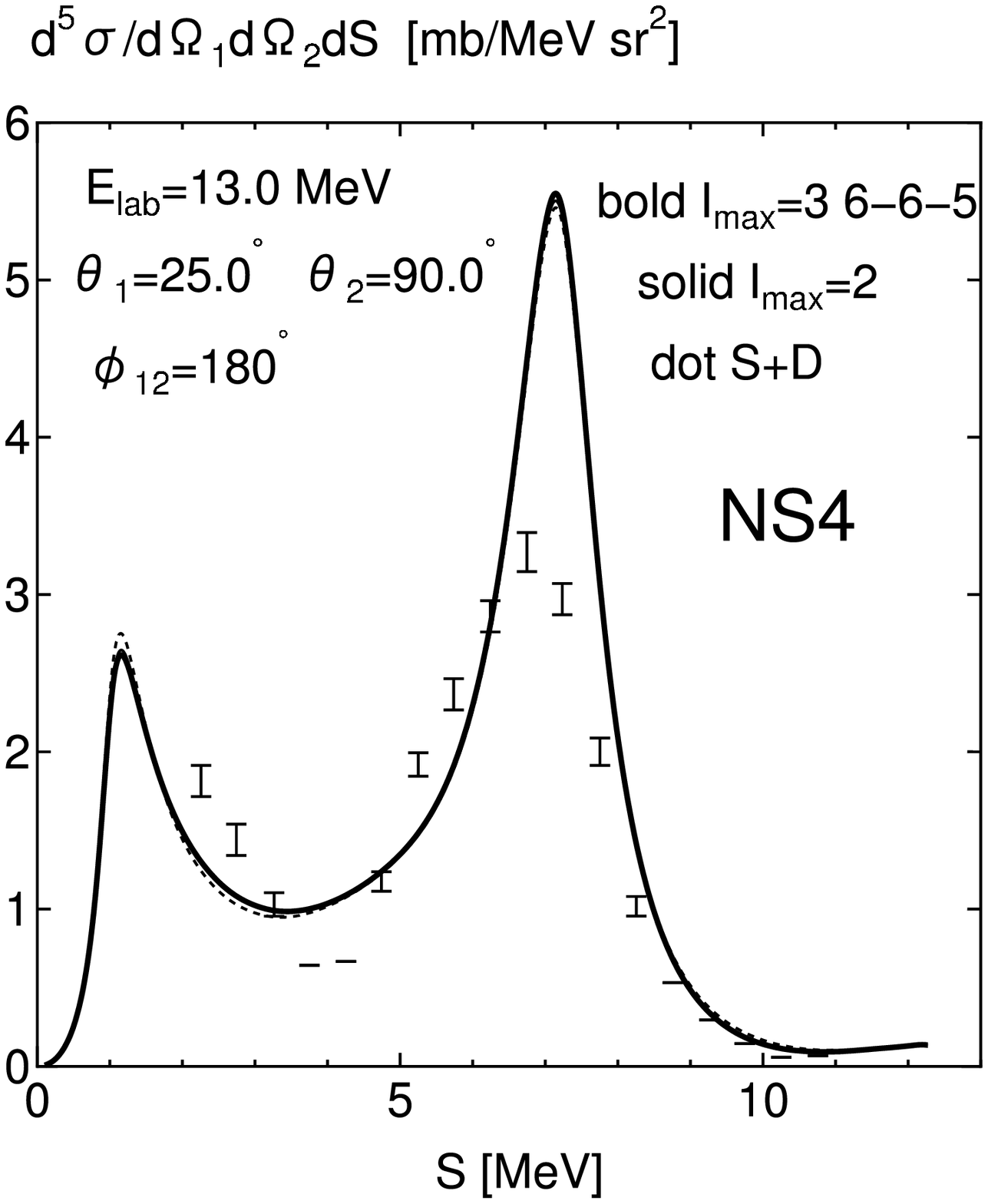}

\includegraphics[angle=0,width=55mm]
{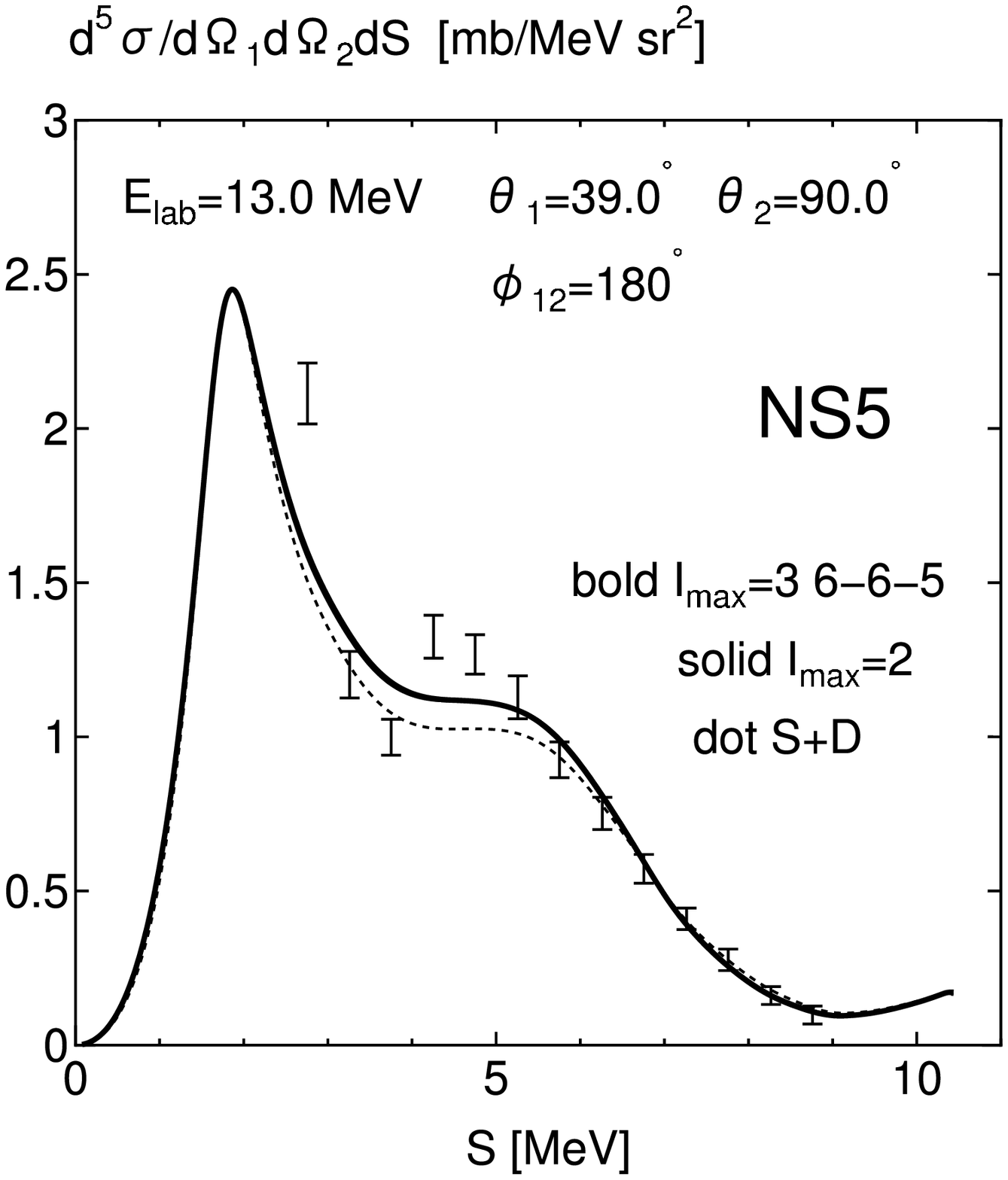}

\includegraphics[angle=0,width=55mm]
{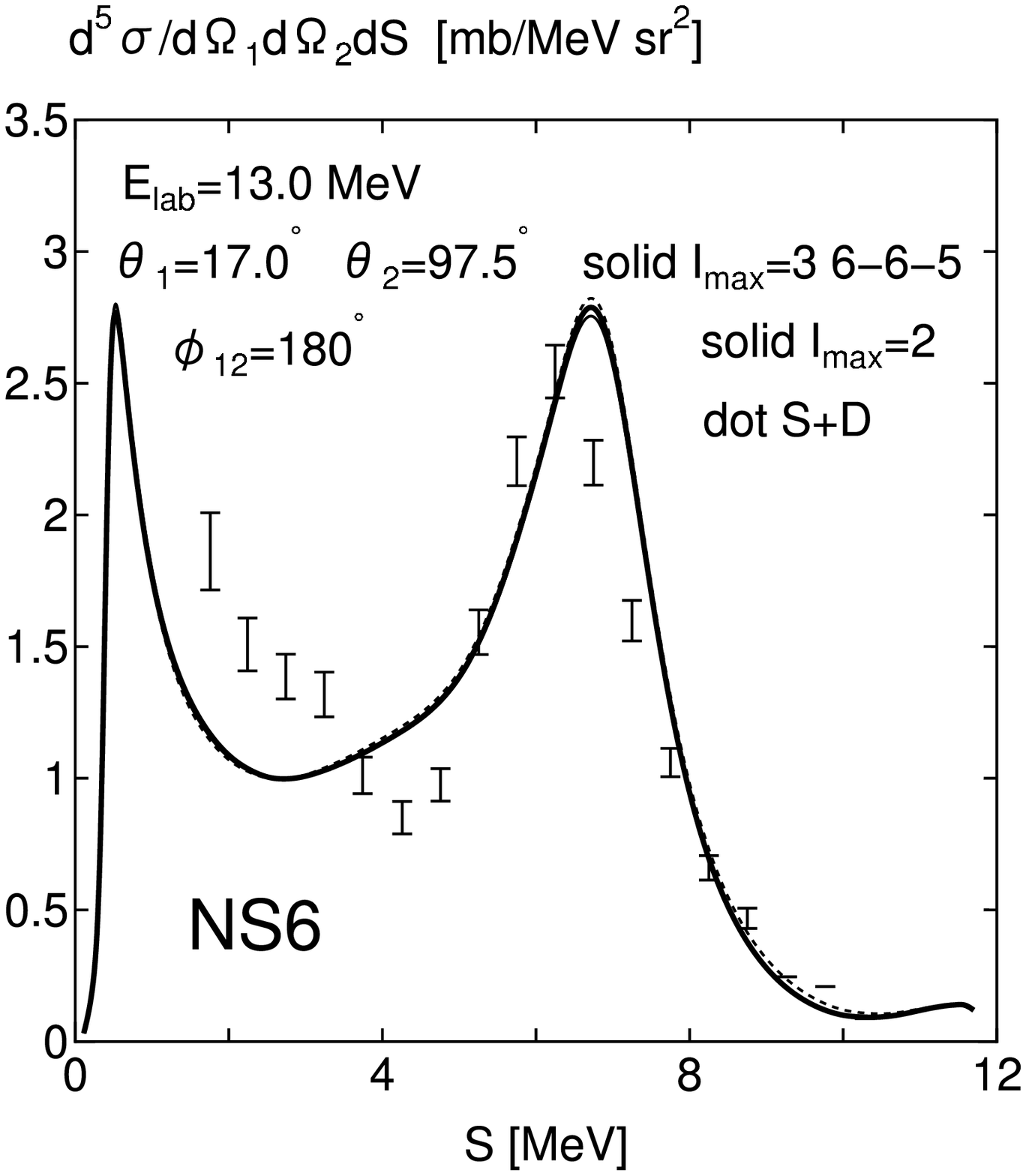}
\end{minipage}
\end{center}
\caption{
Breakup differential cross sections
for the non-standard (NS) configurations.
The experimental data are taken from Ref.\,\citen{St89} ($nd$).
}
\label{fig11}
\end{figure}

\begin{figure}[htb]
\begin{center}
\begin{minipage}{0.48\textwidth}
\includegraphics[angle=0,width=55mm]
{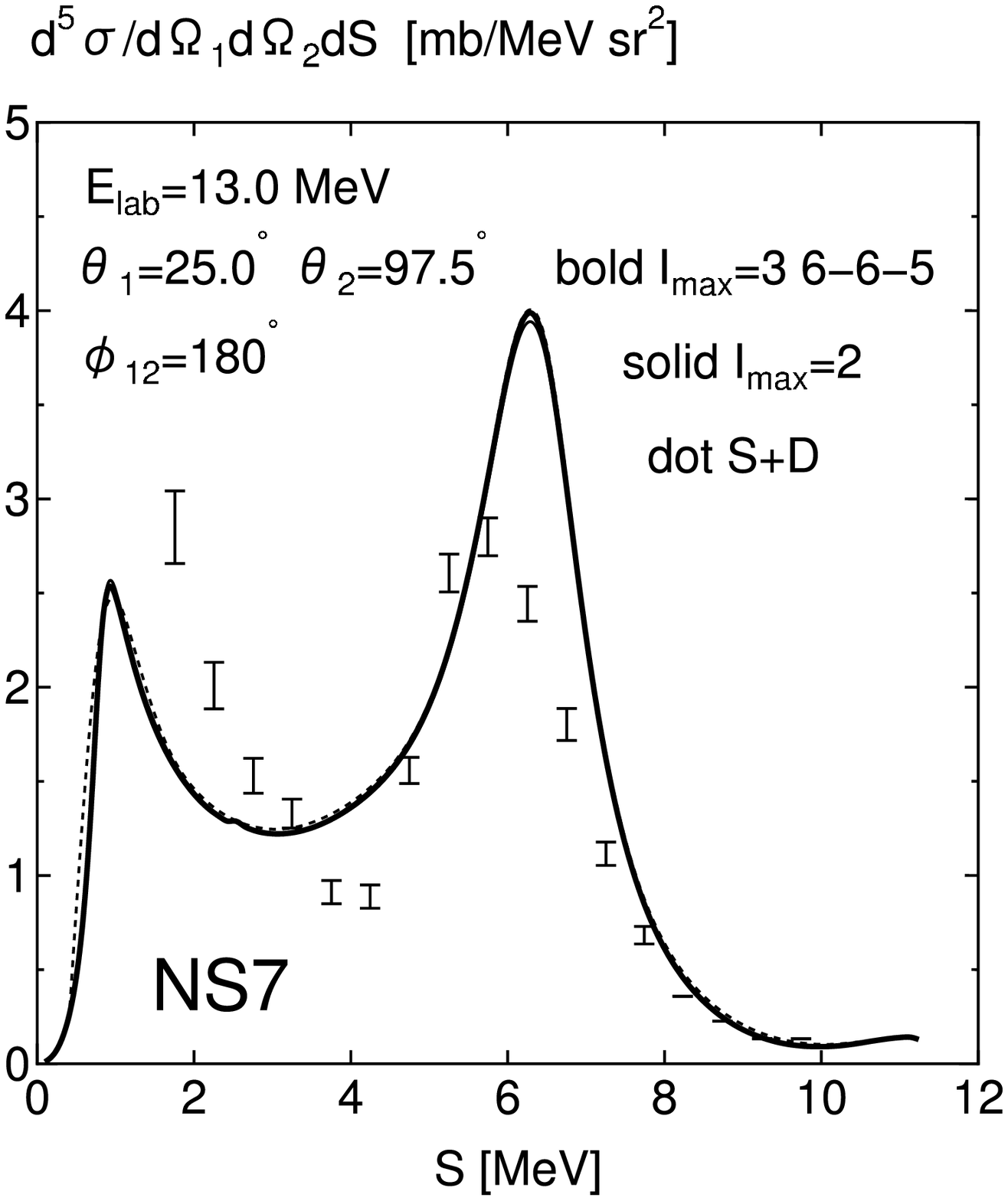}

\includegraphics[angle=0,width=55mm]
{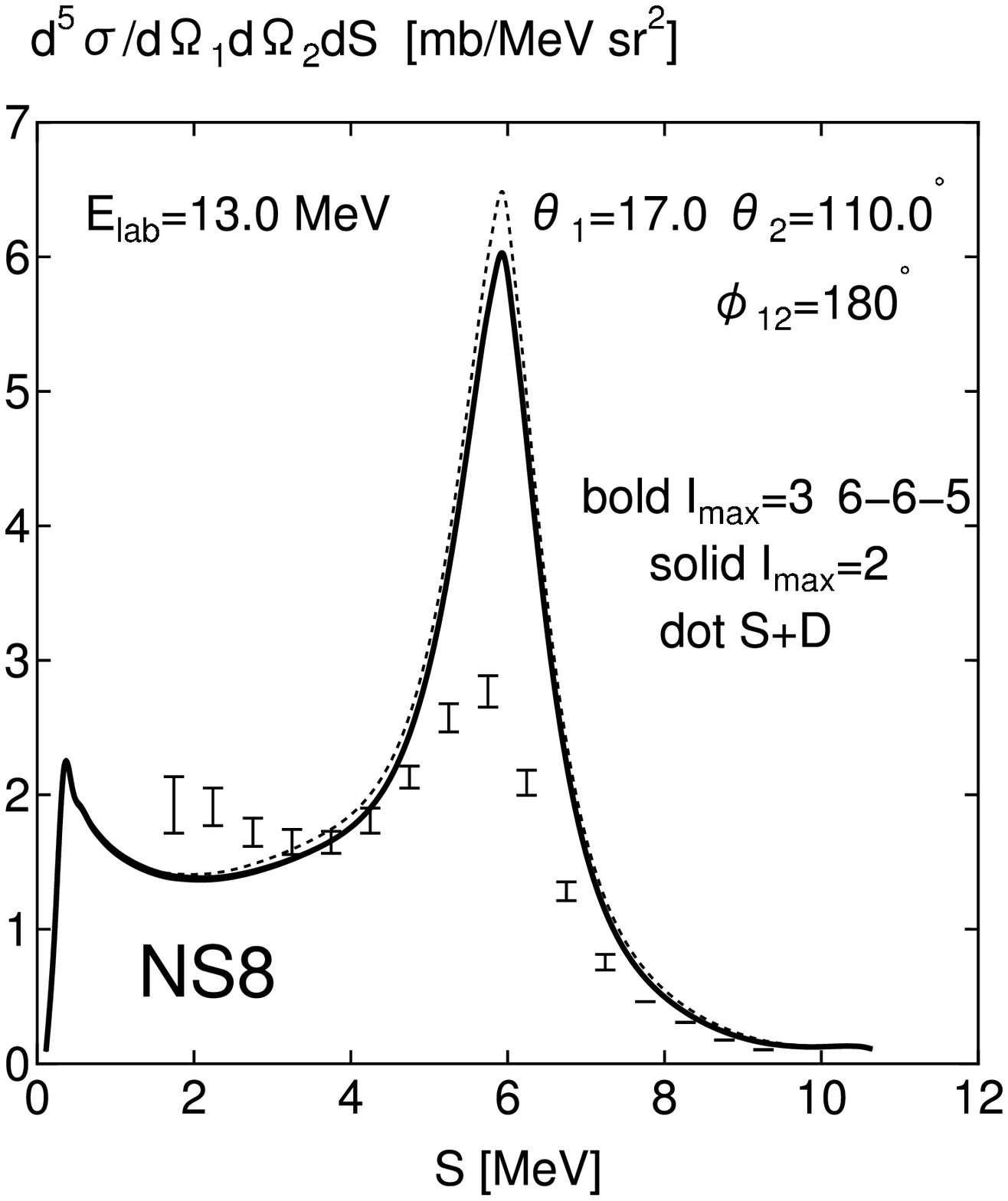}

\includegraphics[angle=0,width=55mm]
{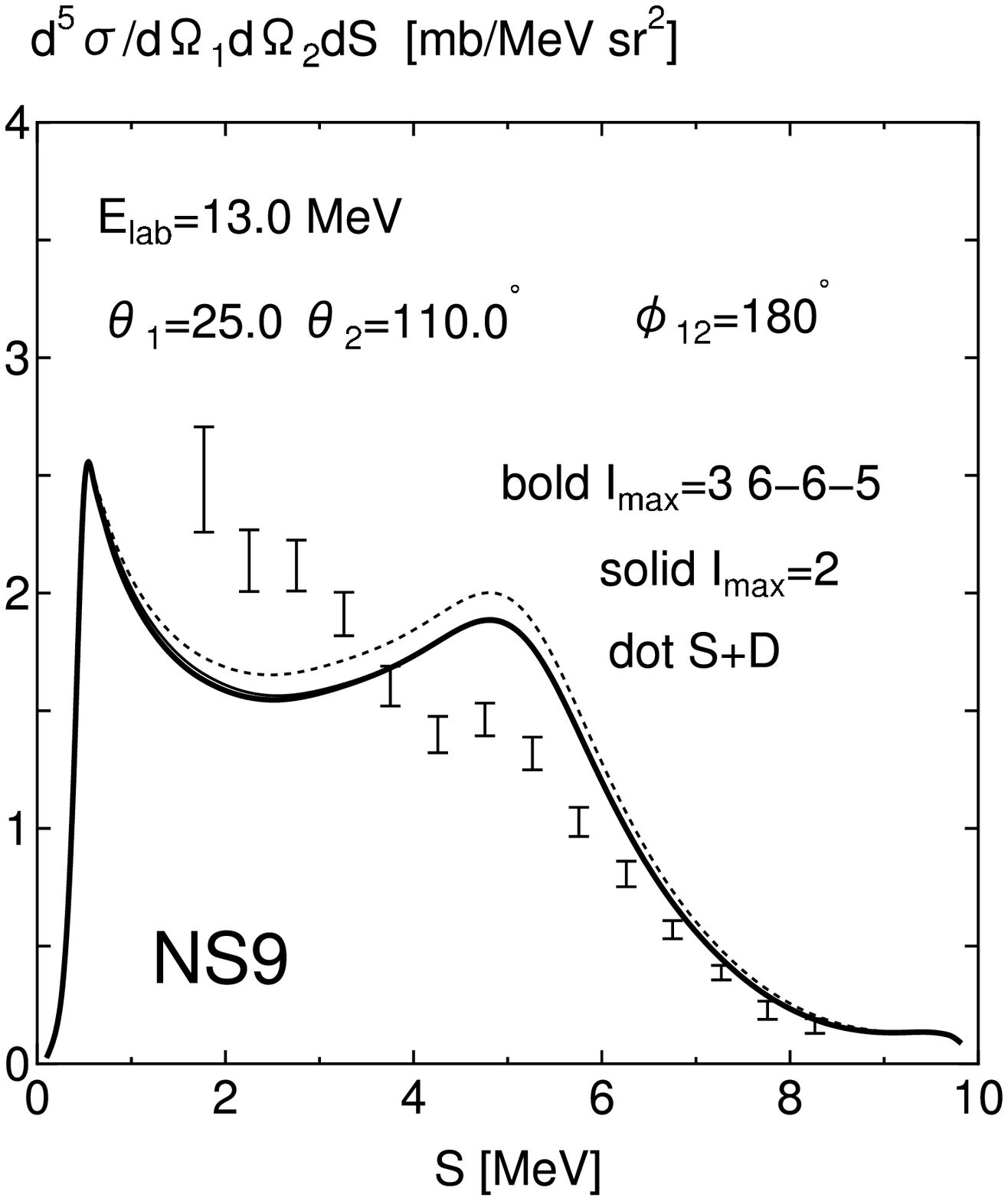}
\end{minipage}~%
\hfill~%
\begin{minipage}{0.48\textwidth}
\includegraphics[angle=0,width=55mm]
{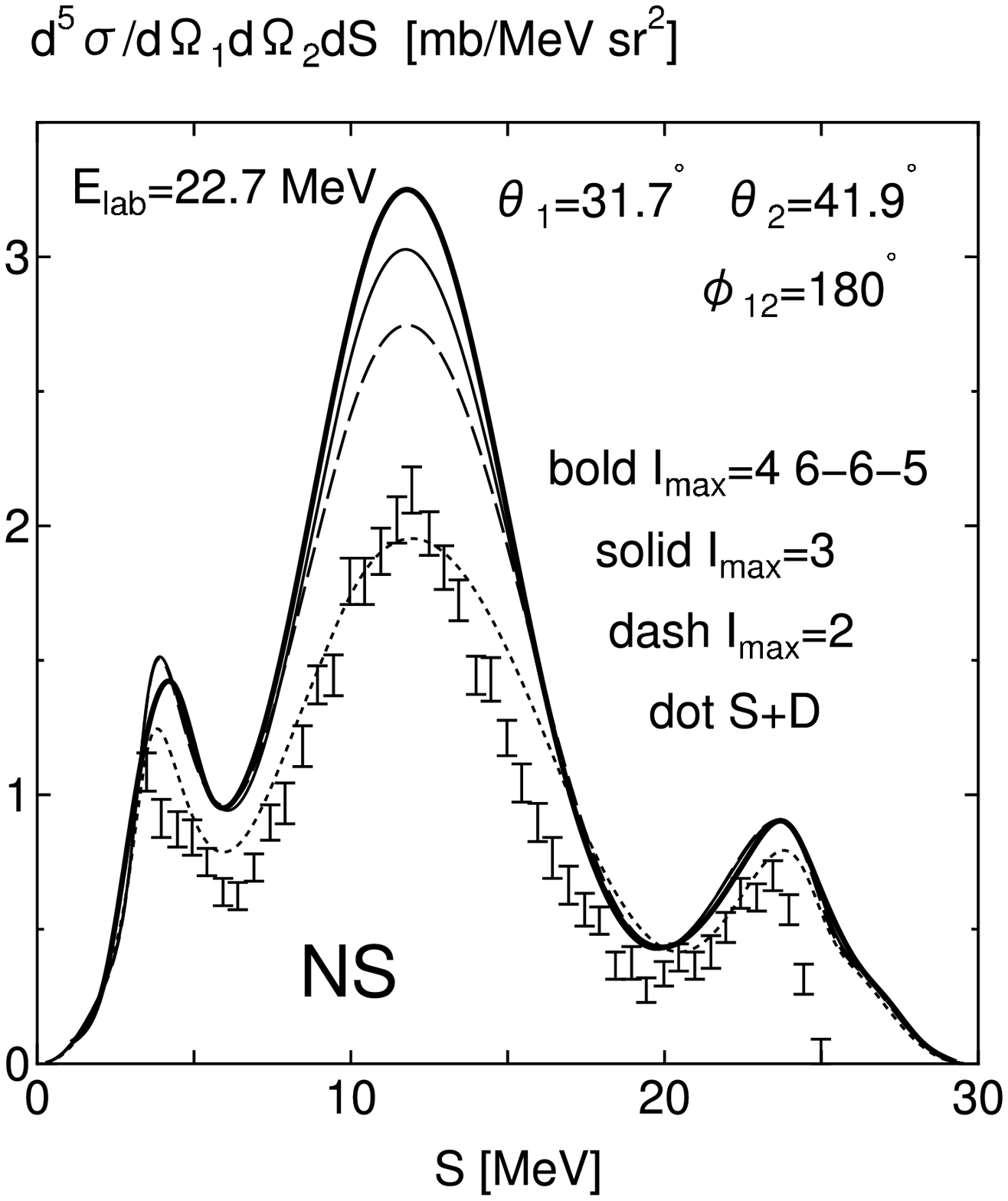}

\includegraphics[angle=0,width=55mm]
{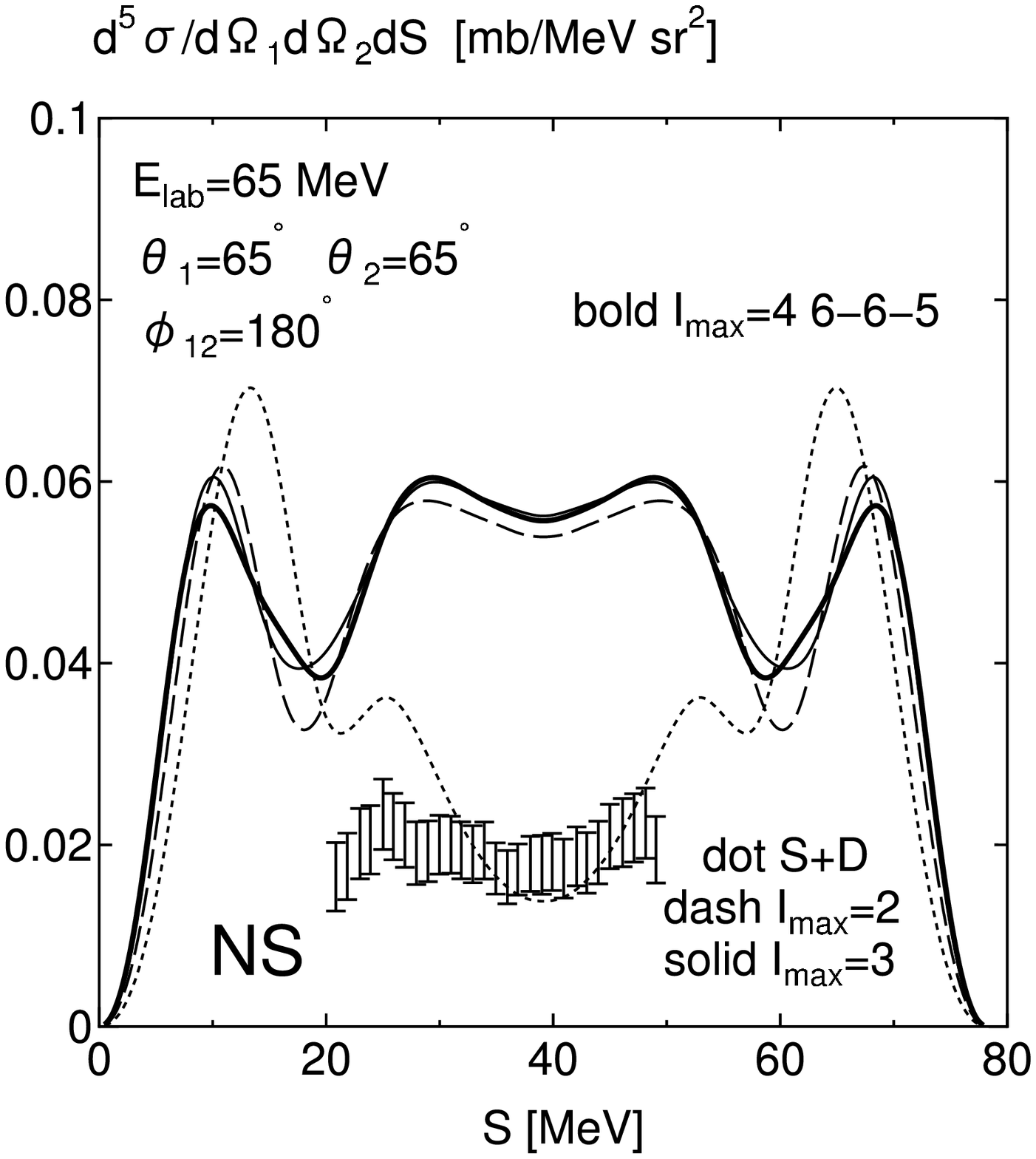}

\vspace{65mm}
\end{minipage}
\end{center}
\caption{
The same as Fig.\,\protect\ref{fig11}, but for other geometrical
configurations and energies.
The experimental data are taken from
Refs.\,\citen{St89} ($nd$) for 13 MeV, \citen{Za94} ($pd$) for 22.7 MeV,
and \citen{PREP} ($pd$) for 65 MeV.
}
\label{fig12}
\end{figure}

The comparison with the experimental data for the non-standard configurations
is shown in Figs.\,\ref{fig11} and \ref{fig12}.
Here we find big deviation from the experimental data in some cases,
again a common feature with the meson-exchange predictions.
These are the NS2 $nd$ scattering of 13 MeV, and the $pd$ scattering
of 22.7 MeV and 65 MeV.
A large number of figures, NS1 - NS9, for 13 MeV are very similar to 
the predictions by the Paris potential in Ref.\,\citen{St89}.
The huge final state interaction peaks in NS4, NS7 and NS8 are
very similar to the results by the Malfliet-Tjon potential.
In 22.7 MeV and 65 MeV cases, the calculated results are completely off the
experimental data.

\section{Summary}

In this and previous papers,\cite{ndscat1,scl10,ndscat2} we have
applied the quark-model (QM) baryon-baryon interaction fss2
to the neutron-deuteron ($nd$) scattering in the Faddeev formalism 
for composite particles.
The main motivation is to investigate the nonlocal effect of the
short-range $NN$ interaction in a realistic model, reproducing
all the two-nucleon properties and yet based on the naive three-quark
structure of the nucleons.
The calculations are carried out by the 15-point Gaussian nonlocal
potential constructed from fss2, which is accurate enough to reproduce
the converged triton binding energy of fss2 with the accuracy
of 15 keV and the $NN$ phase shift parameters with
the difference of less than $0.1^\circ$.\cite{scl10,Fuk09}
The potential keeps all the nonlocal effects of the original fss2,
including the energy-dependent term of the QM resonating-group
method (RGM).
This energy dependence is eliminated by the standard off-shell transformation
utilizing the square root of the normalization kernel
for two three-quark clusters.
It is extremely important to deal with this energy dependence properly, 
since an extra nonlocal kernel from this procedure is crucial
to reproduce all the elastic scattering observables
below $E_n \leq 65$ MeV.\cite{ndscat1,ndscat2}

In this paper, we have studied the neutron-induced deuteron breakup 
differential cross sections for the incident energies $E_n \leq 65$ MeV, 
and compared them with available experimental data 
and the predictions by meson-exchange potentials.
We have found that our calculations reproduce almost all the results
for the breakup differential cross sections
predicted by the meson-exchange potentials, including the disagreement
with the experiment.
This feature is probably related with the structure of the direct 
breakup amplitudes in Eqs.\,(\ref{fm18}) and (\ref{fm19}). 
First, they are constrained by the elastic scattering amplitudes,
$f^J_{(\ell^\prime S^\prime_c)(\ell S_c)}$, in the initial stage.
In the final stage of reactions, only the half-off shell 
two-nucleon $t$-matrix appears owing to the energy conservation 
for outgoing nucleons. The effect of the completely off-shell
$t$-matrix therefore appears only at the stage of solving the
basic AGS equations
for $\widetilde{Q}^{(\ell^\prime S^\prime_c)J}_{i \mu \gamma}$,
for which the present investigations imply that the difference between
our QM $NN$ interaction and the meson-exchange potentials is
rather minor.
On the whole, the agreement with the experimental data is fair,
but there exist some discrepancies in certain particular kinematical
configurations, which are commonly found both for our predictions 
and for meson-exchange predictions.
In particular, the space star anomaly of 13 MeV $nd$ scattering 
is not improved even in our model.
There are severe disagreement of breakup differential cross sections
in some of the non-standard configurations.
In our model, some overestimations of cross sections are found
at the energy $E_n=65$ MeV.
Since these large disagreements can be resolved neither with the Coulomb
effect nor by the introduction of the $3N$ force, 
systematic studies from more basic viewpoints for the $NN$ interaction
are still needed both experimentally and theoretically.

In spite of the apparent disagreement between the theory 
and the experiment in some of the breakup differential cross sections,
our QM baryon-baryon interaction fss2 was very successful to reproduce
almost all other experimental data of the three-nucleon system
without reinforcing it with the three-body force. 
These include: 1) a nearly correct binding energy of the triton,\cite{ren}
2) reproduction of the doublet and quartet $S$-wave scattering 
lengths, $\hbox{}^2a$ and $\hbox{}^4a$,\cite{scl10}
3) not too small differential cross sections of the $nd$ elastic
scattering up to $E_n \sim 65$ MeV 
at the diffraction minimum points,\cite{ndscat1}
4) the improved maximum height of the nucleon analyzing
power $A_y(\theta)$ in the low-energy region $E_n \leq 25$ MeV,
although not sufficient,\cite{ndscat2} and 5) the breakup differential
cross sections with many kinematical configurations, discussed in this paper.
Many of these improvements are related to the sufficiently attractive 
$nd$ interaction in the $\hbox{}^2S$ channel, in which the strong distortion
effect of the deuteron is very sensitive to the treatment of the short-range
repulsion of the $NN$ interaction. In our QM $NN$ interaction,
this part is described by the quark exchange kernel of the color-magnetic term
of the quark-quark interaction.
In the strangeness sector involving the $\Lambda N$ and $\Sigma N$ 
interactions, the effect of the Pauli repulsion on the quark level
appears in some baryon channels.
It is therefore interesting to study $\Sigma^{\pm}$-deuteron
scattering in the present framework, to find the repulsive effect
directly related to the quark degree of freedom. 
Such an experiment is planned at the J-PARC facility.\cite{Mi10}

\section*{Acknowledgements}

The authors would like to thank Professors K. Miyagawa,
H. Wita\l a, H. Kamada and S. Ishikawa for many useful comments.
We also thank Professor K. Sagara for useful information
on the space-star anomaly, which was obtained through
the systematic experiments carried out by the Kyushu university group.
This work was supported by the Grant-in-Aid for Scientific
Research on Priority Areas (Grant No.~20028003), and
by the Grant-in-Aid for the Global COE Program
``The Next Generation of Physics, Spun from Universality and Emergence'' 
from the Ministry of Education, Culture, Sports, Science 
and Technology (MEXT) of Japan.
It was also supported by the core-stage backup subsidies of Kyoto University.
The numerical calculations were carried out on Altix3700 BX2 at YITP 
in Kyoto University and on the high performance computing system,
Intel Xeon X5680, at RCNP in Osaka University.

\appendix

\section{Three-nucleon breakup kinematics} 

In this appendix, we will discuss the $3N$ breakup kinematics
used in this paper. We choose the standard set of Jacobi coordinate
in the momentum space as $\alpha=3$ and set
\begin{eqnarray}
\bp=\frac{1}{2}(\bk_1-\bk_2)\ ,\quad
\bq=\frac{1}{3}(2 \bk_3-\bk_1-\bk_2)\ ,\quad
\bK=\bk_1+\bk_2+\bk_3\ ,
\label{a1}
\end{eqnarray}
with $\bk_\alpha$ being the momentum coordinate of the particle $\alpha$
in the lab system. We also choose the $z$-axis as the direction
of the incident particle and assume that the magnitude of the
incident momentum is $q_0$ in the cm system.
This implies that the incident momentum is
always $\bq_{\rm cm}=q_0 \be_z$ and
the incident energy is $E_{\rm cm}=(3\hbar^2/4M){q_0}^2$,
either the nucleon or the deuteron is the incident particle.
Here, $\be_z$ is the unit vector of the $z$-axis.
In the lab system, the incident momentum $\bk_{\rm lab}$ and
the energy $E_{\rm lab}$ are given by
\begin{eqnarray}
& & \left\{\begin{array}{l}
\bk_{\rm lab}=\frac{3}{2} q_0 \be_z\ , \\ [2mm]
E_{\rm lab}=\frac{\hbar^2}{2M}{\bk_{\rm lab}}^2
=\frac{9\hbar^2}{8M}{q_0}^2=\frac{3}{2} E_{\rm cm} \\
\end{array}\right.
\quad \hbox{for} \quad \hbox{nucleon-incident}\ ,\nonumber \\
& & \left\{\begin{array}{l}
\bk_{\rm lab}=3 q_0 \be_z\ , \\ [2mm]
E_{\rm lab}=\frac{\hbar^2}{4M}{\bk_{\rm lab}}^2
=\frac{9\hbar^2}{4M}{q_0}^2=3 E_{\rm cm}
\end{array}\right.
\quad \hbox{for} \quad \hbox{deuteron-incident}\ .
\label{a2}
\end{eqnarray}
In the following, all quantities in the initial state are
expressed by $k_{\rm lab}=|\bk_{\rm lab}|$ and $E_{\rm cm}$,
which are determined solely by $q_0$.

In the experimental setup to detect two particles 1 and 2,
the three-particle breakup configurations
are uniquely specified with two polar angles $\theta_1$, $\theta_2$,
and a difference of azimuthal angles $\phi_{12}=\phi_1-\phi_2$,
in addition to the energy $S$ discussed below. 
These angles are defined through $\cos\,\theta_\alpha
=(\widehat{\bk}_\alpha \cdot\be_z)$ 
and ${\rm tan}\,\phi_\alpha=(k_{\alpha y}/k_{\alpha x})$.
We will choose the $x$-axis such that $\phi_1=\pi$. \cite{Oh65}
If $k_1=|\bk_1|$ and $k_2=|\bk_2|$ are determined from $S$,
$k_3=|\bk_3|$ and all other angles in the lab system are 
calculated from
\begin{eqnarray}
& & k_3=\left[(k_{\rm lab}-k_1 \cos\,\theta_1-k_2 \cos\,\theta_2)^2
+(k_1 \sin\,\theta_1)^2+(k_2 \sin\,\theta_2)^2 \right. \nonumber \\
& & \left. \qquad +2k_1 k_2\,\sin\,\theta_1\,\sin\,\theta_2
~\cos\,\phi_{12} \right]^{1/2}\ ,\nonumber \\
& & \cos\,\theta_3=\frac{1}{k_3}\left(k_{\rm lab}
-k_1 \cos\,\theta_1-k_2 \cos\,\theta_2\right)\ ,\nonumber \\
& & \sin\,\theta_3=\frac{1}{k_3}\left[ (k_1 \sin\,\theta_1)^2
+(k_2 \sin\,\theta_2)^2+2k_1 k_2\,\sin\,\theta_1\,\sin\,\theta_2
~\cos\,\phi_{12}\right]^{1/2}\ ,\nonumber \\
& & \cos\,\phi_3=\frac{1}{k_3 \sin\,\theta_3}
\left(k_1 \sin\,\theta_1+k_2\,\sin\,\theta_2\,\cos\,\phi_{12}\right)
\ ,\nonumber \\
& & \sin\,\phi_3=-\frac{1}{k_3 \sin\,\theta_3}
k_2\,\sin\,\theta_2\,\sin\,\phi_{12} \ ,
\label{a3}
\end{eqnarray}
with $\phi_1=\pi$ and $\phi_2=\pi-\phi_{12}$.
Once all the momentum vectors in the lab system
are determined, the momentum vectors in the cm system
are easily determined from the relationship,
$\bp=(1/2)(\bk_1-\bk_2)$ and $\bq=(2/3)\bk_{\rm lab}-\bk_1-\bk_2$
for $\alpha=3$, and their cyclic permutations of (123).
The basic magnitude of $q=|\bq|$ is obtained from $k_3$ in \eq{a3}
by a simple replacement of $k_{\rm lab}$ with $(2/3)k_{\rm lab}$:
\begin{eqnarray}
q & = & \left[((2/3)k_{\rm lab}-k_1 \cos\,\theta_1-k_2 \cos\,\theta_2)^2
+(k_1 \sin\,\theta_1)^2+(k_2 \sin\,\theta_2)^2 \right. \nonumber \\
& & \left. +2k_1 k_2\,\sin\,\theta_1\,\sin\,\theta_2
~\cos\,\phi_{12} \right]^{1/2}\ .
\label{a4}
\end{eqnarray}
The two-nucleon momentum $p=|\bp|$ is determined from the
energy conservation in the cm system:
\begin{eqnarray}
E=E_{\rm cm}+\varepsilon_d=\frac{\hbar^2}{M}
\left(p^2+\frac{3}{4}q^2\right)\ ,
\label{a5}
\end{eqnarray}
where $\varepsilon_d=-(3\hbar^2/4M){\kappa_d}^2$ is the deuteron
energy. It is convenient to use the threshold momentum 
$q_M=\sqrt{{q_0}^2-{\kappa_d}^2}$ for the deuteron breakup,
by which we find
\begin{eqnarray}
p=\sqrt{\frac{3}{4}({q_M}^2-q^2)} \equiv p_0\ .
\label{a6}
\end{eqnarray}
The angles of $\widehat{\bp}$ and $\widehat{\bq}$ are
obtained from
\begin{eqnarray}
& & \cos\,\theta_p=\frac{1}{2p}
\left(k_1 \cos\,\theta_1-k_2 \cos\,\theta_2\right)\ ,\nonumber \\
& & \cos\,\phi_p=\frac{1}{2p \sin\,\theta_p}
\left( k_1 \sin\,\theta_1\,\cos\,\phi_1
-k_2 \sin\,\theta_2\,\cos\,\phi_2\right)\ ,\nonumber \\
& & \sin\,\phi_p=\frac{1}{2p \sin\,\theta_p}
\left( k_1 \sin\,\theta_1\,\sin\,\phi_1
-k_2 \sin\,\theta_2\,\sin\,\phi_2\right)\ ,\nonumber \\
& & \cos\,\theta_q=\frac{1}{q}
\left( \frac{2}{3}k_{\rm lab}-k_1 \cos\,\theta_1-k_2 \cos\,\theta_2\right)
\ ,\nonumber \\
& & \cos\,\phi_q=-\frac{1}{q \sin\,\theta_q}
\left( k_1 \sin\,\theta_1\,\cos\,\phi_1
+k_2 \sin\,\theta_2\,\cos\,\phi_2\right)\ ,\nonumber \\
& & \sin\,\phi_q=-\frac{1}{q \sin\,\theta_q}
\left( k_1 \sin\,\theta_1\,\sin\,\phi_1
+k_2 \sin\,\theta_2\,\sin\,\phi_2\right)\ .
\label{a7}
\end{eqnarray}

In order to determine $k_1$ and $k_2$ from $S$,
we start from the energy conservation in the lab
system:
\begin{eqnarray}
E_{\rm lab}+\varepsilon_d=\frac{\hbar^2}{2M}
\left({\bk_1}^2+{\bk_2}^2+{\bk_3}^2 \right)\ .
\label{a8}
\end{eqnarray}
We rewrite this as
\begin{eqnarray}
{k_1}^2+{k_2}^2+k_1 k_2\,\cos\,\theta_{12}
-k_{\rm lab} (k_1\,\cos\,\theta_1+k_2\,\cos\,\theta_2)+\Delta=0\ ,
\label{a9}
\end{eqnarray}
where $\cos\,\theta_{12}=(\widehat{\bk}_1 \cdot \widehat{\bk}_2)$,
and we have defined
\begin{eqnarray}
\Delta \equiv -\frac{M}{\hbar^2}\left(E_{\rm lab}+\varepsilon_d \right)
+\frac{1}{2}{k_{\rm lab}}^2
=\left\{ \begin{array}{l}
\frac{3}{4}{\kappa_d}^2 \\ [2mm]
\frac{1}{4} {k_{\rm lab}}^2+\frac{3}{4}{\kappa_d}^2 \\
\end{array}\right.
\quad \hbox{for} \quad \left\{ \begin{array}{l}
\hbox{nucleon-incident} \\ [2mm]
\hbox{deuteron-incident} \\
\end{array}\right.\ .\nonumber \\
\label{a10}
\end{eqnarray}
We rotate $k_1$-$k_2$ plane by $45^\circ$,\cite{PREP} and parametrize
the ellipse with an angle $\theta$. In this process, it is convenient
to express $\theta_{12}$ by $\theta_0$, which is defined by
\begin{eqnarray}
\theta_0=\frac{1}{2} \hbox{Arccos}
\left(\frac{1}{2} \cos\,\theta_{12}\right)\ .
\label{a11}
\end{eqnarray}
Here, Arccos implies the principal value between 0 and $\pi$.
Note that $\theta_0$ changes from $\pi/6$ to $\pi/3$ for
the change of $\theta_{12}$ from 0 to $\pi$. If we use this $\theta_0$,
the solution of \eq{a9} is parametrized as
\begin{eqnarray}
k_1=\frac{k_{\rm lab}}{2\sqrt{2}}\,\frac{A}{\sin\,2\theta_0}
\,\CC(\theta-\theta_0)\ \ ,\qquad
k_2=\frac{k_{\rm lab}}{2\sqrt{2}}\,\frac{A}{\sin\,2\theta_0}
\,\CS(\theta-\theta_0)\ ,
\label{a12}
\end{eqnarray}
with
\begin{eqnarray}
\CC(\theta)=\cos\,\theta+\CC_0\ \ ,\qquad
\CS(\theta)=\sin\,(\theta+2\theta_0-\pi/2)+\CS_0\ ,
\label{a13}
\end{eqnarray}
and
\begin{eqnarray}
\ \hspace{-10mm}
\CC_0=\frac{\sqrt{2}}{A}\,\frac{
\cos\,\theta_1-\cos\,2\theta_0\,\cos\,\theta_2}{\sin\,2\theta_0}\ ,
\quad \CS_0=\frac{\sqrt{2}}{A}\,\frac{
\cos\,\theta_2-\cos\,2\theta_0\,\cos\,\theta_1}{\sin\,2\theta_0}\ .
\label{a14}
\end{eqnarray}
In \eq{a12}, we have defined
\begin{eqnarray}
A=\sqrt{2}\left\{
\frac{(\cos\,\theta_1)^2+(\cos\,\theta_2)^2-2\cos\,2\theta_0\,\cos\,\theta_1
\,\cos\,\theta_2}{(\sin\,2\theta_0)^2}
-\frac{4\Delta}{{k_{\rm lab}}^2} \right\}^{\frac{1}{2}}\ .
\label{a15}
\end{eqnarray}
We measure the arc length $S$ in the $E_1$-$E_2$ plane counterclockwise,
starting from a certain starting point $\theta_{\rm st}$.
The expression $S(\theta)$ is obtained by integrating 
$d\,S=\sqrt{(d\,E_1)^2+(d\,E_2)^2}$:
\begin{eqnarray}
S(\theta)=\frac{\hbar^2{k_{\rm lab}}^2}{2M} \left(\frac{A}{2\,\sin\,2\theta_0}
\right)^2
\int^\theta_{\theta_{\rm st}} d\,\theta~f(\theta-\theta_0)\ ,
\label{a16}
\end{eqnarray}
with
\begin{eqnarray}
f(\theta)=\sqrt{\left(\CC(\theta)\,\sin\,\theta \right)^2
+\left( \CS(\theta)\,\cos\,(\theta+2\theta_0-\pi/2)\right)^2}\ .
\label{a17}
\end{eqnarray}
Note that $S(\theta)$ is a monotonically increasing function
of $\theta$, satisfying $S(\theta_{\rm st})=0$ and 
$S(\theta+2\pi)=S(2\pi)+S(\theta)$.

For the practical calculation, we first discretize the integral region
of \eq{a16} into small intervals
by
\begin{eqnarray}
\theta_\mu=\theta_{\rm st}+\mu \frac{2\pi}{N}\ ,
\label{a18}
\end{eqnarray}
with $\mu=1$ - $N$. The number of discretization points $N$ is 
typically $N=50$. 
We use third order spline interpolation for $\theta$,
\begin{eqnarray}
f(\theta-\theta_0)=\sum^N_{\nu=1} S^{(\kappa)}_\nu (\theta)
\,f(\theta_\nu-\theta_0)\ ,
\label{a18-1}
\end{eqnarray}
with
\begin{eqnarray}
S^{(\kappa)}_\nu (\theta)=\sum^3_{m=0} \alpha^{\kappa (m)}_\nu
(\theta-\theta_\kappa)^m
\qquad  \hbox{for} \qquad \theta \in [\theta_{\kappa-1}, \theta_\kappa]\ .
\label{a20}
\end{eqnarray}
If we integrate \eq{a16} over $\theta$ from $\theta_{\rm st}$ to $\theta_\mu$
by using Eqs.\,(\ref{a18-1}) and (\ref{a20}), 
we can carry out the $\theta$-integral
analytically and obtain
\begin{eqnarray}
S_\mu & = & S(\theta_\mu)=\frac{\hbar^2{k_{\rm lab}}^2}{2M} 
\left(\frac{A}{2\,\sin\,2\theta_0}\right)^2 \nonumber\\
& & \times 
\sum^3_{m=0} (-1)^{m+1} \frac{1}{m+1}\left(\frac{2\pi}{N}\right)^{m+1}
\sum^N_{\nu=1}\sum^\mu_{\kappa=1} \alpha^{\kappa (m)}_\nu
f(\theta_\nu-\theta_0)\ .
\label{a21}
\end{eqnarray}
To obtain the angle $\theta$ from the arc length $S$, we again
use the spline interpolation technique,
\begin{eqnarray}
\theta (S)=\sum^N_{\mu=1} S_\mu(S)\,\theta(S_\mu)\ .
\label{a22}
\end{eqnarray}
Here, $S_\mu (S)$ is the third order spline function
for the mesh points $\{S_\mu\}=[S_0, S_1, S_2,$
$\cdots , S_N]$ with $S_0=0$.
From \eq{a22} with $\theta(S_\mu)=\theta_\mu$, we obtain
\begin{eqnarray}
\theta (S)=\theta_{\rm st}+\frac{2\pi}{N}\sum^N_{\mu=1} \mu\,S_\mu(S)\ ,
\label{a23}
\end{eqnarray}
where \eq{a18} and $\sum^N_{\mu=1} S_\mu(S)=1$ are used.
We therefore only need to calculate the sum in \eq{a21}
and prepare the coefficients of the spline interpolation
for the mesh points $\{S_\mu\}$.

The starting angle $\theta$ is selected as follows.
Let us first consider the nucleon-incident reaction.
In this case, it is convenient to define the angle $\theta^{(-)}$ 
through \cite{PREP}
\begin{eqnarray}
\cos\,\theta^{(-)}=\frac{2\sqrt{\Delta}}{k_{\rm lab}}
=\frac{2}{\sqrt{3}}\,\frac{\kappa_d}{q_0}\ .
\label{a24}
\end{eqnarray}
If we assume $k_2$=0 in \eq{a9}, we find that there are
two non-negative solutions 
\begin{eqnarray}
k^{(\pm)}_1=\frac{k_{\rm lab}}{2} \left[
\cos\,\theta_1 \pm \sqrt{(\cos\,\theta_1)^2
-(\cos\,\theta^{(-)})^2}\right]\ ,
\label{a25}
\end{eqnarray}
only when $\theta_1 < \theta^{(-)}$.
We choose the larger value $k^{(+)}_1$ as
the starting point to measure $S$. In this case,
we can easily find that the corresponding $\theta$ value
is given by
\begin{eqnarray}
\theta_{\rm st}=\hbox{Arccos}\,\CS_0-\theta_0\ ,
\label{a26}
\end{eqnarray}
where $\CS_0$ is given in \eq{a14}.
When $\theta_1 > \theta^{(-)}$, we have two cases.
In the case of $\theta_2 \leq \theta^{(-)}$, we choose 
$k_1=0$ and 
\begin{eqnarray}
k^{(-)}_2=\frac{k_{\rm lab}}{2} \left[
\cos\,\theta_2 - \sqrt{(\cos\,\theta_2)^2
-(\cos\,\theta^{(-)})^2}\right]\ ,
\label{a27}
\end{eqnarray}
as the starting point with
\begin{eqnarray}
\theta_{\rm st}=\hbox{Arccos}\,\CC_0+\theta_0-\pi\ .
\label{a28}
\end{eqnarray}
In the case of $\theta_2 > \theta^{(-)}$, the ellipse does not
cross over either $k_1$ or $k_2$ axis. We therefore use
the smaller value of $k_1=k_2$ as the starting point.
This condition yields
\begin{eqnarray}
\theta_{\rm st}=\hbox{Arccos}\left(
\frac{1}{A}\,\frac{\cos\,\theta_1-\cos\,\theta_2}{\sqrt{2}\,\sin\,\theta_0}
\right)+\pi\ ,
\label{a29}
\end{eqnarray}
and
\begin{eqnarray}
k^{(-)}_1=k^{(-)}_2=\frac{k_{\rm lab}}{4\,\cos\,\theta_0}
\left[ \frac{\cos\,\theta_1+\cos\,\theta_2}{2\,\cos\,\theta_0}
-\sqrt{ \left(\frac{\cos\,\theta_1+\cos\,\theta_2}{2\,\cos\,\theta_0}
\right)^2-\frac{4\Delta}{{k_{\rm lab}}^2}}\right]\ .\nonumber \\
\label{a30}
\end{eqnarray}

When the deuteron is an incident particle, there is 
no crossing point across either the $k_1$-axis or the $k_2$-axis.
We follow the definition of the Correll et al.'s paper \cite{Co87},
that discusses the deuteron incident reaction around the
collinear configurations, choosing the collinear point as the
starting point to measure $S$. The collinear point $\theta_c$ is defined
as the configuration with $\bq=(2/3) \bk_{\rm lab}-(\bk_1+\bk_2)=0$.
To find the corresponding $\theta_{\rm st}=\theta_c$,
we assume $\phi_{12}=\pi$ ($\phi_1=\pi$, $\phi_2=0$) and
the $x$-$z$ plane as the reaction plane, just as the experimental setup.
Under this assumption, the magnitude $\bq^2$ is expressed as
\begin{eqnarray}
\bq^2=\left(k_1\,\cos\,\theta_1+k_2\,\cos\,\theta_2
-\frac{2}{3} k_{\rm lab}\right)^2
+\left(k_1\,\sin\,\theta_1-k_2\,\sin\,\theta_2\right)^2\ ,
\label{a31}
\end{eqnarray}
resulting in the two conditions
\begin{eqnarray}
k_1\,\cos\,\theta_1+k_2\,\cos\,\theta_2=\frac{2}{3} k_{\rm lab}\ \ ,\qquad
k_1\,\sin\,\theta_1=k_2\,\sin\,\theta_2\ .
\label{a32}
\end{eqnarray}
If $\theta_1=\theta_2$, we assume $k_1=k_2$ and find
$\theta_c=\pi/2$ with
\begin{eqnarray}
k_1=k_2=\frac{k_{\rm lab}}{4}
\left( \frac{\cos\,\theta_1}{(\cos\,\theta_0)^2}
+\frac{A}{\sqrt{2}\,\cos\,\theta_0} \right)\ .
\label{a33}
\end{eqnarray}
This corresponds to the $k^{(+)}_1=k^{(+)}_2$ case of \eq{a30} with
the opposite sign for the second term.
In the general case, the three conditions of \eq{a32} and the energy
conservation in \eq{a9} are not always simultaneously satisfied.
(Note that we only need two conditions to determine $k_1$ and $k_2$.)
We take the following procedure to determine $\theta_c$.
Let us use the notation
\begin{eqnarray}
a =\frac{\cos\,\theta_1-\cos\,\theta_2}
{\sqrt{2}\,\sin\,\theta_0}\ \ ,\qquad
b =\frac{\cos\,\theta_1+\cos\,\theta_2}
{\sqrt{2}\,\cos\,\theta_0}\ ,
\label{a34}
\end{eqnarray}
to simplify the expressions. We have 
$a^2+b^2=A^2+(8\Delta/{k_{\rm lab}}^2)$ and define a new angle
$\alpha$ by \footnote{By using $\alpha$, $\CC_0$ and $\CS_0$
in \protect\eq{a14} are expresses
as $\CC_0=\sqrt{1+\varepsilon^2}\,\cos\,(\alpha-\theta_0)$
and $\CS_0=\sqrt{1+\varepsilon^2}\,\sin\,(\alpha+\theta_0-\pi/2)$ 
with $\varepsilon=(2\sqrt{2\Delta}/k_{\rm lab} A)$.}
\begin{eqnarray}
\cos\,\alpha=\frac{a}{\sqrt{a^2+b^2}}\ \ ,\qquad
\sin\,\alpha=\frac{b}{\sqrt{a^2+b^2}}\ .
\label{a35}
\end{eqnarray}
We consider, $I(\theta)=k_1 \cos\,\theta_1
+k_2 \cos\,\theta_2$, as a function of $\theta$,
by using \eq{a12} and others. We find
\begin{eqnarray}
\ \hspace{-10mm} I(\theta) & = & k_1 \cos\,\theta_1
+k_2 \cos\,\theta_2=\frac{1}{4} k_{\rm lab} \sqrt{a^2+b^2}
\left[ A\,\cos\,(\theta-\alpha)+\sqrt{a^2+b^2}\right]\ .
\label{a36}
\end{eqnarray}
The crossing point with $I(\theta)=(2/3)k_{\rm lab}$ is
found only when the condition
\begin{eqnarray}
\frac{1}{A}\left| \frac{8}{3}\frac{1}{\sqrt{a^2+b^2}}
-\sqrt{a^2+b^2} \right| \leq 1\ ,
\label{a37}
\end{eqnarray}
is satisfied. The solution $\theta=\theta_c$ is found as
\begin{eqnarray}
\theta_c=\alpha \pm {\rm Arccos}\,\frac{1}{A}
\left[ \frac{8}{3}\frac{1}{\sqrt{a^2+b^2}}
-\sqrt{a^2+b^2} \right]\ ,
\label{a38}
\end{eqnarray}
and a unique point is determined when the equality is satisfied
in \eq{a37}.
Next, we examine the condition, $k_1\,\sin\,\theta_1=k_2\,\sin\,\theta_2$, is
satisfied or not for the two solutions of \eq{a38}.
The one satisfying this condition is the collinear point with $\bq=0$
from \eq{a31}. If both solutions satisfy the condition, we choose
the smaller one for $\theta_c$. If neither of the solution satisfies 
the condition, there is no exact collinear point. In this case,
we minimize \eq{a31} with respect to $\theta$ in the interval bounded by 
the two solutions of \eq{a38}.


\section{Isospin factors for the breakup amplitudes}

In this appendix, we extend the definition of the spin factors
\begin{eqnarray}
\langle \widetilde{\Gamma}_\sigma |(P^\sigma_{(123)})^\alpha|
\Gamma_\sigma \rangle
= \delta_{\widetilde{S}, S}\,\delta_{\widetilde{S}_z, S_z}
\left\{
\begin{array}{c}
(-1)^{1+s} X^S_{\widetilde{s}, s} \\ [2mm]
(-1)^{1+\widetilde{s}} X^S_{\widetilde{s}, s} \\ [2mm]
\delta_{\widetilde{s}, s} \\
\end{array} \right. \quad 
\hbox{for} \quad \alpha= \left\{ \begin{array}{c}
1 \\ [2mm]
2 \\ [2mm]
3 \\
\end{array} \right.\ ,
\label{b1}
\end{eqnarray}
to the isospin factors and calculate the matrix elements
$\langle \widetilde{\Gamma_\tau}|\CO^{\alpha \beta}_\tau|\Gamma_\tau \rangle$
in \eq{fm13}.
The most convenient definition of the isospin factors is probably
\begin{eqnarray}
& & \langle \widetilde{\Gamma}_\tau |
\CO^{\alpha \beta}_\tau| \Gamma_\tau \rangle
=\langle (\widetilde{t}\H)\H T_z|
(P^\tau_{(123)})^{3-\alpha} \CO_\tau
(P^\tau_{(123)})^\beta|
(t\H)\H T_z \rangle \nonumber \\
& & = \left\{
\begin{array}{c}
(-1)^{1+t} X^{\tau (\alpha \beta)}_{\widetilde{t}, t} \\ [2mm]
(-1)^{1+\widetilde{t}} X^{\tau (\alpha \beta)}_{\widetilde{t}, t} \\ [2mm]
X^{\tau (\alpha \alpha)}_{\widetilde{t}, t } \\
\end{array} \right. \quad
\hbox{for} \quad \beta-\alpha=\left\{ \begin{array}{c}
1 \\ [2mm]
2 \\ [2mm]
3 \\
\end{array} \right. \quad \hbox{in~(mod~3)}\ ,
\label{b2}
\end{eqnarray}
which yields the results in \eq{fm14}.
If we set $\CO_\tau=1$, all the factors
$X^{\tau (\alpha \beta)}_{\widetilde{t}, t}$ in \eq{b2}
are reduced to $X^{1/2}_{\widetilde{t}, t}$ since
$\CO^{\alpha \beta}_\tau=(P^\tau_{(123)})^{\beta-\alpha}$.
Here, $X^{1/2}$ is the common matrix with
the spin factors given by\cite{scl10}
\begin{eqnarray}
\left(X^\H_{s, s^\prime}\right)=\left( \begin{array}{cc}
\frac{1}{2} & -\frac{\sqrt{3}}{2} \\ [2mm]
-\frac{\sqrt{3}}{2} & -\frac{1}{2} \\
\end{array}\right)\ \ ,\qquad
\left(X^\3H_{s, s^\prime}\right)=\left( \begin{array}{cc}
0 & 0 \\
0 & 1 \\
\end{array}\right)\ .
\label{b3}
\end{eqnarray}
In \eq{b3}, the upper row (the left-most column) corresponds
to $s=0$ ($s^\prime=0$) and the second row (the right-most column)
corresponds to $s=1$ ($s^\prime=1$).

In order to calculate $X^{\tau (\alpha \beta)}_{\widetilde{t}, t}$,
we only need $\langle \widetilde{t}|\CO_\tau|t \rangle$ with
$|t\rangle=|(t\H)\H T_z \rangle$, since $P^\tau_{(123)}$ does not
change the total isospin $T=1/2$. Furthermore, $\CO_\tau$ in \eq{fm10} are
expressed by the symmetric isospin operator $t_z=(\tau_z(1)+\tau_z(2))/2$ and
the antisymmetric operator $t^a_z=(\tau_z(1)-\tau_z(2))/2$ for
the two-particle states $\eta_t(1,2)$.
The former does not change $t=0$ or 1, while the latter flips the isospin
value. For $pp$ or $nn$, the non-zero matrix elements are only
for $\widetilde{t}=t=1$, but $np$ and $pn$ contains $t^a_z$.
However, we only need to calculate the sum of $np$ and $pn$ contributions. 
Furthermore, \eq{fm5} tells us that $np$ and $pn$ give the
same contribution owing to the permutation operator $(1+P)$.
We can therefore assume $\CO_\tau$ in \eq{b2} as
\begin{eqnarray}
\CO^{pp}=\frac{1}{2}t_z(1+t_z)\ ,\quad 
\CO^{nn}=\frac{1}{2}t_z(-1+t_z)\ ,\quad
\CO^{pn}+\CO^{np}=1-t^2_z\ .
\label{b4}
\end{eqnarray}
If we decompose $t^2_z$ into the rank 0, 1, and 2 tensors
as $t^2_z=(1/3)\bt^2+\sqrt{2/3}[t\,t]^{(2)}_0$,
we immediately find that the rank 2 tensor does not contribute
since the total isospin in our case is $T=1/2$.
We can therefore replace $t^2_z$ in \eq{b4} with $(1/3)\bt^2$,
resulting in 
\begin{eqnarray}
& & \langle \widetilde{t}|\CO^{pp}|t \rangle
=\langle \widetilde{t}|\CO^{nn}|t \rangle
=\delta_{\widetilde{t}, t}\,\delta_{t,1}\frac{2}{3}\ ,\nonumber \\
& & \langle \widetilde{t}|\CO^{pn}|t \rangle
+\langle \widetilde{t}|\CO^{np}|t \rangle
=\delta_{\widetilde{t}, t}\,
\left(\delta_{t,0}+\frac{1}{3}\delta_{t,1}\right)\ .
\label{b5}
\end{eqnarray}
It is convenient to introduce the isospin projection operators
$P_0=(1-\bftau(1)\cdot \bftau(2))/4$ and
$P_1=(3+\bftau(1)\cdot \bftau(2))/4$, and define
\begin{eqnarray}
& & X^{\tau (\alpha \beta)}_{\widetilde{t},t} \nonumber \\
& & =\left\{ \begin{array}{c}
(-1)^{1+t} \langle \widetilde{t}|(P^\tau_{(123)})^{3-\alpha}
P_\tau (P^\tau_{(123)})^\beta|t \rangle \\ [2mm]
(-1)^{1+\widetilde{t}} \langle \widetilde{t}|(P^\tau_{(123)})^{3-\alpha}
P_\tau (P^\tau_{(123)})^\beta|t \rangle \\ [2mm]
\langle \widetilde{t}|(P^\tau_{(123)})^{3-\alpha}
P_\tau (P^\tau_{(123)})^\alpha|t \rangle \\
\end{array}\right. \quad
\hbox{for} \quad \beta-\alpha=\left\{\begin{array}{c}
1 \\ [2mm]
2 \\ [2mm]
3 \\
\end{array} \right. \quad \hbox{in~(mod~3)}\ ,\nonumber \\
\label{b6}
\end{eqnarray}
for $\tau=0$ and 1. Since $P_\tau$ is expressed as
$P_\tau=|\tau \rangle \langle \tau|$ in our model space, 
the matrix elements $\langle \widetilde{t}|(P^\tau_{(123)})^{3-\alpha}
P_\tau (P^\tau_{(123)})^\beta|t \rangle
=\langle \widetilde{t}|(P^\tau_{(123)})^{3-\alpha}|\tau \rangle
\langle \tau|(P^\tau_{(123)})^\beta|t \rangle$ can be easily
calculated from \eq{b1}.
The correspondence
\begin{eqnarray}
\CO^{pp},~\CO^{nn} \sim \frac{2}{3} P_1\ ,\quad 
\CO^{pn}+\CO^{np} \sim P_0+\frac{1}{3} P_1\ ,
\label{b7}
\end{eqnarray}
from \eq{b5} yields
\begin{eqnarray}
X^{pp},~X^{nn}=\frac{2}{3} X^{1(\alpha \beta)}\ ,\quad 
X^{pn}+X^{np}=X^{0(\alpha \beta)}
+\frac{1}{3} X^{1(\alpha \beta)}\ ,
\label{b8}
\end{eqnarray}
in the matrix form.
Here $X^{\tau(\alpha \beta)}$ with $\tau=0$ and 1 are given by

\bigskip

\noindent
($\tau=0$ factors)
\begin{eqnarray}
& & X^{0(11)}=\left( \begin{array}{cc}
\frac{1}{4} & \frac{\sqrt{3}}{4} \\ [2mm]
\frac{\sqrt{3}}{4} & \frac{3}{4} \\
\end{array}\right)\ ,\quad
X^{0(22)}=\left( \begin{array}{cc}
\frac{1}{4} & -\frac{\sqrt{3}}{4} \\ [2mm]
-\frac{\sqrt{3}}{4} & \frac{3}{4} \\
\end{array}\right)\ ,\quad
X^{0(33)}=\left( \begin{array}{cc}
1 & 0 \\
0 & 0 \\
\end{array}\right)\ ,\nonumber \\ [2mm]
& & X^{0(12)}=X^{0(21)}=\left( \begin{array}{cc}
-\frac{1}{4} & -\frac{\sqrt{3}}{4} \\ [2mm]
-\frac{\sqrt{3}}{4} & -\frac{3}{4} \\
\end{array}\right)\ ,\nonumber \\ [2mm]
& & X^{0(23)}=\hbox{}^t X^{0(32)}=X^{0(13)}=\hbox{}^t X^{0(31)}
=\left( \begin{array}{cc}
\frac{1}{2} & 0 \\ [2mm]
-\frac{\sqrt{3}}{2} & 0 \\
\end{array}\right)\ .
\label{b9}
\end{eqnarray}
%


\noindent
($\tau=1$ factors)
\begin{eqnarray}
& & X^{1(11)}=\left( \begin{array}{cc}
\frac{3}{4} & -\frac{\sqrt{3}}{4} \\ [2mm]
-\frac{\sqrt{3}}{4} & \frac{1}{4} \\
\end{array}\right)\ ,\quad
X^{1(22)}=\left( \begin{array}{cc}
\frac{3}{4} & \frac{\sqrt{3}}{4} \\ [2mm]
\frac{\sqrt{3}}{4} & \frac{1}{4} \\
\end{array}\right)\ ,\quad
X^{1(33)}=\left( \begin{array}{cc}
0 & 0 \\
0 & 1 \\
\end{array}\right)\ ,\nonumber \\ [2mm]
& & X^{1(12)}=X^{1(21)}=\left( \begin{array}{cc}
\frac{3}{4} & -\frac{\sqrt{3}}{4} \\ [2mm]
-\frac{\sqrt{3}}{4} & \frac{1}{4} \\
\end{array}\right)\ ,\nonumber \\ [2mm]
& & X^{1(23)}=\hbox{}^t X^{1(32)}=X^{1(13)}=\hbox{}^t X^{1(31)}
=\left( \begin{array}{cc}
0 & -\frac{\sqrt{3}}{2} \\ [2mm]
0 & -\frac{1}{2} \\
\end{array}\right)\ .
\label{b10}
\end{eqnarray}
%


\newcommand{\etal}{{\em et al.}}

\end{document}